\documentclass[a4paper,12pt]{article}
\usepackage{amsmath,amssymb,amsthm}
\usepackage{latexsym,graphicx,color,bbm,subfigure}

\newcommand{\NF}{N_{\rm f}}

\newcommand{\beq}{\begin{eqnarray}}
\newcommand{\eeq}{\end{eqnarray}}
\newcommand{\non}{\nonumber\\}
\newcommand{\D}{\mathcal{D}}
\newcommand{\p}{\partial}
\newcommand{\Tr}{{\rm Tr}}
\newcommand{\diag}{{\rm diag}}

\begin{document}

\thispagestyle{empty}
\begin{flushright}
IFUP-TH/2010-15
\end{flushright}
\vspace{10mm}
\begin{center}
{\Large \bf Fractional and semi-local\\[5pt]
non-Abelian Chern-Simons vortices} 
\\[15mm]
{Sven Bjarke~{\sc Gudnason}}\footnote{\it e-mail address:
gudnason(at)df(dot)unipi(dot)it}
\vskip 6 mm

\bigskip\bigskip
{\it
Department of Physics, E.~Fermi, University of Pisa, \\
INFN, Sezione di Pisa,\\
Largo Bruno Pontecorvo, 3, Ed. C, 56127 Pisa, Italy
}

\bigskip
\bigskip

{\bf Abstract}\\[5mm]
{\parbox{14cm}{\hspace{5mm}
\small
In this paper we study fractional as well as semi-local
Chern-Simons vortices in $G=U(1)\times SO(2M)$ and $G=U(1)\times
USp(2M)$ theories. 
The master equations are solved numerically using appropriate
Ans\"atze for the moduli matrix field. In the fractional case the
vortices are solved in the transverse plane due to the broken axial 
symmetry of the configurations (i.e.~they are non-rotational
invariant). 
It is shown that unless the fractional vortex-centers are all
coincident (i.e.~local case) the ring-like flux structure,
characteristic of Chern-Simons vortices, will become bell-like fluxes
-- just as those of the standard Yang-Mills vortices. 
The asymptotic profile functions are calculated in all cases and the
effective size is identified. 
}}
\end{center}
\newpage
\pagenumbering{arabic}
\setcounter{page}{1}
\setcounter{footnote}{0}
\renewcommand{\thefootnote}{\arabic{footnote}}
%%%%%%%%%%%%%%%%%%%%%%%%%%%%%%%%

\section{Introduction}

Solitons play a crucial role in a vast area of physics, from solid
state physics, through high-energy physics, string theory, condensed
matter physics to cosmology. 
One of the most well-known solitons is the Abrikosov-Nielsen-Olesen
(ANO) vortex \cite{Abrikosov:1956sx,Nielsen:1973cs} which is
measurable in the laboratory forming a so-called Abrikosov lattice in
type II superconducting materials in the presence of an external
magnetic field. 

In the recent years, there has been developed a genuine non-Abelian
generalization of this vortex \cite{Hanany:2003hp,Auzzi:2003fs}, which
possesses orientational modes and a moduli space of solutions.
These vortices correspond to 1/2 BPS objects in ${\cal N}=2$
gauge theories and the BPS properties have as a consequence that all
static inter-vortex forces cancel out exactly -- just as in the
Abelian case. 
This development started out with the unitary gauge
groups, which first in the last three years has been extended to
arbitrary gauge groups \cite{Eto:2008yi} and in particular the
orthogonal \cite{Ferretti:2007rp,Eto:2009bg} and symplectic groups
\cite{Eto:2009bg}, viz.~$G=U(1)\times SO(N)$ and 
$G=U(1)\times USp(2M)$. The overall $U(1)$ factor is of utter 
importance for the topological construction, giving rise to the
stability supported by $\pi_1(G)\sim \mathbb{Z}$.\footnote{In the
  $SO$ case, there is an additional
  $\mathbb{Z}_2=\pi_1\left((U(1)\times SO(N))/\mathbb{Z}_{n_0}\right)$
  factor (with $n_0=2,1$ for even and odd 
  $N$, respectively), i.e.~a topological charge not being the vortex
  number, which has important consequences for the connectedness
  properties of the moduli spaces of vortices \cite{Eto:2009bg}. } 
The moduli space of these vortices has been studied much in detail in
Refs.~\cite{Hanany:2003hp,Auzzi:2003fs,Eto:2005yh,Auzzi:2005gr,Eto:2006cx,Auzzi:2010jt} 
for $U(N)$ and in Ref.~\cite{Eto:2009bg} for $U(1)\times
SO(N),U(1)\times USp(2M)$. 
The vacuum of the $U(N)$ theory is a unique color-flavor locked phase
when the minimal number of flavors (in order to break completely the
gauge symmetry) is present: $\NF=N$, in this case
the vacuum is simply just a point. On the other hand, when the number
of flavors is increased, the vacuum becomes in general a manifold and
it depends on the details of the matter content, such as the charges
etc. A crucial difference between the aforementioned theories, which
possess local vortices\footnote{By local vortices, we define the
  localized topological objects which has exponential transverse
  cut-offs, as opposed to polynomial tales, which will be termed
  semi-local.}, and this theory with extra matter, is that so-called
semi-local zero-modes appear. They bear the property that they are not
normalizable -- they tend to zero obeying a power-law instead of
exponentially as the local profile functions. They were first found in
the (extended) Abelian Higgs 
model 
\cite{Vachaspati:1991dz,Achucarro:1992hs,Achucarro:1999it} and later
also in its non-Abelian extension
\cite{Hanany:2003hp,Shifman:2006kd,Eto:2007yv}. An 
interesting fact about the vortices in 
$U(1)\times SO(N)$ and $U(1)\times USp(2M)$ theories, is that they are 
in general of the semi-local type \cite{Eto:2008qw,Eto:2009bg}. There
is a very interesting relation between the vortices of the semi-local
kind and non-linear sigma model (NL$\sigma$M) lumps. The gauge
theories become NL$\sigma$Ms and the vortices become lumps
\cite{Hindmarsh:1991jq,Hindmarsh:1992yy,Preskill:1992bf,Eto:2008qw,Eto:2007yv}. 
This is technically done by taking the strong gauge coupling limit: 
$g\to\infty$, which corresponds to very low energies or very long
distances. The lumps are supported topologically by
$\pi_2(\mathcal{M})$ with $\mathcal{M}$ being the vacuum manifold. 

The theories with fundamental matter content all charged with respect
to an overall $U(1)$ factor of the gauge group have been embedded into
a high-energy theory with a simple gauge group for $SU(N+1)\to U(N)$
in Ref.~\cite{Auzzi:2003fs}. For $SO(N)$ theories this embedding has
been made in Ref.~\cite{Ferretti:2007rp} using adjoint matter content
at high energies giving rise to fundamental matter with the following
breaking $SO(N+2)\to U(1)\times SO(N)$. Similarly, for $USp(2M)$
theories, the embedding has been made explicit in
Ref.~\cite{Gudnason:2010jq} with the breaking 
$USp(2M+2)\to U(1)\times USp(2M)$
again using adjoint matter at high energies as in the $SO$ case.
These symmetry breaking patterns are relevant for non-Abelian
monopoles. Eventually, at much lower energies, all these theories get
their gauge symmetry completely broken by a Fayet-Iliopoulos term. The
interest in these systems lies in the exact homotopy sequences that
relate the monopole properties to the vortex properties in this kind
of system. For a review see Ref.~\cite{Konishi:2007dn}.

Some of the recent results and studies on the topic of non-Abelian
vortices include the following: A multiple layer structure has been
recognized in the nature of the standard non-Abelian vortex with
a $U(N)$ gauge group in Ref.~\cite{Eto:2009wq}. Non-Abelian global
vortices still with a $U(N)$ group has been studied in
Refs.~\cite{Nitta:2007dp,Nakano:2007dq,Eto:2009wu} giving rise to a
somewhat surprising non-rational $U(1)$-winding going like
$1/\sqrt{N}$. 
Non-Abelian vortices have also been studied on a torus in
Ref.~\cite{Lozano:2007bk}. 
Non-Abelian vortices in dense QCD have been studied in a series of
papers \cite{Nakano:2007dr,Eto:2009kg,Eto:2009bh,Eto:2009tr,Yasui:2010yw}. 
The metric of the moduli space of the non-Abelian vortices with a
$U(N)$ gauge group has been calculated in
Ref.~\cite{Baptista:2010rv} and for well-separated vortices in
Ref.~\cite{Fujimori:2010fk}.
Using a non-linear realization method the low-energy effective action
including the spatial fluctuations of the string as well as its
internal orientations and their higher-order mixed terms have been
derived in Ref.~\cite{Liu:2009rz}. 
In Ref.~\cite{Kimura:2009gn} a vortex description of a quantum Hall
ferromagnet has been made, considering an effective theory for an
incompressible fluid corresponding to the moduli space of a vortex
theory. 
The Refs.~\cite{Auzzi:2007wj,Auzzi:2007iv} considered the non-BPS
corrections to the non-Abelian vortices finding new types of vortices
with interactions depending on both relative distance and internal
group orientation. 
A non-Abelian vortex in the mass deformed ABJM model has also been
considered in Refs.~\cite{Kim:2009ny,Auzzi:2009es}.
The quantum phases of the vortex in ${\cal N}=1^\star$ theory have
been studied in Ref.~\cite{Auzzi:2009yw}.
A new duality between a $U(N)$ theory at strong coupling and a
$U(N-\NF)$ theory at weak coupling, relating extended objects
(solitons) to elementary objects of either side of the duality, has
been found in 
Refs.~\cite{Shifman:2009ay,Shifman:2009mb,Shifman:2010id}.

Many results for the non-Abelian vortices, with especially the
$U(N)$ gauge group, are summarized in the excellent reviews
\cite{Tong:2005un,Eto:2006pg,Shifman:2007ce}.

One property of the vortices and lumps which has not been noticed
until recently and is only seen when there is a non-trivial vacuum
manifold present in the theory -- namely that the minimal vortex or
lump can be composed by seemingly unconfined sub-objects which,
however, cannot be infinitely separated. If one tries to do so, the
configuration will either dilute completely or become singular
\cite{Eto:2009bz}. This type has been named fractional vortices and
lumps.  
They can be engineered in many ways. For instance, one can alter the
charge assignment of the different flavors. Another possibility is
that there resides a singular submanifold in the vacuum manifold,
which is exactly the case in the theories with $U(1)\times SO(N)$ and
$U(1)\times USp(2M)$ gauge groups \cite{Eto:2008qw}. In fact, the
first fractional lump solution was found in Ref.~\cite{Eto:2008qw}. 

Soon after a fractional lump was constructed in the Taub-NUT space
\cite{Collie:2009iz} with a smooth interpolation from a fractional
lump to the standard $\mathbb{C}P^{1}$ lump. It was made by turning on
a finite gauge coupling in an appropriate quiver theory which in the
strong gauge coupling limit reduces to the standard $\mathbb{C}P^1$
NL$\sigma$M. 

Subsequently, a classification of fractional vortices was made in
Ref.~\cite{Eto:2009bz}. The first type is constructed with a conical 
singularity on the vacuum manifold, which in practice is easily made
by a non-equal relatively prime and rational charge assignment for the
different flavors of squarks in the theory. The second type, however,
does not rely on a singularity on the vacuum manifold, but it suffices
to deform the geometry in some way, and it has been conjectured that a
strong positive scalar curvature in at least two (separate) regions
of the vacuum manifold gives rise to a fractional vortex or lump
\cite{Eto:2009bz}.  

So far the story has nothing to do with the Chern-Simons vortices. 
The vortices can be thought of as string-like objects in for instance
four space-time dimensions or as particle-like objects in three
space-time dimensions. The latter option is interesting per se for
several reasons. The physics has radically different characteristics,
first of all, because the spin is not quantized in half-integers as in
four space-time dimensions, admitting possibility for anyons, objects
possessing fractional charge and statistics
\cite{Wilczek:1983cy}. This can be realized by introducing the
Chern-Simons term, which has been widely used, e.g.~in the theory of
the fractional quantum Hall effect \cite{Zhang:1988wy}. Furthermore,
Chern-Simons theories provide a topological and gauge invariant
mechanism for mass generation, not relying on the Higgs mechanism
\cite{Deser:1981wh}.  
Abelian vortices in the Chern-Simons Higgs model were studied in 
Refs.~\cite{Hong:1990yh,Jackiw:1990pr} (see
e.g.~Refs.~\cite{Dunne:1998qy,Horvathy:2008hd} for excellent reviews). 
In the non-Abelian case, the Chern-Simons vortices of genuinely
non-Abelian kind, that is, possessing orientational moduli and a
moduli space of solutions were found in the
Refs.~\cite{Aldrovandi:2007nb,Lozano:2007yz} in the case of a
$U(N)$ gauge group. Aldrovandi and Schaposnik identified the moduli
space of a single vortex solution as
$\mathbb{C}\times\mathbb{C}P^{N-1}$ (which equals that of a single
vortex in Yang-Mills-Higgs theory with a $U(N)$ gauge group) and
furthermore found the vortex world-line theory, i.e.~the sigma model
living on the vortex. 
Furthermore, Refs.~\cite{Collie:2008mx,Collie:2008za,Buck:2009pd}
have considered packaging together the Yang-Mills and the non-Abelian 
Chern-Simons terms for $U(N)$ gauge groups. 
In Ref.~\cite{Collie:2008mx} the dynamics of the vortices has been 
studied and they found a modification to the adiabatic motion of
linear order proportional to the Chern-Simons coupling as well as the
usual free kinetic term. 
In Ref.~\cite{Collie:2008za}, in addition to the topological charge,
conserved Noether charges associated with a $U(1)^{N-1}$ flavor
symmetry of the theory, due to inclusion of a mass term for the
squarks, were found. A dimensional reduction of this theory to $1+1$
dimensions gives rise to so-called trions \cite{Eto:2010am}.
Numerical solutions were provided in 
Ref.~\cite{Buck:2009pd} for the Yang-Mills-Chern-Simons $U(N)$ theory
and in Ref.~\cite{Lozano:2007yz} for the Chern-Simons $U(N)$ theory
and in both cases they have found that the semi-local moduli
parameter, when non-zero, destroys the ring-like characteristic of the 
magnetic flux that the vortex solutions in Chern-Simons theories
usually possess. Thus the magnetic flux turns into bell-like
structures similar to those of Yang-Mills theories.  

In Ref.~\cite{Gudnason:2009ut}, the program of non-Abelian vortices
with arbitrary gauge groups has been extended to the Chern-Simons 
vortices and furthermore, the identification of the moduli spaces
using the moduli matrix formalism was pursued. It was conjectured that
the $k$ moduli space of the vortices with gauge group 
$G=U(1)\times G'$, with $G'$ being a simple gauge group, is the same
in the case of only the Yang-Mills kinetic term as in the case of only
the Chern-Simons kinetic term for the gauge fields. 
The vortices constructed in Ref.~\cite{Gudnason:2009ut} and studied
numerically with gauge group $U(1)\times SO(2M)$ and $U(1)\times
USp(2M)$, provide the base for fractional vortices
due to the singular submanifolds in the vacuum manifolds. Their
existence is quite easily guessed, as the moduli matrix generating
them, is exactly the one giving rise to the fractional lumps of
Ref.~\cite{Eto:2008qw}.   

Now we want to pose a very simple question, what happens to the
planar structure of the flux when the fractional sub-vortices are
moved apart? Do we get $M$ rings of flux or $M$ Gauss bells of 
flux. The analysis of the holomorphic invariants tells us that the
position moduli of the fractional sub-vortices should be interpreted
as a kind of semi-local moduli. Then we would guess the answer is the
last option. The flux spreads out to $M$ Gaussian bells of flux as we
confirm with a numerical study. We furthermore study the semi-local
vortex with gauge groups $G=U(1)\times SO(2M)$ and $G=U(1)\times
USp(2M)$ and in addition to the previous studies, we change the
relative coupling constants $\kappa,\mu$ of the $U(1)$ and $G'$ part
of the gauge fields, respectively. In the local case studied in
Ref.~\cite{Gudnason:2009ut}, this gave rise to negative Abelian
magnetic flux and positive non-Abelian flux at the origin of the
vortex in the case of $\kappa>\mu$ (and vice versa for
$\kappa<\mu$). We expect and confirm numerically, that this effect
vanishes when semi-local size moduli are turned on. For completeness,
we give also an example of the fractional semi-local vortex in our
model.

%%%%%%%%%%%%%%%%%%%%%%%
\section{The model}

We will consider the model studied in Ref.~\cite{Gudnason:2009ut} in the
limit that it reduces to the non-Abelian Chern-Simons-Higgs theory. 
It is an ${\cal N}=2$ supersymmetric gauge theory in $2+1$ dimensions
with gauge group $G=U(1)\times G'$ where $G'$ is a simple group. The
theory contains the gauge fields, an adjoint scalar and $\NF$ complex
scalar squark fields (hypermultiplets). 
Taking the strong coupling limit of the Yang-Mills
couplings $e,g\to\infty$, being the Maxwell and Yang-Mills
coupling respectively, we obtain the following non-Abelian 
${\cal N}=2$ Chern-Simons-Higgs theory 
\begin{align}
\mathcal{L}_{\rm CSH} &=
-\frac{\mu}{8\pi}\epsilon^{\mu\nu\rho}
  \left(A_{\mu}^a\partial_\nu A_\rho^a  
  -\frac{1}{3}f^{abc}A_{\mu}^a A_{\nu}^b A_{\rho}^c\right)
-\frac{\kappa}{8\pi}\epsilon^{\mu\nu\rho}
  \left(A_{\mu}^0\p_\nu A_\rho^0\right)
+\Tr\left(\D_\mu H\right)^\dag\left(\D^\mu H\right) \non&\phantom{=\ }
-4\pi^2\Tr\left|\left\{
  \frac{\mathbf{1}_N}{N\kappa}\left(\Tr\left(H H^\dag\right)
  -\xi\right)
  +\frac{2}{\mu}\Tr\left(H H^\dag t^a\right)t^a\right\} H\right|^2
  \ .
\end{align}
The adjoint scalars are infinitely massive in the limit considered
here, so they have been integrated out. The remaining fields are the
complex scalar fields $H$ which are combined into a $\dim(R_G)\times
\NF$ matrix, consisting of $\NF$ matter multiplets, 
$A_\mu = A_\mu^\alpha t^\alpha,\ \alpha =0,1,2,\ldots,\dim(G')$, is
the gauge potential and finally $t^a$ and $f^{abc}$ are the generators
and the structure constants, respectively, of the non-Abelian gauge
group $G'$. $N\equiv \dim(R_G)$ and the Abelian generator is defined
as $t^0=\mathbf{1}_N/\sqrt{2N}$. The index $0$ is for the Abelian part
of the gauge group, 
while $a,b,c=1,2,\ldots,\dim(G')$ are the non-Abelian indices. The
space-time indices are denoted by Greek letters $\mu,\nu=0,1,2$ where 
we adopt the metric $\eta^{\mu\nu} = \diag(+,-,-)$.  
We will use the following conventions 
\begin{align}
F_{\mu\nu} = \p_\mu A_\nu - \p_\nu A_\mu + i\left[A_\mu,A_\nu\right] \ ,
\quad
\D_\mu H = \left(\p_\mu + i A_\mu\right)H \ , 
\quad
\D_\mu\phi &= \p_\mu\phi + i \left[A_\mu,\phi\right]
\ .  
\end{align}
In this paper we will fix the gauge group to $G=U(1)\times
SO(2M)$ and $G=U(1)\times USp(2M)$ and treat them on equal footing
with just the change of the invariant rank-two tensor
$J$ (in the following $N=2M$). It is defined as $J^{\rm T}=\epsilon
J,\  J^\dag J = 
\mathbf{1}_{2M}$ 
and explicitly, we will choose the basis
\beq J = \begin{pmatrix}
\mathbf{0} & \mathbf{1}_M \\
\epsilon\mathbf{1}_M & \mathbf{0}
\end{pmatrix} \ , \eeq
where $\epsilon=+1$ for $SO$, while $\epsilon=-1$ for $USp$.
Furthermore, we will only consider the fundamental representation
of the matter fields here, hence $R_G:=\square$. The remaining
parameters of the theory are 
the Chern-Simons couplings $\kappa\in\mathbb{R}, \mu\in\mathbb{Z}$,
being the Abelian 
(tracepart) and the non-Abelian (traceless part) couplings,
respectively. $\xi>0$ is the Fayet-Iliopoulos parameter, putting the
theory on the Higgs branch. The last parameter governing the vortex
solutions, which is not explicitly present in the Lagrangian density,
is $k$ being the vorticity or winding number. The $U(1)$ winding, is
however given by $\nu=k/n_0$ \cite{Eto:2008yi}, where $n_0$
denotes the greatest common divisor (gcd) of the Abelian charges of
the holomorphic invariants of $G'$, see \cite{Eto:2008yi}. For simple 
groups this coincides with the center as $\mathbb{Z}_{n_0}$. We will
take $k>0$, corresponding to vortices as opposed to anti-vortices. 

There are three different phases of the theory at hand. An unbroken
phase with $\langle H\rangle=0$ and a broken phase with 
$\langle H\rangle=\mathbf{1}_{2M}\sqrt{\xi/(2M)}$. In between there are
partially broken phases. Furthermore, there are plenty of other
possible vacua breaking the gauge symmetry, which however will break
(partially) also the global color-flavor symmetry, which is only
present (in its entirety) in the latter vacuum. This will be our
reason for choosing this particular vacuum.  

The masses are generated by the Higgs mechanism (not topologically, as
in the case with also a Yang-Mills term present in the Lagrangian) and
are given by 
\beq 
m_\kappa = \frac{2\pi\xi}{M\kappa} \ , \quad
m_\mu = \frac{2\pi\xi}{M\mu} \ ,
\eeq
which by supersymmetry are the same for the gauge fields as the scalar
fields, Abelian and non-Abelian, respectively. 

The tension, defined by the integral on the plane over the time-time
component of the energy-momentum tensor, is given by
\begin{align}
T =&\ \int_{\mathbb{C}}\Tr\left\{\left|\D_0 H\right|^2
+\left|\D_i H\right|^2 
+4\pi^2\left|\left(
  \frac{\mathbf{1}_{2M}}{2M\kappa}\left(\Tr\left(H H^\dag\right)
  -\xi\right)
  +\frac{2}{\mu}\Tr\left(H H^\dag t^a\right)t^a\right)
  H\right|^2\right\} \ .
\end{align}
As shown in more detail in Ref.~\cite{Gudnason:2009ut}, by a
Bogomol'nyi completion, the BPS-equations 
\begin{align}
\bar{\D}H = 0 \ , \quad
\D_0 H = i2\pi
  \left(
  \frac{\mathbf{1}_{2M}}{2M\kappa}\left(\Tr\left(H H^\dag\right)
  -\xi\right)
  +\frac{2}{\mu}\Tr\left(H H^\dag t^a\right)t^a\right)
  H \ ,
\end{align}
can be combined with the
Gauss law 
\begin{align}
F_{12}^a =
-\frac{i4\pi}{\mu}\Tr\left[H^\dag t^a \D_0 H - \left(\D_0 H\right)^\dag
  t^a H \right] \ , \quad
F_{12}^0 =
-\frac{i4\pi}{\kappa}\Tr\left[H^\dag t^0 \D_0 H - \left(\D_0 H\right)^\dag
  t^0 H \right] \ ,\label{eq:Gausslaw}
\end{align}
yielding the following system
\begin{align}
\bar{\D}H &= 0 \ , \\
F_{12}^a t^a &= \frac{2\pi^2}{M\kappa\mu}
  \left(\Tr\left(H H^\dag\right) -\xi\right)
  \left(H H^\dag - J^\dag\left(H H^\dag\right)^{\rm T}J\right)
+\frac{2\pi^2}{\mu^2}
  \left[\left(H H^\dag\right)^2
  -J^\dag\left(\left(H H^\dag\right)^2\right)^{\rm T}J\right] \ , 
  \nonumber \\ 
F_{12}^0 t^0 &= \frac{2\pi^2}{M^2\kappa^2}\Tr\left(H H^\dag\right)
  \left(\Tr\left(H H^\dag\right)-\xi\right)\mathbf{1}_{2M}
+\frac{2\pi^2}{M\kappa\mu}
  \Tr\left(H H^\dag\left(H H^\dag 
  -J^\dag \left(H H^\dag\right)^{\rm T}J\right)\right)\mathbf{1}_{2M}
  \ ,
  \nonumber
\end{align}
and by using the moduli matrix Ansatz $H=S^{-1}H_0(z)$, $S=sS'$, the
master equations can be written down
\begin{align}
\bar{\p}\left[\Omega'\p{\Omega'}^{-1}\right] &= 
  \frac{\pi^2}{M\kappa\mu}\frac{1}{\omega}
  \left(\frac{1}{\omega}\Tr\left(\Omega_0{\Omega'}^{-1}\right) -\xi\right)
  \left(\Omega_0{\Omega'}^{-1} 
  -J^\dag\left(\Omega_0{\Omega'}^{-1}\right)^{\rm T}J\right)
  \non&\phantom{=\ }
+\frac{\pi^2}{\mu^2}\frac{1}{\omega^2}
  \left[\left(\Omega_0{\Omega'}^{-1}\right)^2
  -J^\dag\left(\left(\Omega_0{\Omega'}^{-1}\right)^2\right)^{\rm T}J\right] \ , 
\label{eq:masterequationsSOUSp1}%\\  
\end{align}
\begin{align}
\bar{\p}\p\log\omega &= -\frac{\pi^2}{M^2\kappa^2}
  \frac{1}{\omega}\Tr\left(\Omega_0{\Omega'}^{-1}\right)
  \left(\frac{1}{\omega}
  \Tr\left(\Omega_0{\Omega'}^{-1}\right)-\xi\right)
  \non&\phantom{=\ }
-\frac{\pi^2}{M\kappa\mu}\frac{1}{\omega^2}
  \Tr\left(\Omega_0{\Omega'}^{-1}\left(\Omega_0{\Omega'}^{-1} 
  -J^\dag \left(\Omega_0{\Omega'}^{-1}\right)^{\rm T}J\right)\right)
  \ , \label{eq:masterequationsSOUSp2}
\end{align} 
where we have defined $\Omega_0\equiv H_0H_0^\dag$ as well as the
gauge invariant quantity $\Omega=S S^\dag=\omega\Omega'$, which splits
into the Abelian part $\omega=|s|^2$ and the non-Abelian part
$\Omega'=S'({S'})^\dag$. These fields are conjectured to be uniquely
determined by the master equations applying the appropriate boundary
conditions for a given moduli matrix $H_0(z)$ -- up to $V$-equivalence
\cite{Eto:2005yh} 
\beq \left\{H_0,S\right\} \sim V \left\{H_0,S\right\} \ , \eeq
with the transformation matrix
\beq V=vV' \ , \quad v\in\mathbb{C}^\star \ , \quad
V'\in {G'}^{\mathbb{C}} \ . \eeq

%%%%%%%%%%%%%%%%%%%%%%%

Writing the energy density in terms of our new variables, including
the boundary term
\beq \mathcal{E} = 2\xi\bar{\p}\p\log\omega 
  + 2\bar{\p}\p\left(\frac{1}{\omega}\Tr\Omega_0{\Omega'}^{-1}\right)
  \ , \eeq 
we obtain by integration, the total energy
\beq E = \int_{\mathbb{C}} \mathcal{E} = 2\pi\xi\nu = \frac{2\pi\xi
  k}{n_0} \ , \eeq
which is simply proportional to the topological charge.
The Abelian and non-Abelian magnetic flux densities are, respectively
\beq \mathcal{B} = F_{12}^0 = -4\sqrt{M}\;\bar{\p}\p\log\omega \ ,
\quad F_{12}^a t^a =
2{S'}^{-1}\bar{\p}\left[\Omega'\p{\Omega'}^{-1}\right]S' \ , 
\eeq 
while the Abelian and non-Abelian electric field densities read,
respectively 
\begin{align}
E_i &= F_{i0}^0 = \frac{2\pi}{\kappa\sqrt{M}}\p_i
\left(\frac{1}{\omega}\Tr\left(\Omega_0{\Omega'}^{-1}\right)\right)
\ , \\
E_i^a t^a &= F_{i0}^a t^a = \frac{\pi}{\mu}\p_i\left[\frac{1}{\omega}
{S'}^{-1}\left(\Omega_0{\Omega'}^{-1} 
-J^\dag\left(\Omega_0{\Omega'}^{-1}\right)^{\rm T}J\right)S'\right] \ .
\end{align}
Finally, we need only to specify the boundary conditions, which have
been obtained in Refs.~\cite{Eto:2008qw,Eto:2009bg}
\beq \Omega' = H_0(z)\frac{\mathbf{1}_{2M}}{\sqrt{{\cal M}^\dag
    {\cal M}}}H_0^\dag(\bar{z}) 
\ , \qquad \omega = \frac{1}{\xi}\Tr\sqrt{{\cal M}^\dag {\cal M}} \ ,
\label{eq:weakSOUSp}\eeq
with ${\cal M}\equiv H_0^{\rm T}(z)J H_0(z)$ being the meson field.

%%%%%%%%%%%%%

For concreteness, we will consider the cases $G'=SO(4)$ and
$G'=USp(4)$ as the main result of the paper will be made using
numerical methods. However, when the generalization to $G'=SO(2M),
USp(2M)$ is straightforward, we will work out the equations in the
generalized case and at the end of the day set $M=2$. 
The question we are addressing, is the study of the 
semi-local and ``fractional'' moduli parameters of this single
non-Abelian Chern-Simons vortex ($k=1$). 
Now, consider the given moduli matrix
\beq
H_0(z) = \begin{pmatrix}
z\mathbf{1}_M - \mathbf{A} & \mathbf{C}_{S,A} \\
\mathbf{B}_{A,S} & \mathbf{1}_M
\end{pmatrix} \ , \label{eq:H0-SOUSp4}
\eeq
which is the most general matrix for $k=1$ in a particular patch of
the moduli space. 
The subscript $S,A$ denotes symmetric in the case of $\epsilon=+1$
($SO$ case) and anti-symmetric for $\epsilon=-1$ ($USp$ case) (and
vice versa for $A,S$). 
The matrix obeys the weak condition on the holomorphic invariants 
\beq H_0^{\rm T}(z) J H_0(z) = (z - z_0) J +\mathcal{O}\left(z^0\right) \ ,
\eeq
and by insertion of the moduli matrix (\ref{eq:H0-SOUSp4}) into the
weak condition, we obtain
\beq H_0^{\rm T}(z) J H_0(z) = (z-a_0) J +
\begin{pmatrix}
\mathbf{B}_{A,S}\hat{\mathbf{A}} - \hat{\mathbf{A}}^{\rm T}\mathbf{B}_{A,S} &
-\hat{\mathbf{A}}^{\rm T} - \mathbf{B}_{A,S}\mathbf{C}_{S,A} \\
\epsilon\mathbf{B}_{A,S}\mathbf{C}_{S,A} -\epsilon\hat{\mathbf{A}} &
\epsilon\, 2\mathbf{C}_{S,A}
\end{pmatrix} \ , \label{eq:semilocalmoduliSO4}
\eeq
where we have defined 
$\mathbf{A}\equiv a_0\mathbf{1}_M+\hat{\mathbf{A}}$ with 
$\Tr\hat{\mathbf{A}}=0$. 
The presence of $\hat{\mathbf{A}}$ (even for $\mathbf{B}=\mathbf{0}$)
in the constant matrix (i.e.~not proportional to $J$) tells us that
the fractional position moduli are semi-local moduli.  
The moduli parameters are classified into two types. Normalizable
orientational zero-modes and non-normalizable semi-local ``size''
parameters. $\mathbf{B}$ contains orientational modes and does not
change the Abelian flux density contribution to the energy
density. The center of mass is given by $a_0=\Tr\mathbf{A}/M$ which is
the sum of the eigenvalues of the complex matrix $\mathbf{A}$. We will
fix this parameter to the origin. The eigenvalues of $\mathbf{A}$ are
the positions of the ``fractional vortices'' and if they all (in this
case both) coincide, they compose a normal non-Abelian vortex, but not
necessarily local. It becomes local only if
$\hat{\mathbf{A}}=\mathbf{C}=\mathbf{0}$. Finally, there are the
semi-local ``size'' parameters $\mathbf{C}$. 
What we would like to show is that the fractional vortex has
semi-independent sub-structures which should be interpreted as an
$M$-th vortex (flux) with a certain size parameter (even when
$\mathbf{C}=\mathbf{0}$).  

We will proceed in three steps. First we will in Sec.~\ref{sec:frac}
study the fractional vortices, choosing $\mathbf{A}$ to be diagonal
and setting $\mathbf{B}=\mathbf{C}=\mathbf{0}$. Subsequently, in 
Sec.~\ref{sec:semilocal} we will study the semi-local parameters
$\mathbf{C}$, setting $\mathbf{B}=\mathbf{A}=\mathbf{0}$. 
Finally, in Sec.~\ref{sec:frac-semilocal} we will give an example of
non-vanishing $\mathbf{A}$ and $\mathbf{C}$.
We will conclude with a discussion in Sec.~\ref{sec:disc} and finally
give a brief review of the asymptotic properties of the Abelian 
semi-local vortex in Appendix \ref{app:abeliansemilocalreview}.

%%%%%%%%%%%%%
\section{Fractional vortices\label{sec:frac}}

In this Section, we will study the moduli matrix (\ref{eq:H0-SOUSp4})
with $\mathbf{B}=\mathbf{C}=\mathbf{0}$ and take $\mathbf{A}$ to be
diagonal
\beq \mathbf{A} = \diag\left(z_1,\ldots,z_M\right) 
\ , \quad
a_0 = \frac{1}{M}\sum_{n=1}^M z_n = 0 \ . \eeq
With this moduli matrix there will be no difference between
$G'=SO(2M)$ and $G'=USp(2M)$, so we will apply the same yardstick to
both of them. As the matrix $\mathbf{A}$ is not proportional to the
unit matrix $\mathbf{1}_M$, the vortex will be of the semi-local type,
but a special semi-local type -- viz.~the fractional vortex. We will
denote it the pure fractional vortex.
Hence, we are left with the moduli matrix of the following form
\beq H_0(z) = \diag\left(z-z_1,\ldots,z-z_M,\mathbf{1}_M\right) \ ,
\eeq
for which we can choose the Ansatz
\beq \Omega' =
\diag\left(e^{\chi_1},\ldots,e^{\chi_M},e^{-\chi_1},\ldots,e^{-\chi_M}\right)
\ , \label{eq:omegapAnsatz}
\eeq
with the $\det\Omega'=1$ being manifest and we define $\omega \equiv
e^{\psi}$.  
Inserting this Ansatz into the master equations
(\ref{eq:masterequationsSOUSp1})-(\ref{eq:masterequationsSOUSp2}) for
the non-Abelian  
Chern-Simons vortex leaves us with the following system of partial
differential equations
\begin{align}
\bar{\p}\p\chi_m &= -\frac{\pi^2}{M\kappa\mu}
\left[\sum_{n=1}^M\left(\left|z-z_n\right|^2e^{-\psi-\chi_n}+e^{-\psi+\chi_n}\right)-\xi\right]
\left[\left|z-z_m\right|^2e^{-\psi-\chi_m}-e^{-\psi+\chi_m}\right]\non
&\phantom{=\ }
-\frac{\pi^2}{\mu^2}\left[\left(\left|z-z_m\right|^2e^{-\psi-\chi_m}\right)^2
-\left(e^{-\psi+\chi_m}\right)^2\right] \ , \quad 
m=1,2,\ldots,M \ , 
\label{eq:frac-masterequation1}\\
\bar{\p}\p\psi &= -\frac{\pi^2}{M^2\kappa^2}
\left[\sum_{n=1}^M\left(\left|z-z_n\right|^2e^{-\psi-\chi_n}+e^{-\psi+\chi_n}\right)-\xi\right]
\times
\sum_{n'=1}^M\left(\left|z-z_{n'}\right|^2e^{-\psi-\chi_{n'}}+e^{-\psi+\chi_{n'}}\right)
\non &\phantom{=\ }
-\frac{\pi^2}{M\kappa\mu}
\sum_{n=1}^M\left(\left|z-z_n\right|^2e^{-\psi-\chi_n}-e^{-\psi+\chi_n}\right)^2
\ . 
\label{eq:frac-masterequation2}
\end{align}
The boundary conditions for $|z|\to\infty$ are
\begin{align}
\psi^{\infty} = \log\left(\frac{2}{\xi}\sum_{n=1}^M\left|z-z_n\right|\right)
\ , \quad
\chi_m^{\infty} = \log\left|z-z_m\right| \ .
\label{eq:frac-boundary-conditions}
\end{align}
The energy density comprises the contribution from the Abelian
magnetic field strength
\begin{align}
F_{12}^0 = - 4\sqrt{M}\bar{\p}\p\psi \ , 
\label{eq:frac-Abelianmagflux}
\end{align}
and a boundary term summing up to
\begin{align}
\mathcal{E} = 2\xi\bar{\p}\p\psi + 
2\sum_{n=1}^M\bar{\p}\p\left(\left|z-z_n\right|^2 
e^{-\psi-\chi_n} + e^{-\psi+\chi_n}\right) \ ,
\end{align}
however, when integrated, only the Abelian magnetic flux contributes as
it is the topological charge of the vortex. 
The non-Abelian magnetic field strength is given by
\begin{align}
F_{12}^{m} = - 4\bar{\p}\p\chi_{m} \ , \quad m=1,\ldots,M \ .
\label{eq:frac-nonAbelianmagflux}
\end{align}
The Abelian and non-Abelian electric field strengths are, respectively
\begin{align}
E_i^0 = \frac{2\pi}{\kappa\sqrt{M}}\p_i
\left[\sum_{n=1}^M\left(\left|z-z_n\right|^2 
e^{-\psi-\chi_n} + e^{-\psi+\chi_n}\right)\right] \ , \quad
E_i^{m} = \frac{2\pi}{\mu}\p_i
\left[\left|z-z_{m}\right|^2 e^{-\psi-\chi_{m}} 
- e^{-\psi+\chi_{m}}\right] \ , \nonumber
\end{align}
where we conveniently have defined the relevant generators for the
non-Abelian field strengths as
\begin{align}
\left(t^m\right)_i^{\phantom{i}j} = 
\frac{1}{2}\left(\delta^{m}_{\phantom{m}i}\delta^{m,j} -
\delta^{m+M}_{\phantom{m+M}i}\delta^{m+M,j}\right) \ ,
\label{eq:frac_generators}
\end{align}
and the labels $m=1,\ldots,M$ denote the generators corresponding to 
normalized generators spanning the Cartan subalgebra of $SO(2M)$ and
$USp(2M)$.  

First let us make some qualitative calculations. We consider some
small fluctuations around the boundary conditions
(\ref{eq:frac-boundary-conditions}) as follows
\beq \chi_m = \chi_m^{\infty} + \delta\!\chi_m \ , \quad
\psi = \psi^{\infty} + \delta\psi \ . \eeq
Plugging them into the master equations
(\ref{eq:frac-masterequation1})-(\ref{eq:frac-masterequation2}) 
yields to linear order in the fluctuations
\begin{align}
\bar{\p}\p\delta\!\chi_m &= \frac{M^2m_\mu^2}{4}
\frac{|z-z_m|^2}{\left(\sum_{n=1}^{M}|z-z_n|\right)^2}\;\delta\!\chi_m
\ ,
\label{eq:chifluc_frac}
\\
\bar{\p}\p\delta\psi + \bar{\p}\p\psi^{\infty} &= 
\frac{m_\kappa^2}{4}\delta\psi \ .
\label{eq:psifluc_frac}
\end{align}
The lump solution for $\chi_m$ vanishes when acted upon by the
Laplacian operator, hence the fluctuation is in some sense local, up
to the corrections of the rational function multiplying the right hand
side of Eq.~(\ref{eq:chifluc_frac}) which asymptotically 
will be of order $1$. Thus asymptotically, the solution will be the
modified Bessel function of the second kind $K_0(m_\mu |z|)$. This is
not the case for the profile function $\psi$, governing the Abelian
magnetic flux. The Laplacian operator on the lump solution does not
vanish. Expanding in $z^{-1},\bar{z}^{-1}$ (and using an appropriate
K\"ahler transformation)
\begin{align}
&\bar{\p}\p\log\left(\sum_{n=1}^M|z-z_n|\right) \simeq 
\frac{1}{4M}\Bigg(\sum_{n=1}^M|z_n|^2
-\frac{1}{M}\left|\sum_{n=1}^M z_n\right|^2\Bigg) |z|^{-4} \non
&+\frac{1}{4M}\Bigg[\frac{1}{2}\sum_{n=1}^M|z_n|^2
  \left(\frac{z_n}{z}+\frac{\bar{z}_n}{\bar{z}}\right)
  -\frac{1}{2M}\left(\sum_{n=1}^M\frac{z_n^2}{z}\sum_{n'=1}^M\bar{z}_{n'}
  +\sum_{n=1}^M z_n\sum_{n'=1}^M\frac{\bar{z}_{n'}^2}{\bar{z}}\right)
\non
&\phantom{+\frac{1}{4M}\Bigg[}
  +\frac{1}{M}\Bigg(\sum_{n=1}^M|z_n|^2
  -\frac{1}{M}\left|\sum_{n=1}^M z_n\right|^2\Bigg)
  \sum_{n'=1}^M
  \left(\frac{z_{n'}}{z}+\frac{\bar{z}_{n'}}{\bar{z}}\right)
\Bigg] |z|^{-4}
+ \mathcal{O}\left(|z|^{-6}\right) \ ,
%\\
%&\bar{\p}\p\log\left(1 
%-\frac{1}{2M}\sum_{n=1}^M\left(\frac{z_n}{z}+\frac{\bar{z}_n}{\bar{z}}\right)
%-\frac{1}{8M}\sum_{n=1}^M\left(\frac{z_n^2}{z^2}+\frac{\bar{z}_n^2}{\bar{z}^2}\right)
%+\frac{1}{4M}\sum_{n=1}^M\frac{|z_n|^2}{|z|^2} + 
%\mathcal{O}\left(|z|^{-3}\right)\right) \ , 
\label{eq:lump_expansion}
\end{align}
we obtain the power behavior for $\delta\psi$, well-known for the
semi-local vortex profile functions:
\begin{align}
\delta\psi &= 
%\frac{1}{M m_\kappa^2}\left(\sum_{n=1}^M|z_n|^2
%-\frac{1}{M}\left|\sum_{n=1}^M z_n\right|^2\right) |z|^{-4}
%+ \mathcal{O}\left(|z|^{-5}\right) \ . 
\frac{1}{M m_\kappa^2}\Bigg(\sum_{n=1}^M|z_n|^2
-\frac{1}{M}\left|\sum_{n=1}^M z_n\right|^2\Bigg) |z|^{-4} \non
&\phantom{=\ }
+\frac{1}{M m_\kappa^2}\Bigg[\frac{1}{2}\sum_{n=1}^M|z_n|^2
  \left(\frac{z_n}{z}+\frac{\bar{z}_n}{\bar{z}}\right)
  -\frac{1}{2M}\left(\sum_{n=1}^M\frac{z_n^2}{z}\sum_{n'=1}^M\bar{z}_{n'}
  +\sum_{n=1}^M z_n\sum_{n'=1}^M\frac{\bar{z}_{n'}^2}{\bar{z}}\right)
\non
&\phantom{=\ \ +\frac{1}{4M}\Bigg[}
+\frac{1}{M}\Bigg(\sum_{n=1}^M|z_n|^2
  -\frac{1}{M}\left|\sum_{n=1}^M z_n\right|^2\Bigg)
  \sum_{n'=1}^M
  \left(\frac{z_{n'}}{z}+\frac{\bar{z}_{n'}}{\bar{z}}\right)
\Bigg] |z|^{-4}
+ \mathcal{O}\left(|z|^{-6}\right) \ ,
\label{eq:fractional_asymptotic}
\end{align}
This clearly demonstrates the semi-local nature of the fractional
vortex also in the Chern-Simons theory. 
It is easily seen that when \emph{all} the centers of the fractional
vortices coincide, $z_n=z_0\forall\, n$, the solution
(\ref{eq:fractional_asymptotic}) vanishes and should be replaced by
the ``local'' solution $K_0(m_\kappa|z|)$ which is easily found from
Eq.~(\ref{eq:psifluc_frac}). Note also that the solution is radially
symmetric only to lowest order: $|z|^{-4}$.
In Eq.~(\ref{eq:chifluc_frac}) it is easy
to see that the rational function becomes simply $1/M^2$ when the
centers $z_n$ coincide.  

We will now solve the equations numerically. For simplicity, as
already mentioned, we will focus on the groups $G'=SO(4)$ and
$G'=USp(4)$, which will give rise to a fractional vortex possessing
two subpeaks. The solution is local in the sense that there are no
size moduli when $z_1 = z_2$, however when the ``centers'' are
non-coincident: $z_1\neq z_2$ the vortex should be interpreted as a
semi-local vortex. 

An important difference between this configuration and the
configurations we normally consider, is that the rotational symmetry
in the $\mathbb{C}$-plane has been lost and we are forced to consider
the full partial differential equations in the $\mathbb{C}$-plane
instead of simplified ones in the radial direction which reduce to 
ordinary differential equations that we easily can solve. 

Furthermore, the fact that the non-Abelian Chern-Simons vortex has
some crucial differences with respect to the corresponding lump
obtained in the weak Chern-Simons coupling limit, for instance the
Abelian magnetic flux being a ring instead of a Gaussian bell,
suggests us to find the finite Chern-Simons coupling solutions in the 
$\mathbb{C}$-plane. Thus, we will now compute the numerical solutions
in the $\mathbb{C}$-plane using a relaxation method. 

\begin{figure}[!p]
\begin{center}
\includegraphics[height=2.9cm]{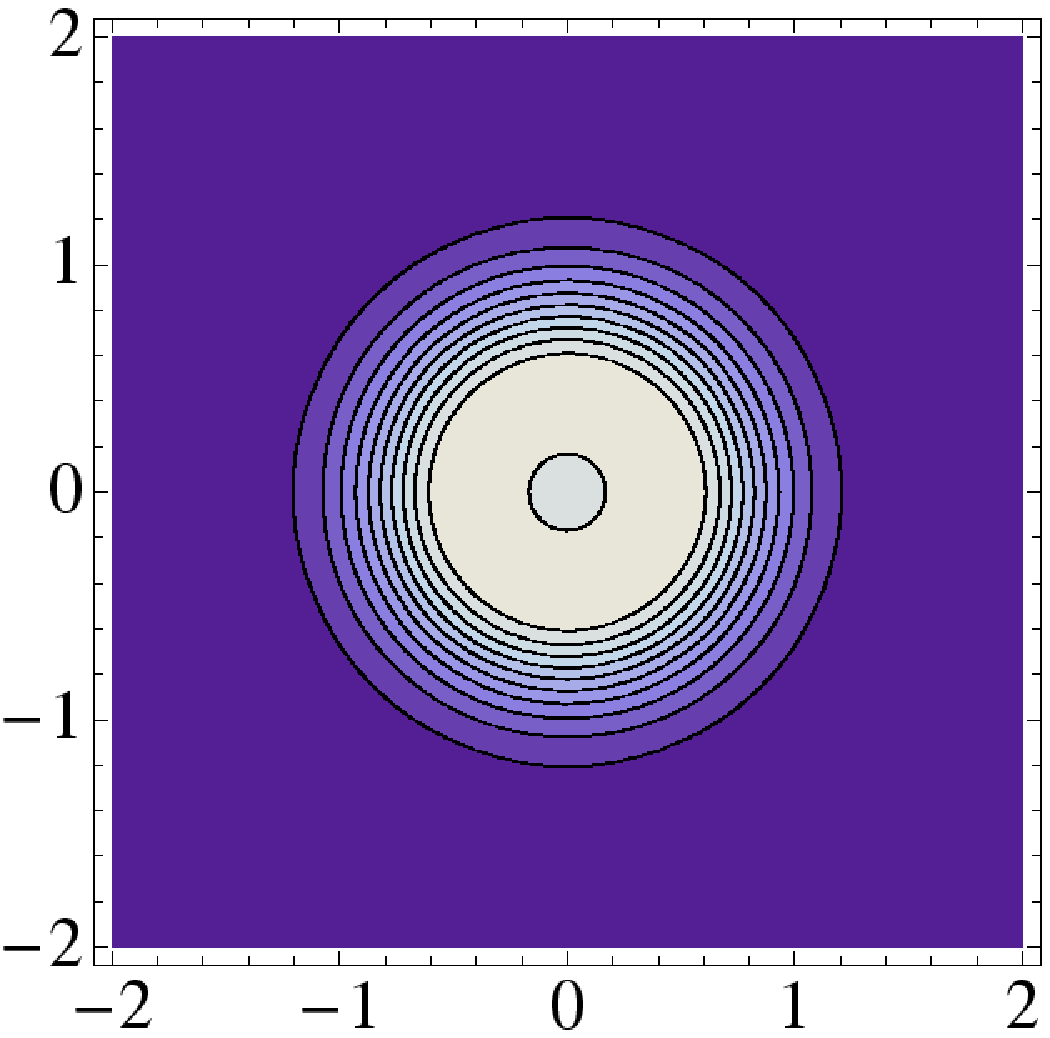}
\includegraphics[height=2.9cm]{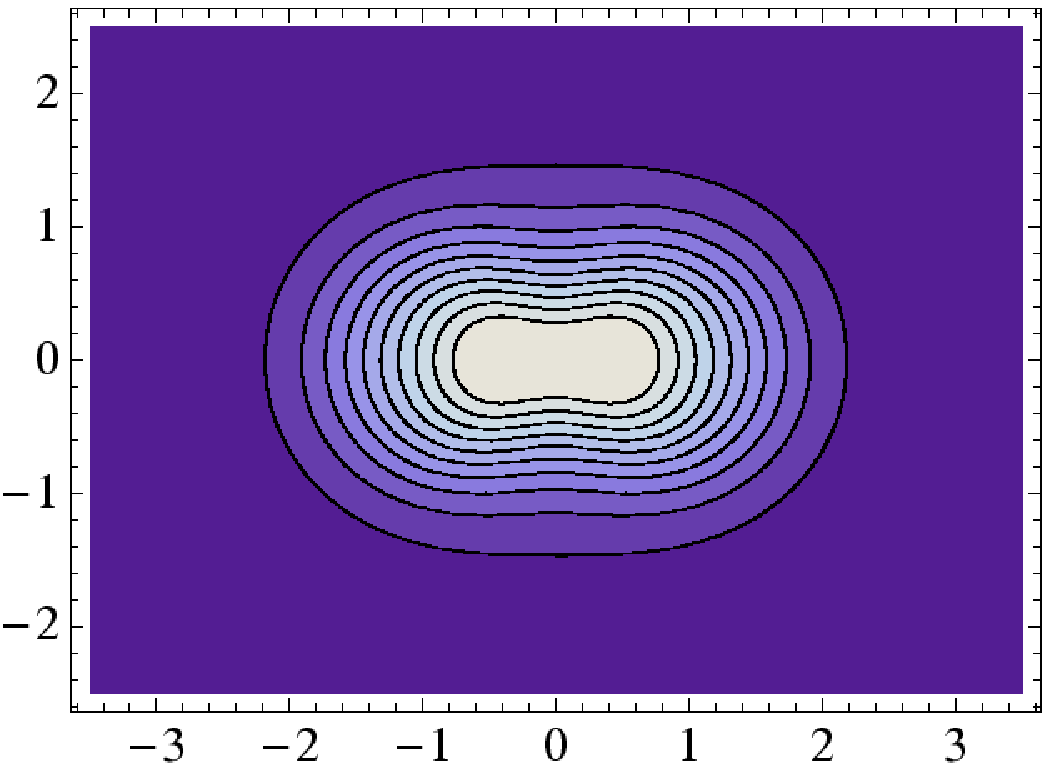}
\includegraphics[height=2.9cm]{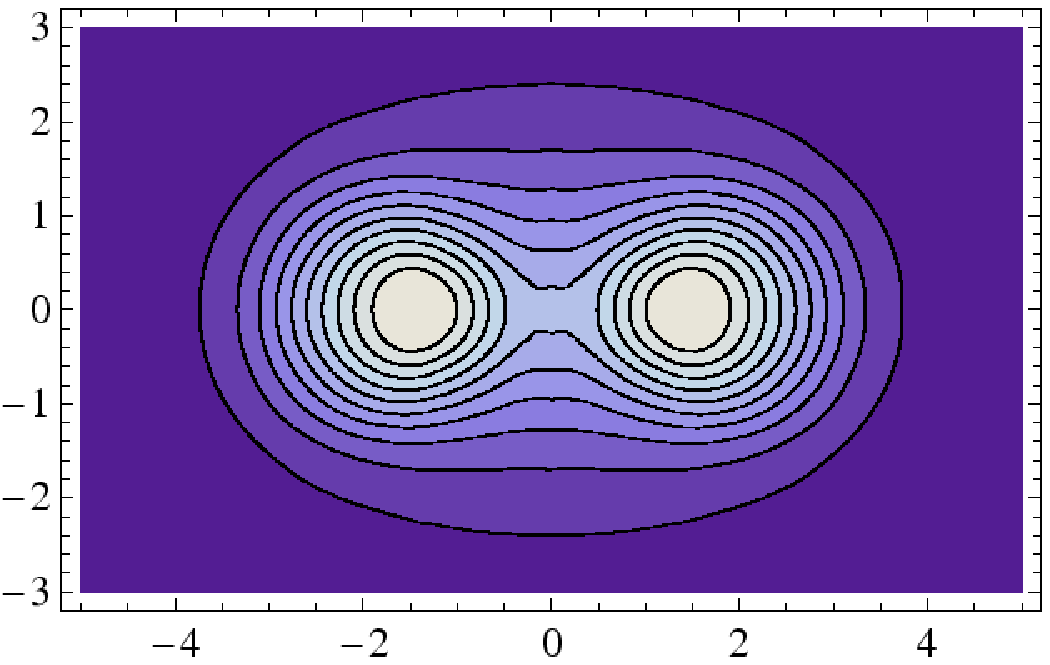}
\includegraphics[height=2.9cm]{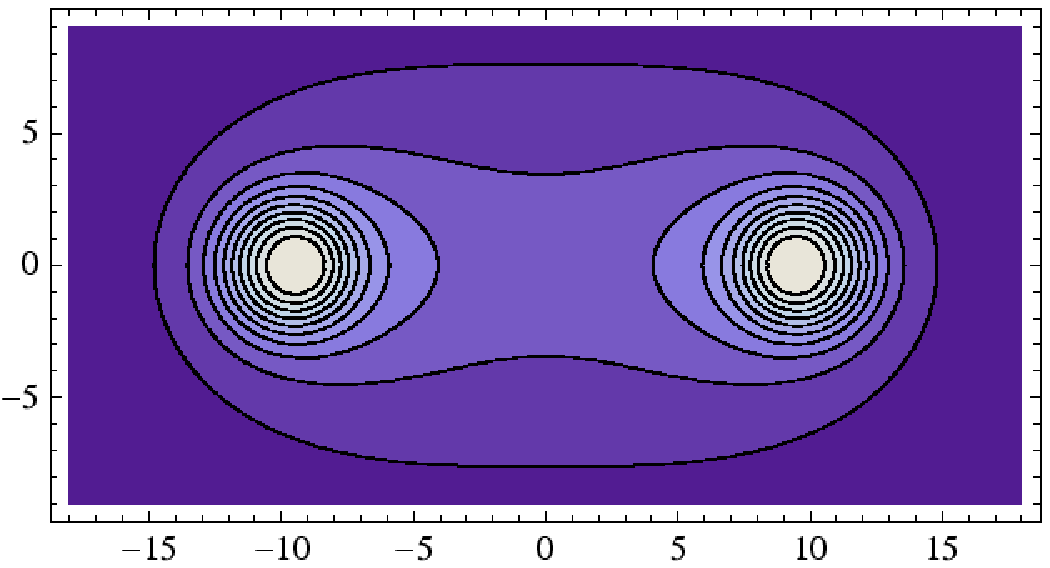}\\
\includegraphics[height=2.9cm]{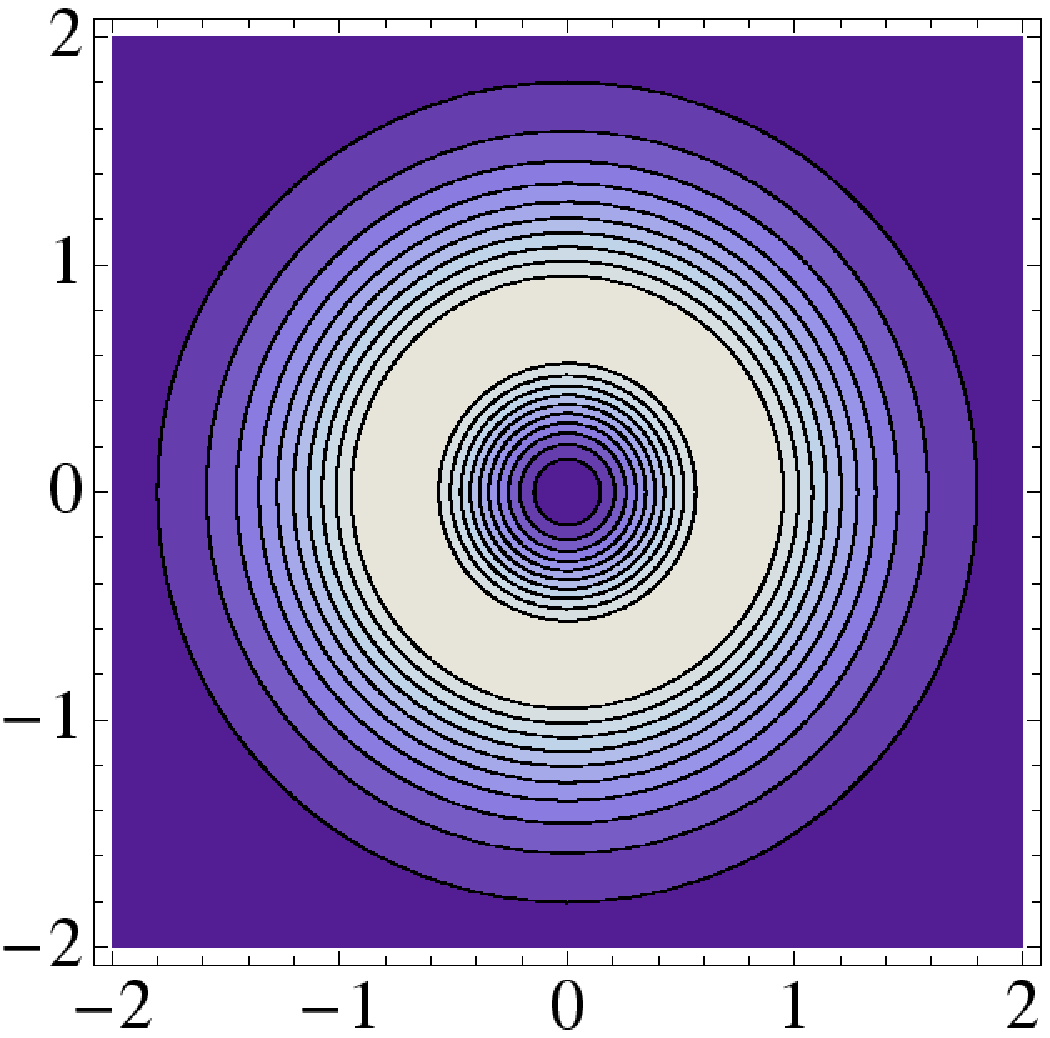}
\includegraphics[height=2.9cm]{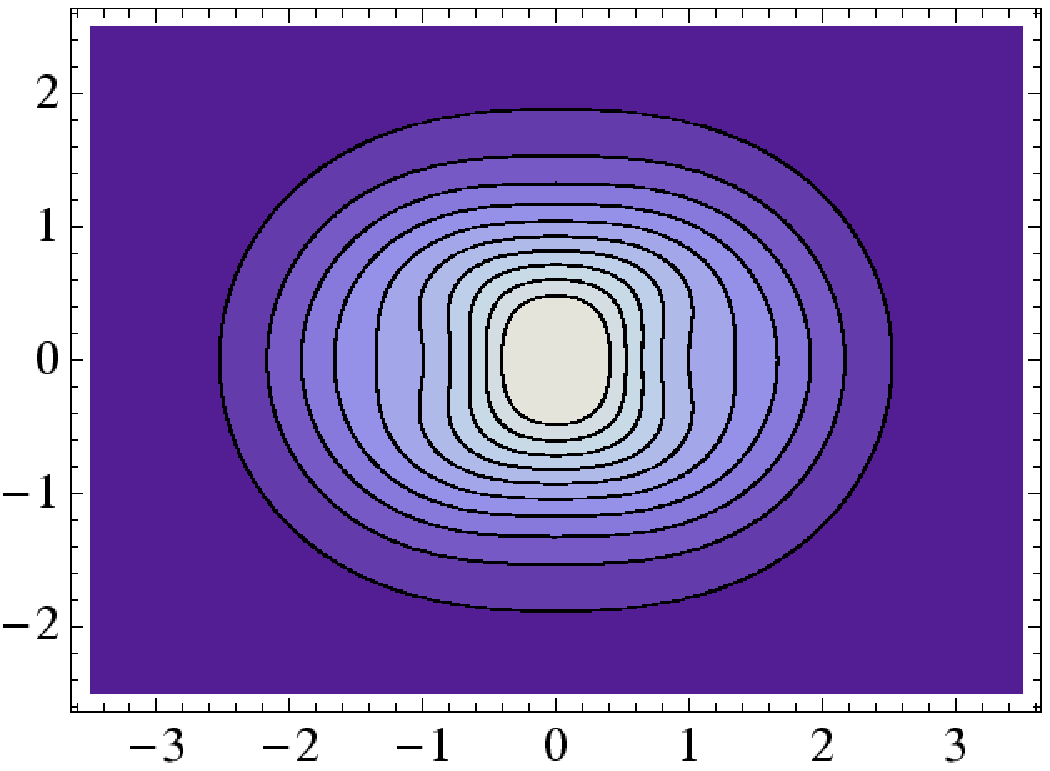}
\includegraphics[height=2.9cm]{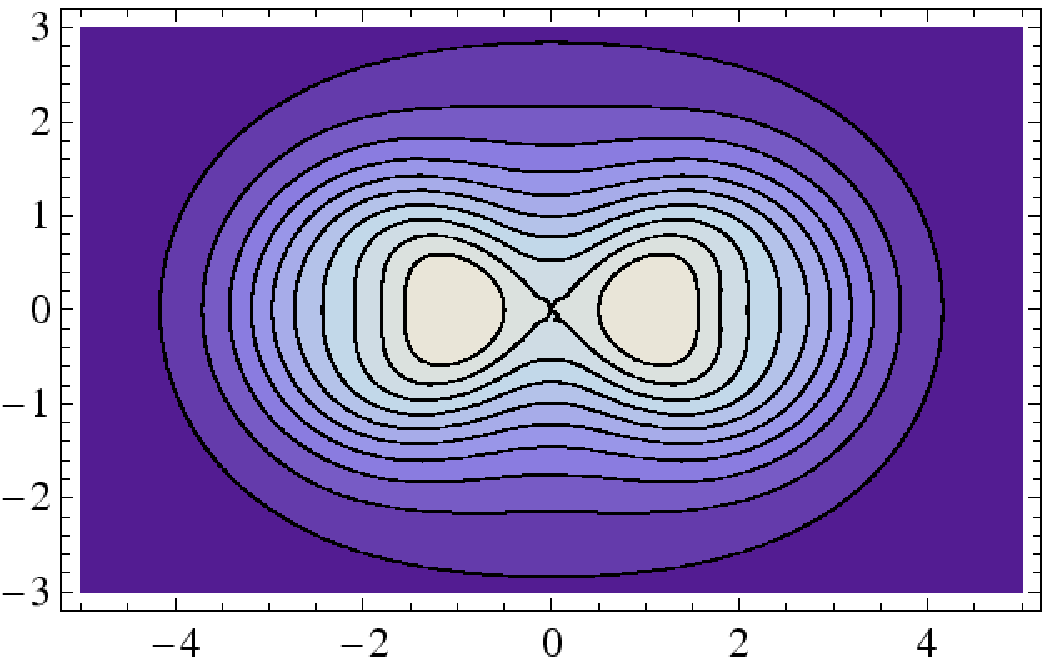}
\includegraphics[height=2.9cm]{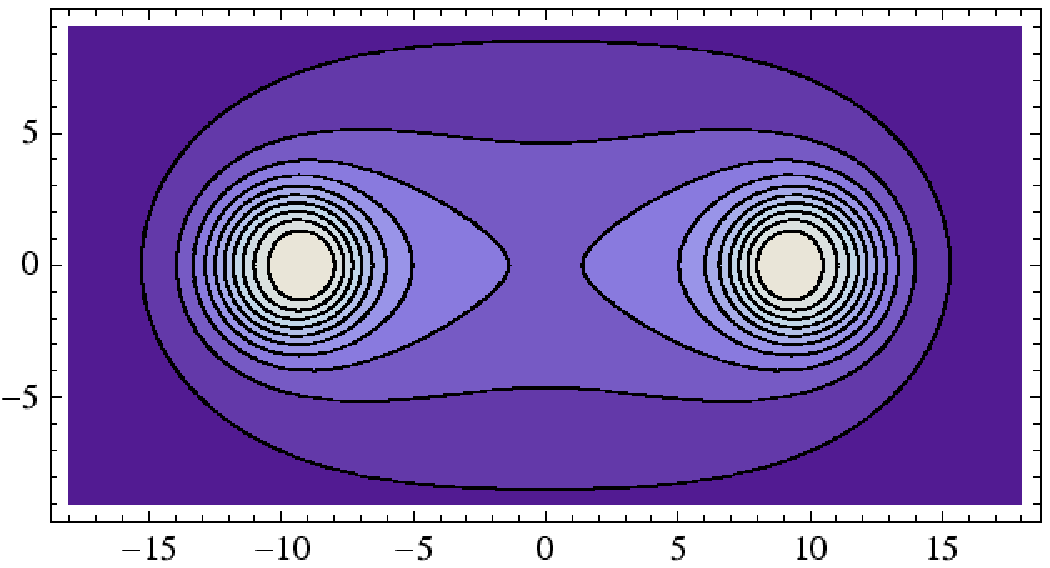}\\
\includegraphics[height=2.9cm]{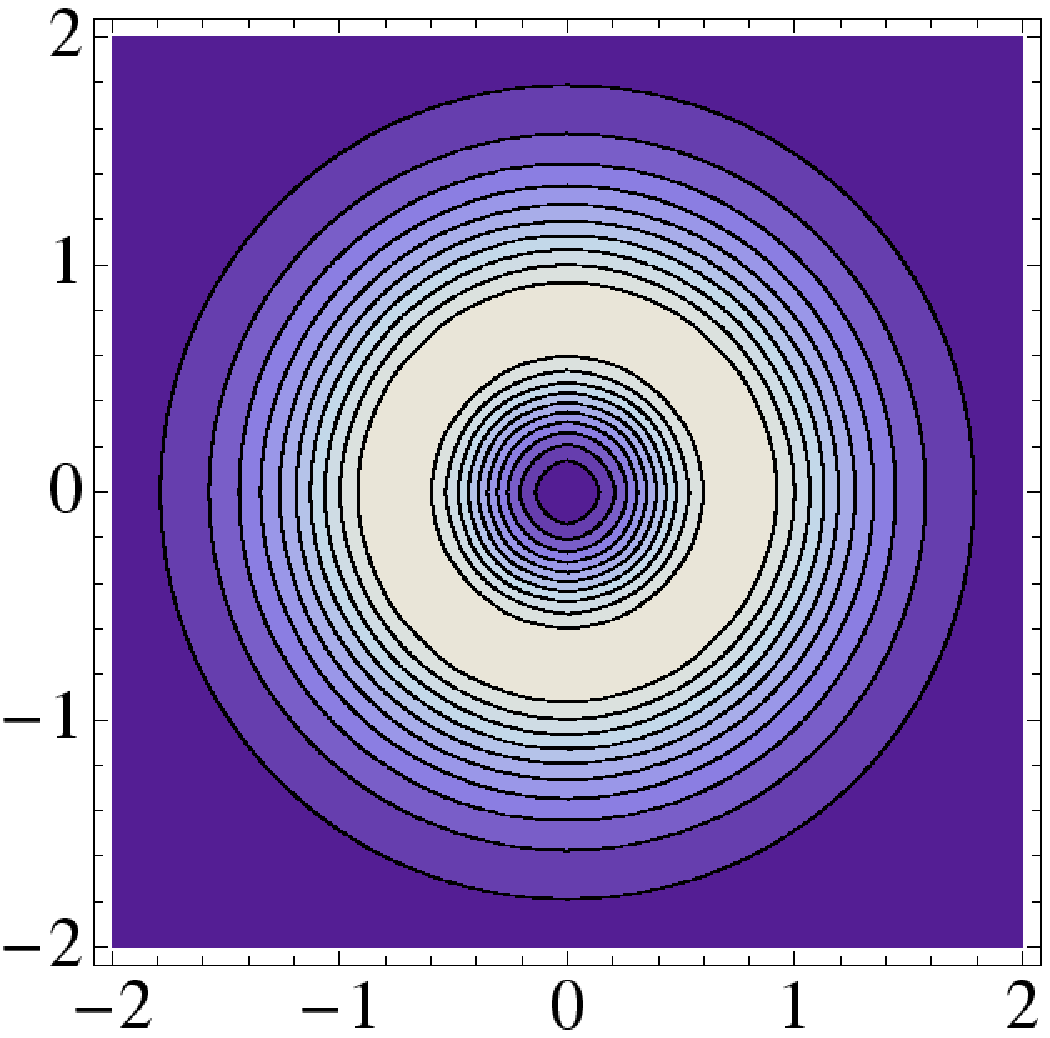}
\includegraphics[height=2.9cm]{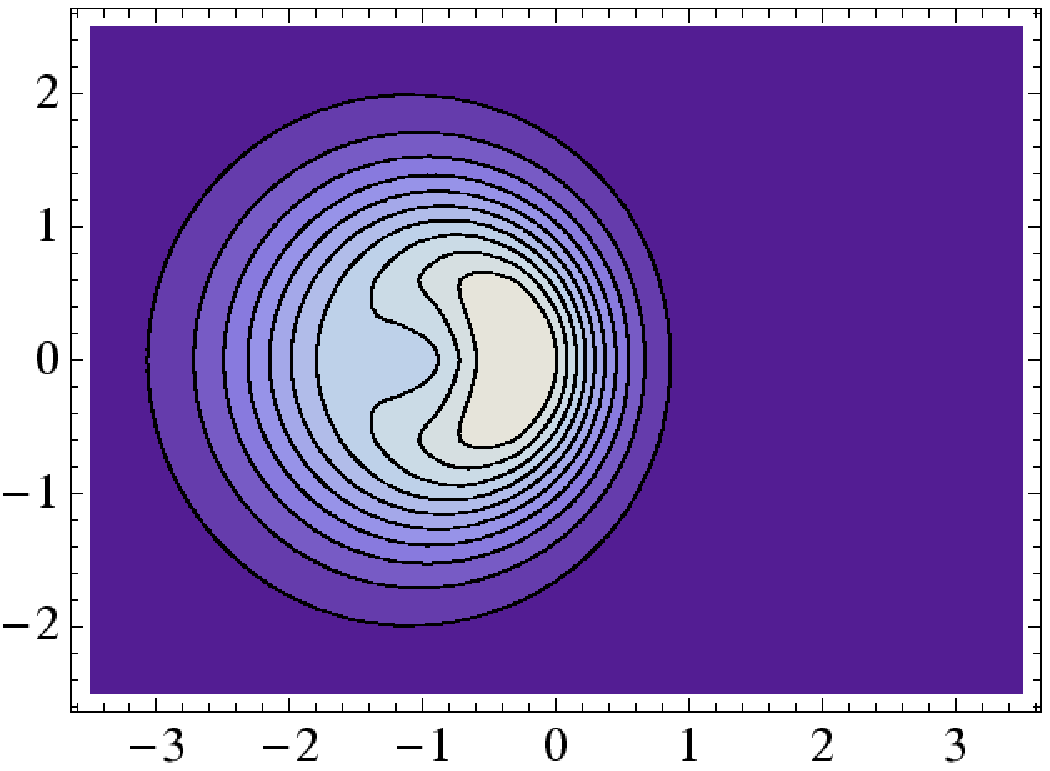}
\includegraphics[height=2.9cm]{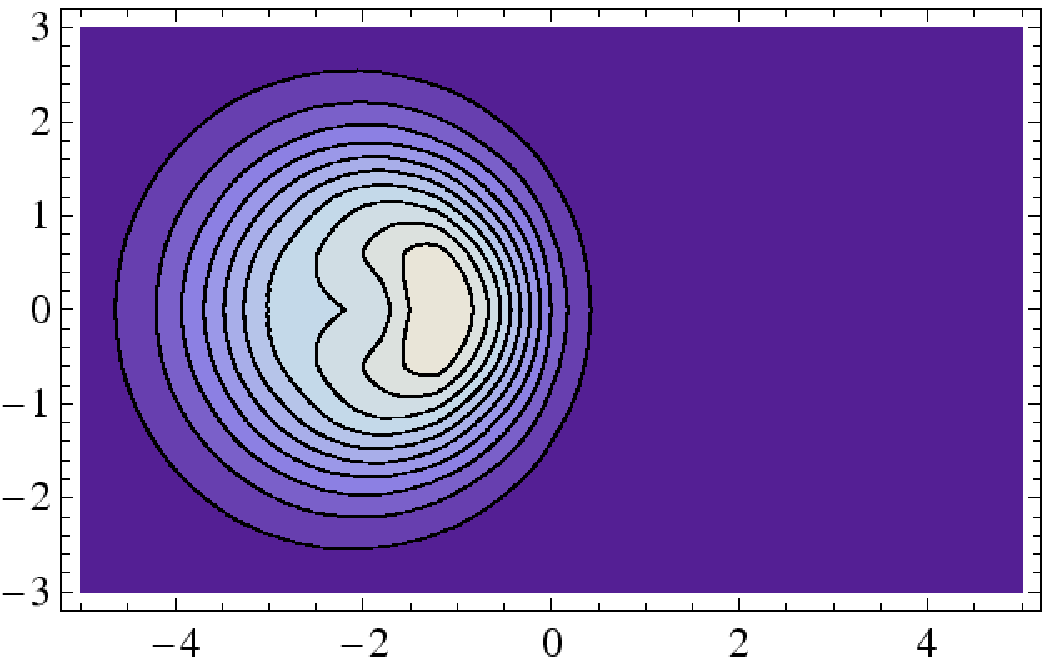}
\includegraphics[height=2.9cm]{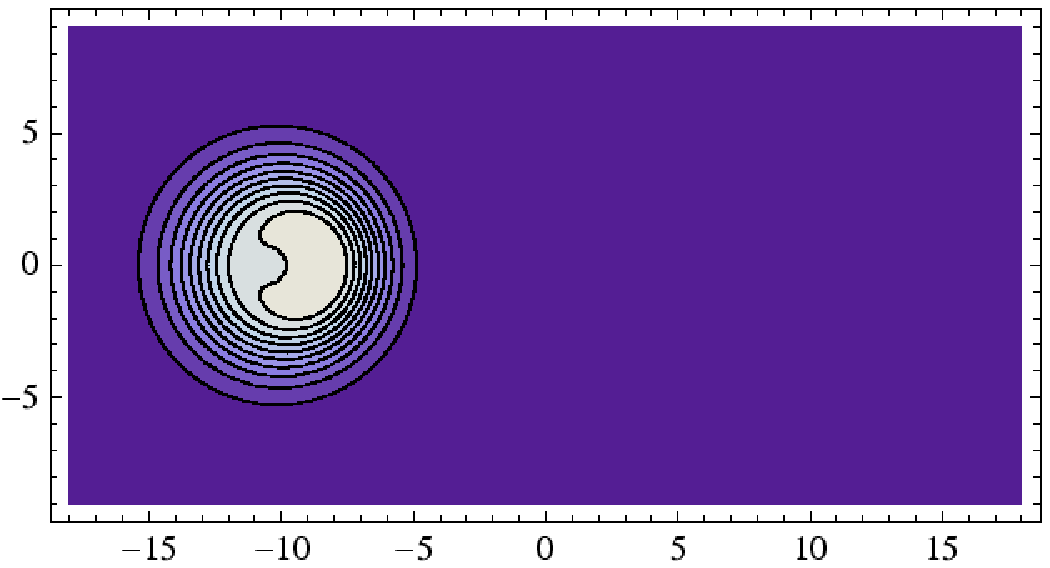}\\
\includegraphics[height=2.9cm]{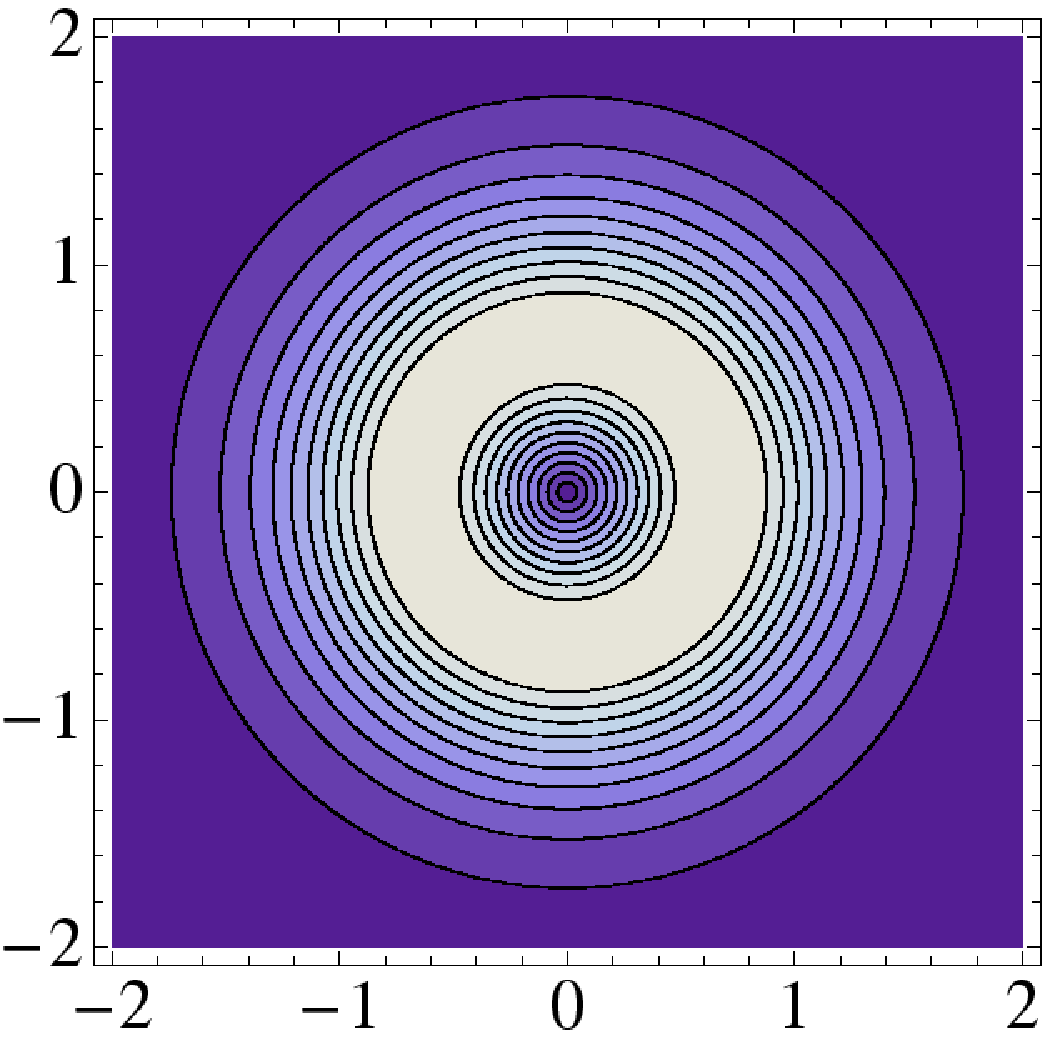}
\includegraphics[height=2.9cm]{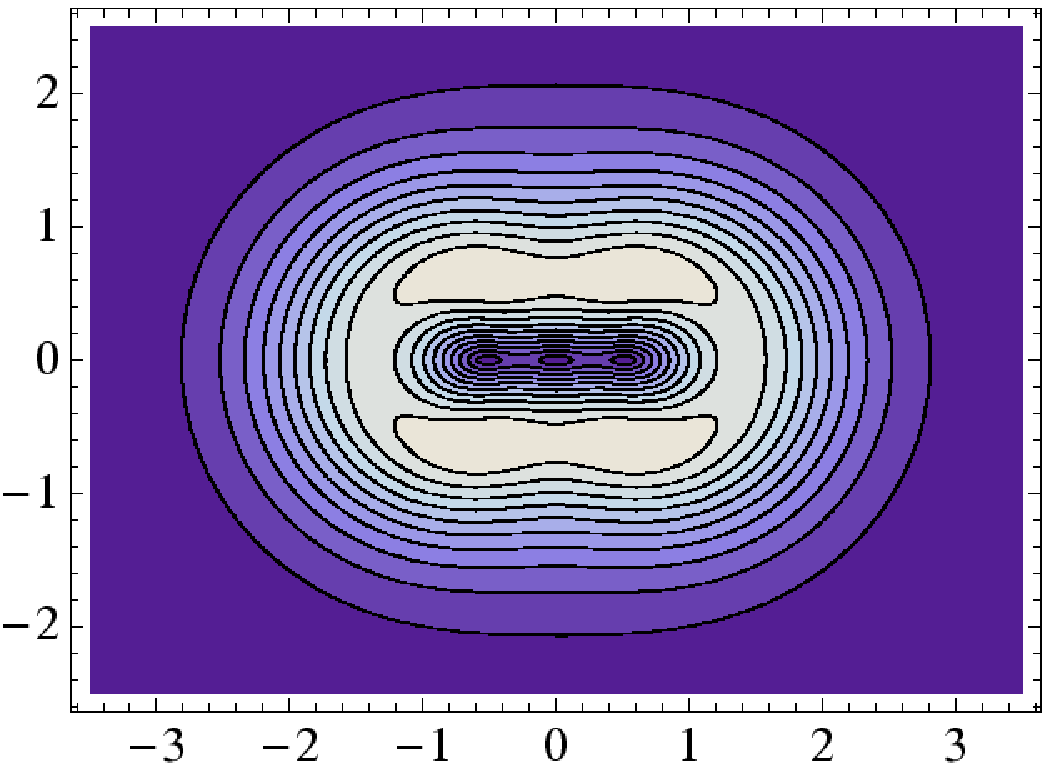}
\includegraphics[height=2.9cm]{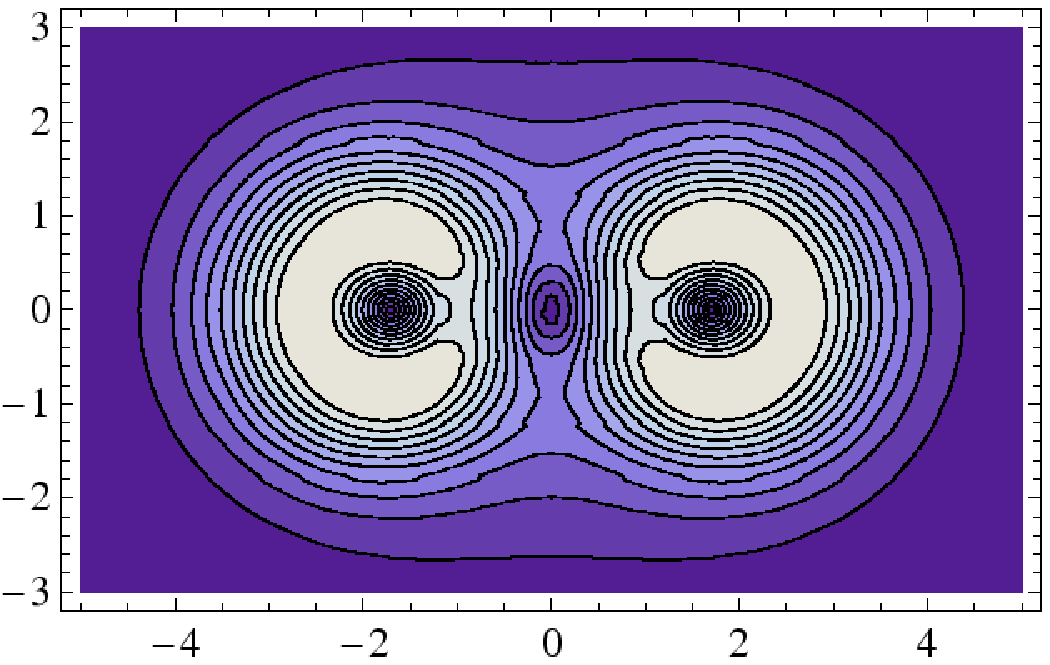}
\includegraphics[height=2.9cm]{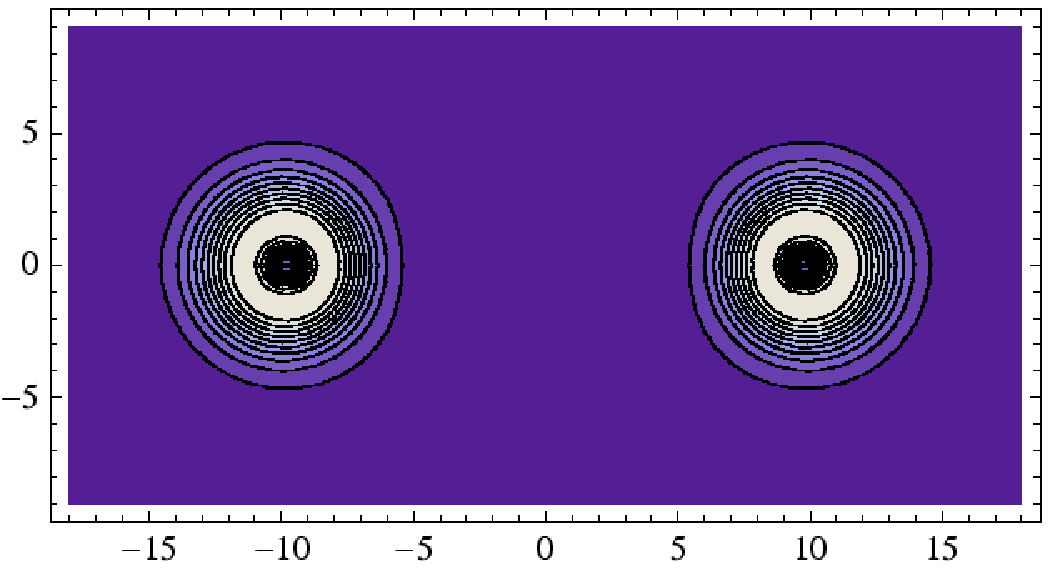}\\
\includegraphics[height=2.9cm]{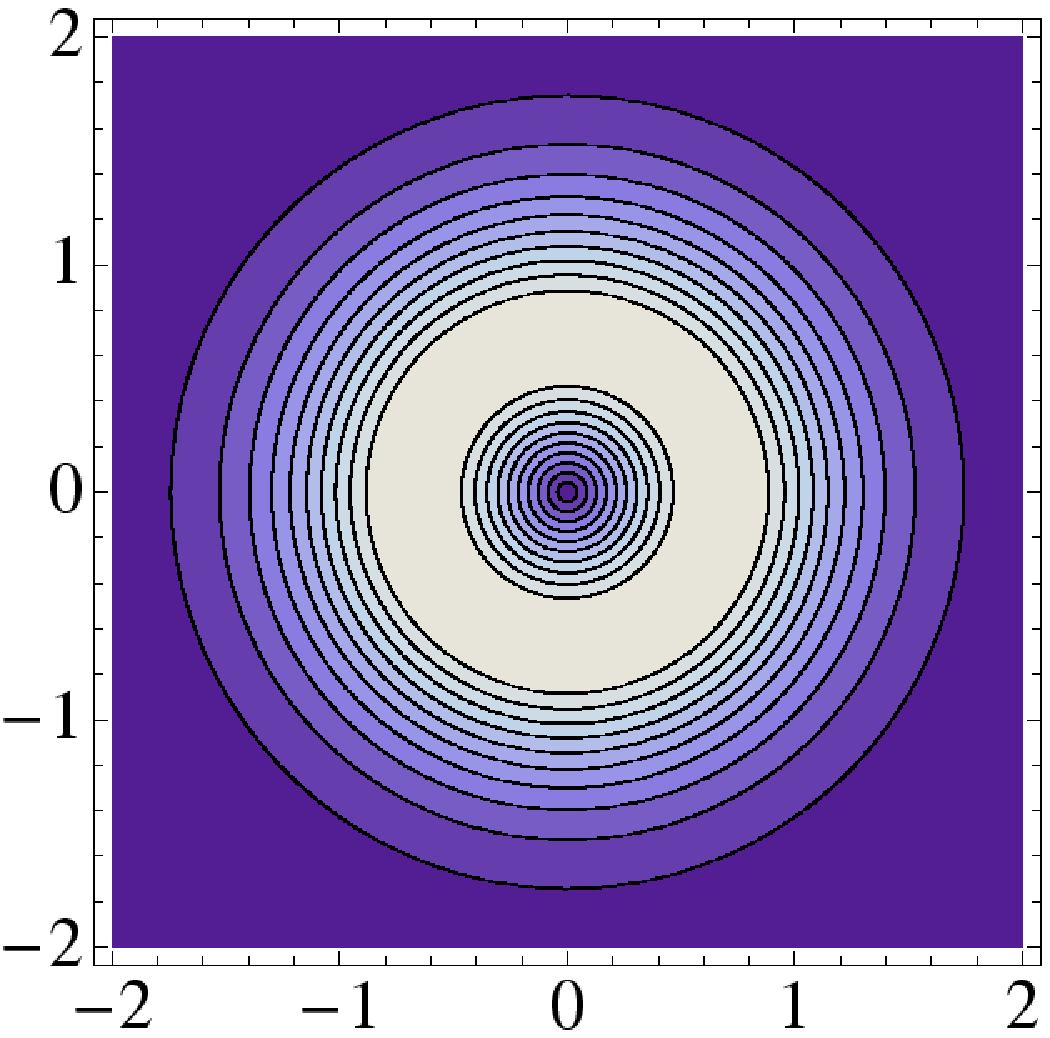}
\includegraphics[height=2.9cm]{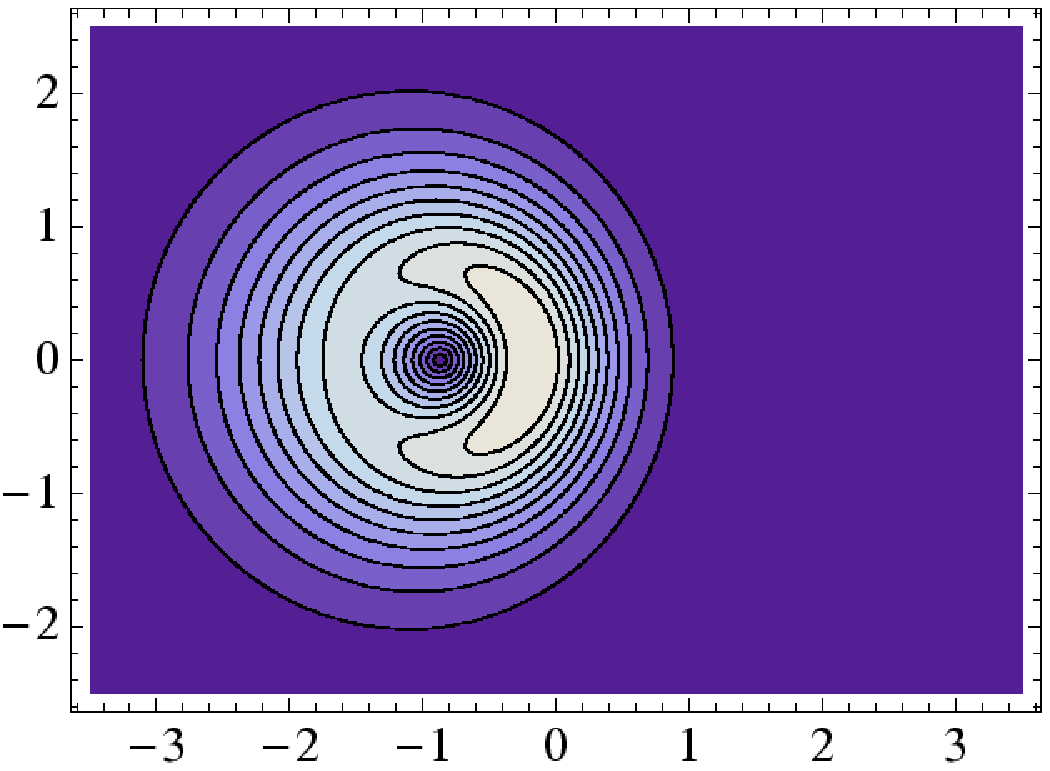}
\includegraphics[height=2.9cm]{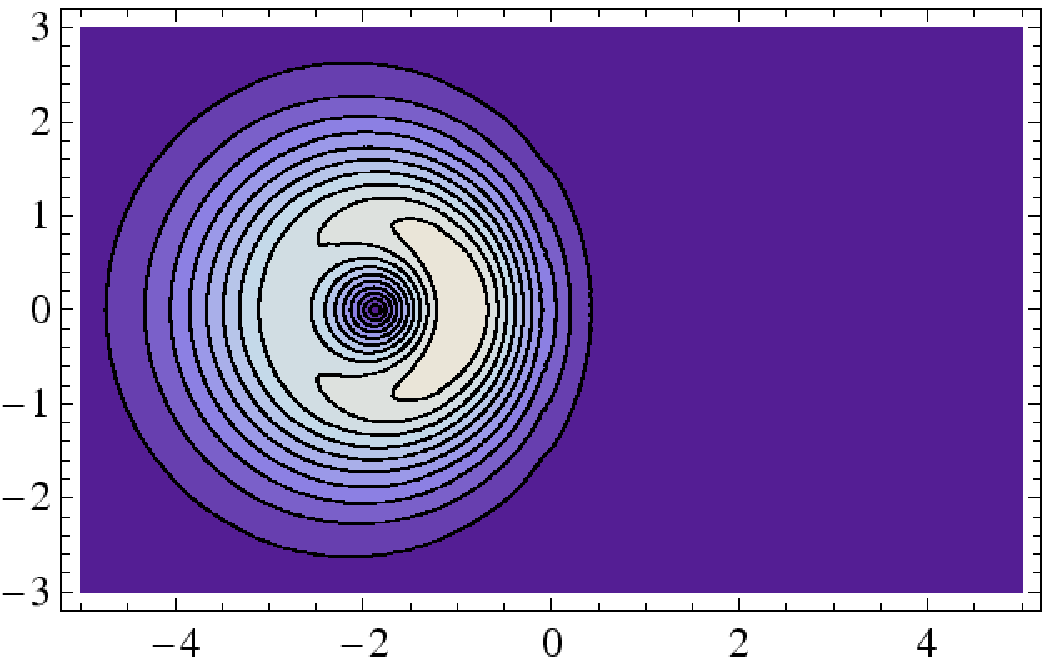}
\includegraphics[height=2.9cm]{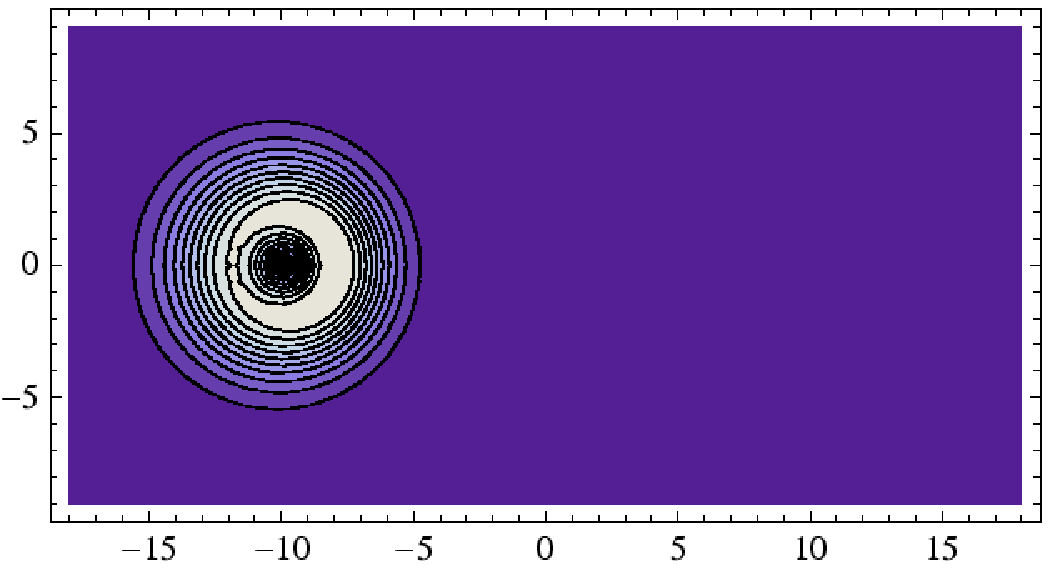}
\end{center}
\caption{The non-Abelian Chern-Simons fractional vortex with
  $G'=SO(4)$ and 
  $G'=USp(4)$ for $\kappa=\mu=2$, where the rows of the figure
  correspond to the energy density, the Abelian magnetic flux
  $F_{12}^0$, the non-Abelian  magnetic flux $F_{12}^2$, the magnitude
  of the Abelian electric field $|E_i^0|$ and finally the magnitude of
  the non-Abelian electric field $|E_i^2|$, whereas the columns of the
  figure correspond to the separation distance $2d=2\{0,1,2,10\}$. We
  have set $\xi=2$. } 
\label{fig:frac_equalcouplings}
\end{figure}
In Fig.~\ref{fig:frac_equalcouplings} is shown a matrix of graphs of
the non-Abelian Chern-Simons vortex configuration for $\kappa=\mu=2$, 
where the rows correspond to the energy density, the Abelian magnetic
flux $F_{12}^0$, the non-Abelian magnetic flux $F_{12}^2$, the
magnitude of the Abelian electric field $|E_i^0|$ and finally the
magnitude of the non-Abelian electric field $|E_i^2|$, whereas the
columns correspond to the separation distance $2d=2\{0,1,2,10\}$. The
centers of the fractional sub-peaks are placed at $z_1=d$ and
$z_2=-d$, respectively. Note that the symmetry of the configuration
has allowed us to show only one of the non-Abelian fields,
viz.~$F_{12}^2$ for which the other field is given by the reflection
in the $x$-axis $F_{12}^1(x,y)=F_{12}^2(-x,y)$. Analogously for the
non-Abelian electric field strengths. If we look at the $d=0$ column,
we see that all fields are ring-like, except the energy density which
has a bell-like shape with a tiny valley on the top. 
For the non-zero but small separation $2d=2$ (the second column in the
Figure), we see that the magnetic flux densities are no more ring-like
structures but (distorted) bell-like ones. The energy density has also
become a distorted bell-like structure and finally the Abelian
electric field strength is a squashed ring, while the non-Abelian
electric field strength is a ring-like structure (a bit distorted
though). The tendency for larger separation, is that the energy is
described by two bell-like shapes and so is the Abelian magnetic flux
density. Each Abelian peak of magnetic flux has a single corresponding
peak of non-Abelian flux (each with a different Cartan generator). The
electric field strengths, Abelian and non-Abelian retain their
ring-like status. The non-Abelian ones however split into rings at the
same positions as the non-Abelian magnetic flux peaks, whereas the
Abelian electric field has a ring on all the sites (i.e.~at each point
where the determinant of $H_0$ vanishes). 

\begin{figure}[!tbp]
\begin{center}
\mbox{\resizebox{!}{5.2cm}{\includegraphics{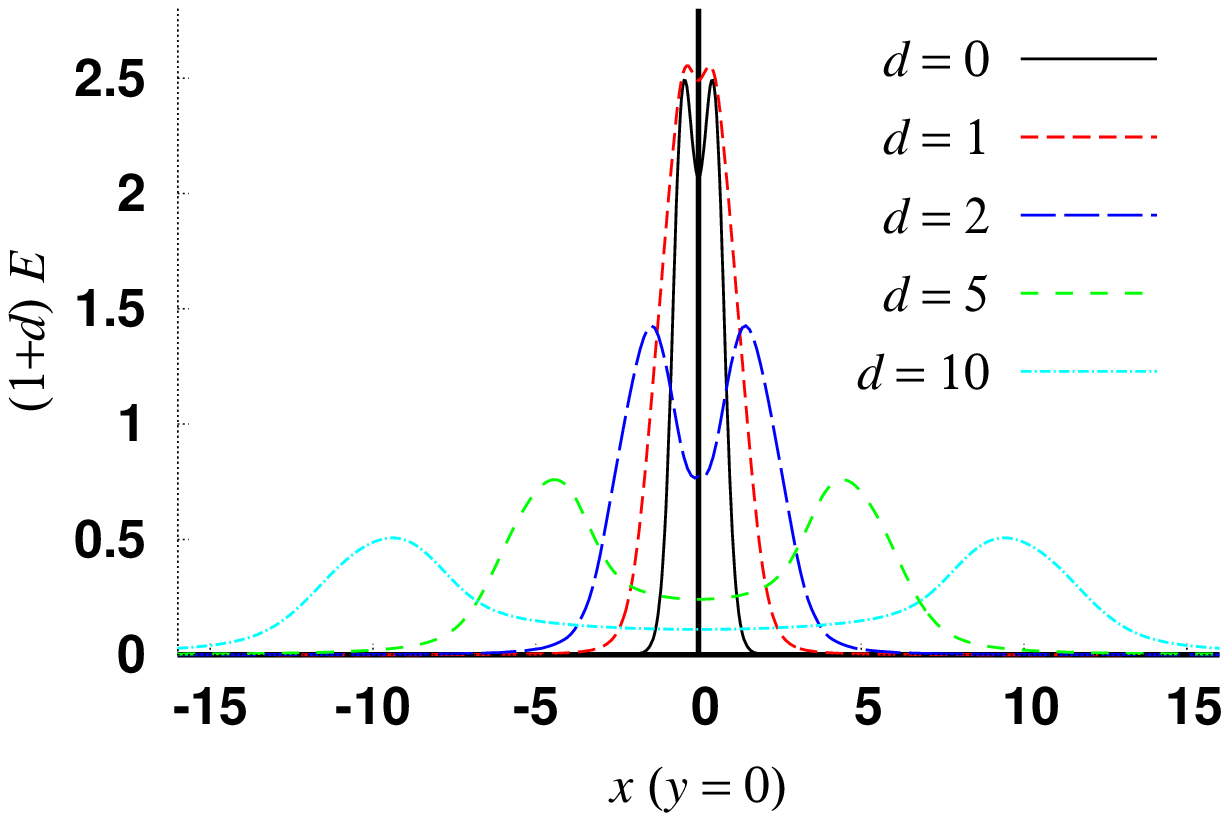}}}
\caption{The energy density, rescaled by $(1+d)$ with $2d$ being the
  separation distance as function of $d$. We have here set
  $\kappa=\mu=2$ and $\xi=2$. } 
\label{fig:frac_energyba_profile}
\end{center}
\end{figure}

\begin{figure}[!tbp]
\begin{center}
\mbox{
\subfigure[]{\resizebox{!}{5.2cm}{\includegraphics{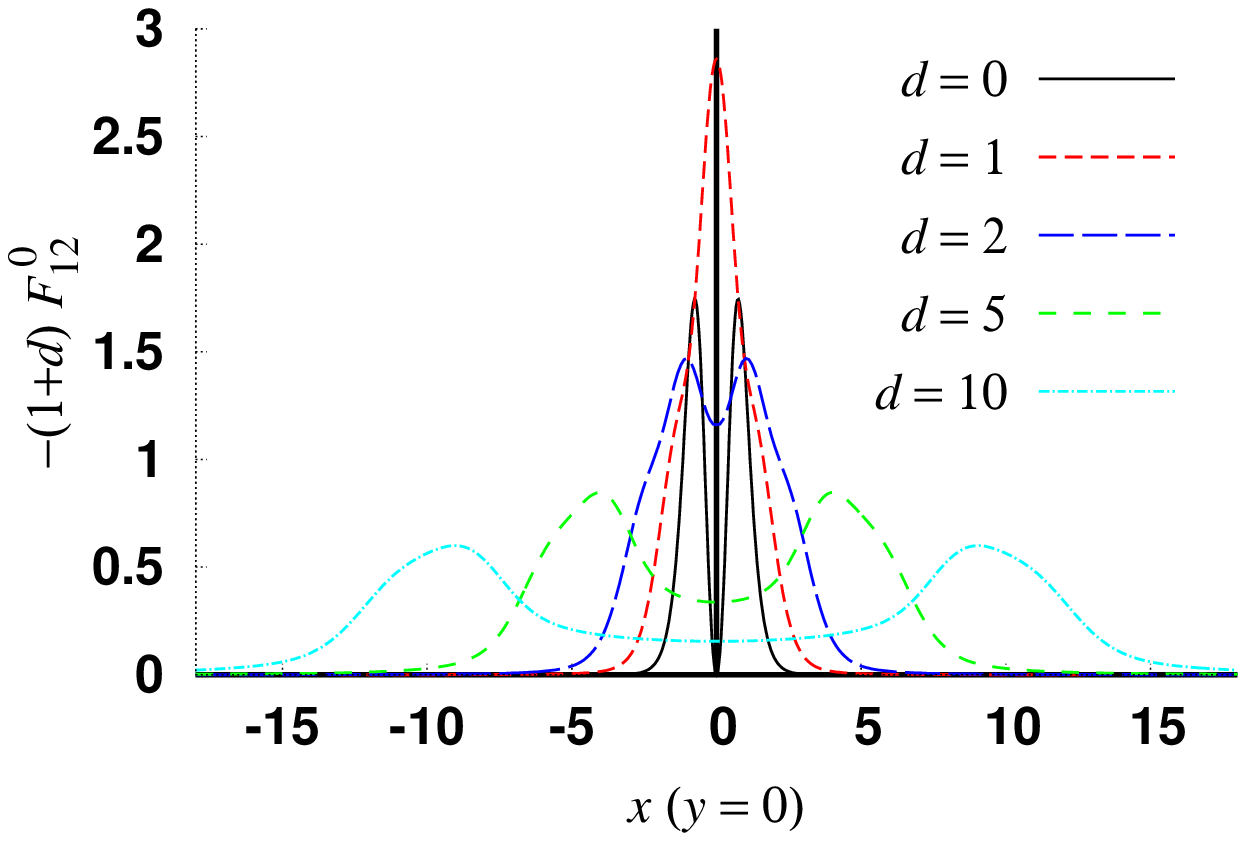}}}\quad
\subfigure[]{\resizebox{!}{5.2cm}{\includegraphics{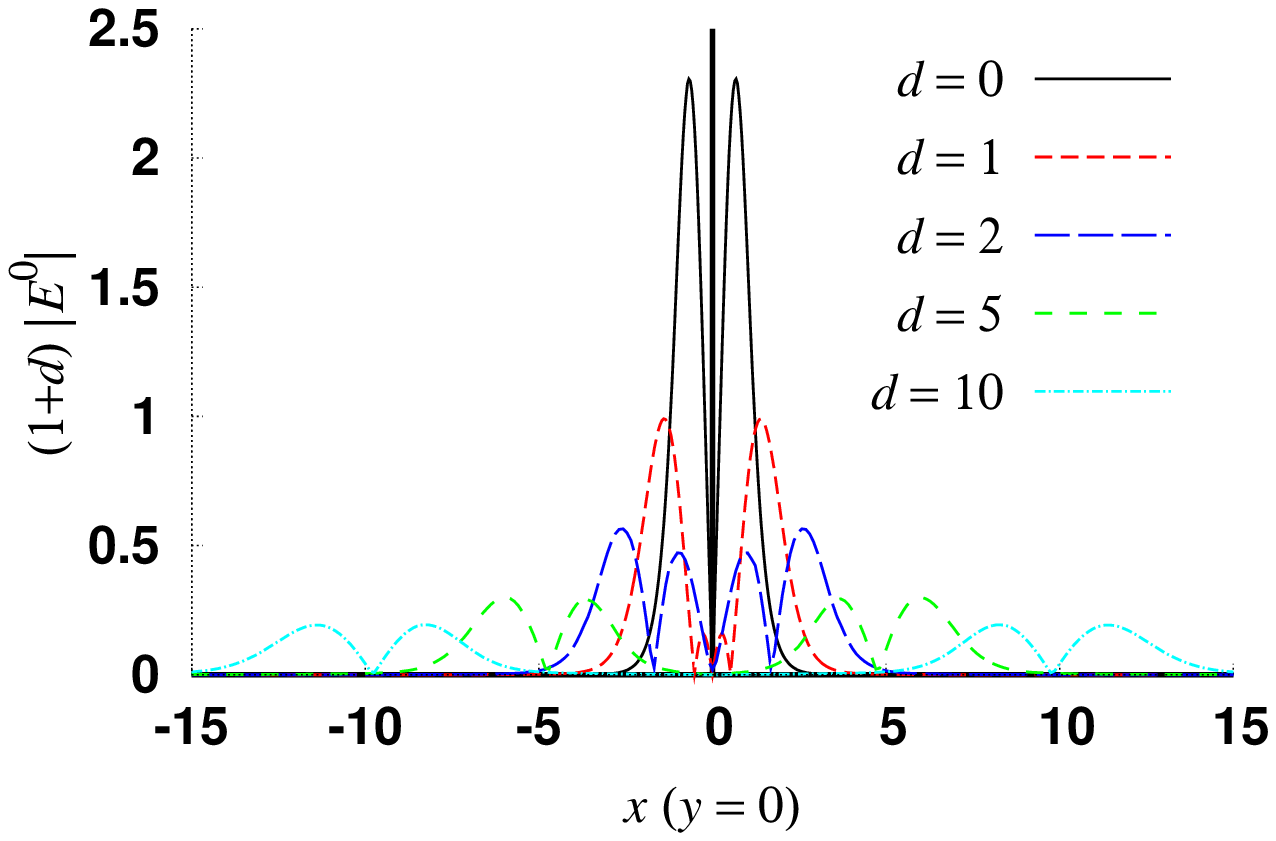}}}}
\caption{The Abelian (a) magnetic and (b) electric field strength
  density, rescaled by $(1+d)$ with $2d$ being the separation distance
  as function of $d$. We have set $\kappa=\mu=2$ and $\xi=2$. } 
\label{fig:frac_baea_profile}
\end{center}
\end{figure}

\begin{figure}[!tbp]
\begin{center}
\mbox{
\subfigure[]{\resizebox{!}{5.2cm}{\includegraphics{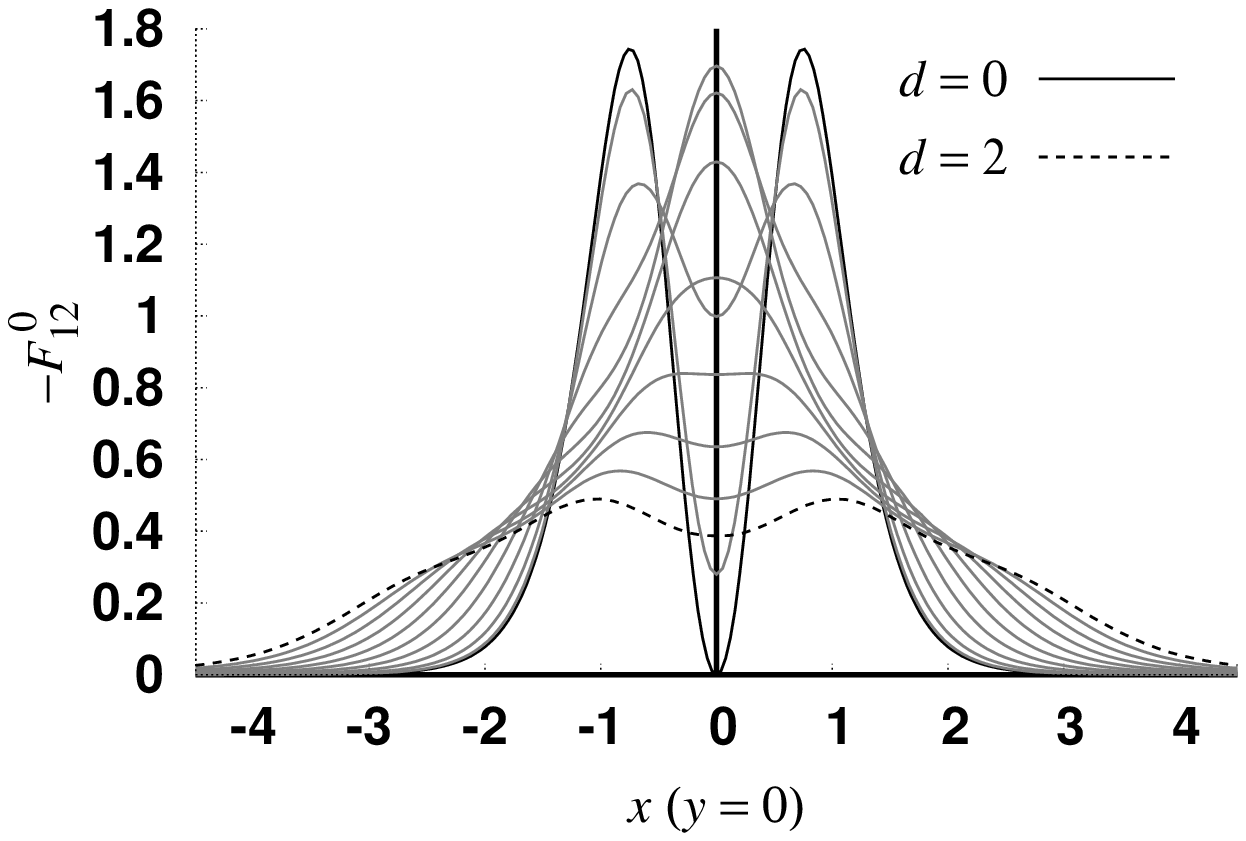}}}\quad
\subfigure[]{\resizebox{!}{5.2cm}{\includegraphics{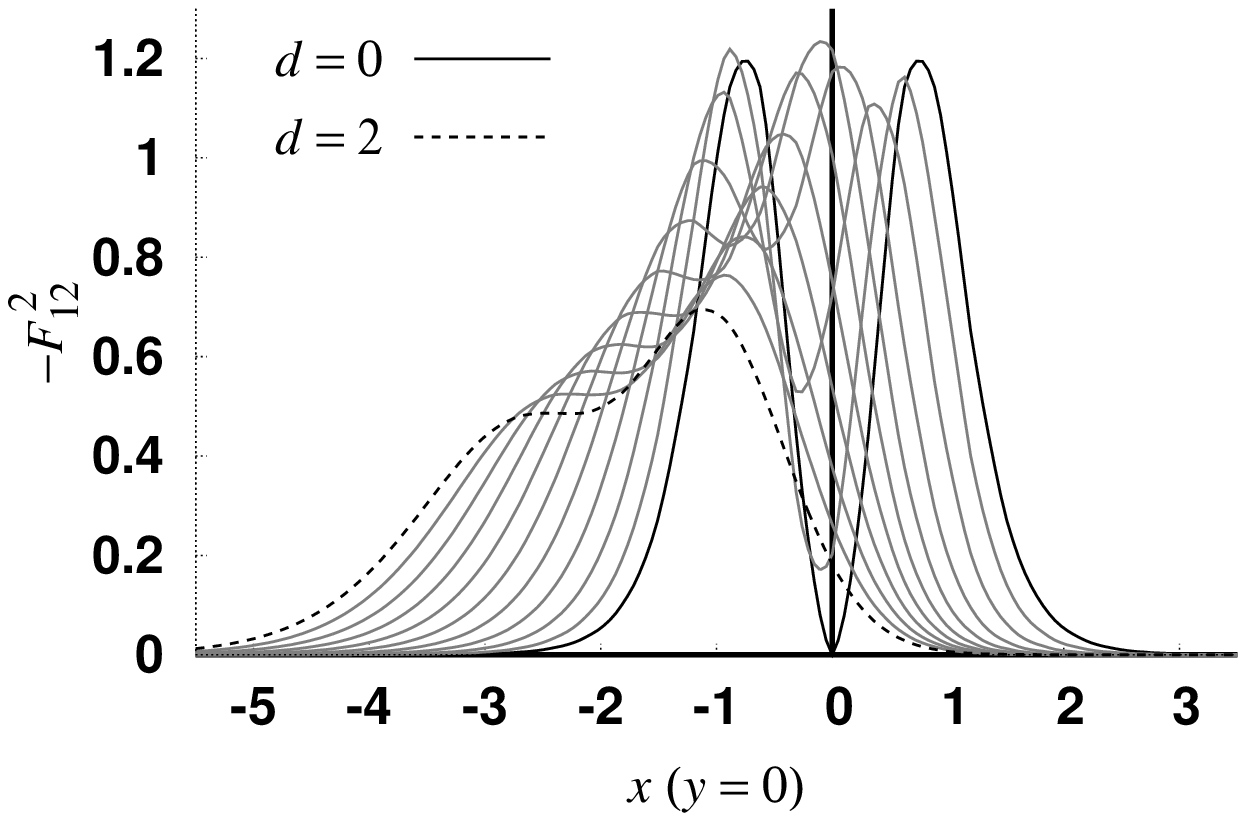}}}}
\caption{The (a) Abelian and (b) non-Abelian magnetic field strength
  density with $2d$ being the separation distance as function of 
  $d=\{0,0.2,0.4,0.6,0.8,1,1.2,1.4,1.6,1.8,2\}$. We have set
  $\kappa=\mu=2$ and $\xi=2$. }
\label{fig:frac_ba_profile_details}
\end{center}
\end{figure}

To understand the structure of the configuration, the contour plots of
Fig.~\ref{fig:frac_equalcouplings} are convenient, however, to get a
more detailed look at the solution, we show slices of the
configuration as function of $x$ for $y=0$ for the values of the
separation modulus parameter $d=\{0,1,2,5,10\}$. In
Fig.~\ref{fig:frac_energyba_profile} is shown the energy density as
function of $x$ for $y=0$, while in Fig.~\ref{fig:frac_baea_profile}
are shown the Abelian magnetic and electric field strengths on the
same slice. 
There is a transition from a ring-like structure to separate bell-like
structures in the Abelian magnetic flux density. We show in
Fig.~\ref{fig:frac_ba_profile_details}a, a detailed graph of the
Abelian magnetic flux with various values of the separation modulus
parameter $d$ from $d=0$ to $d=2$ and in
Fig.~\ref{fig:frac_ba_profile_details}b the corresponding non-Abelian
magnetic field strength $F_{12}^2$. 
We observe that the Abelian magnetic flux goes quite fast from being a
ring-like structure to become a single peak, which eventually spreads
into two sub-peaks that will depart from each other and dilute as the
separation modulus parameter is increased. For the non-Abelian
magnetic flux a similar situation is happening. The ring-structure 
becomes a single peak at the origin, which then moves (and gets
distorted) as $d$ is increased.

\subsection{Negative Abelian magnetic flux density at the origin:
  $\kappa>\mu$} 

For completeness, let us repeat the calculation with the different
Chern-Simons couplings chosen as in Ref.~\cite{Gudnason:2009ut}. First
we consider the case with $\kappa=4,\mu=2$. The main difference
between the configurations with $\kappa=\mu$ and those with
$\kappa\neq\mu$ lies in the magnetic field strengths. Therefore we will
focus on those. 
\begin{figure}[!tbp]
\begin{center}
\mbox{\resizebox{!}{5.2cm}{\includegraphics{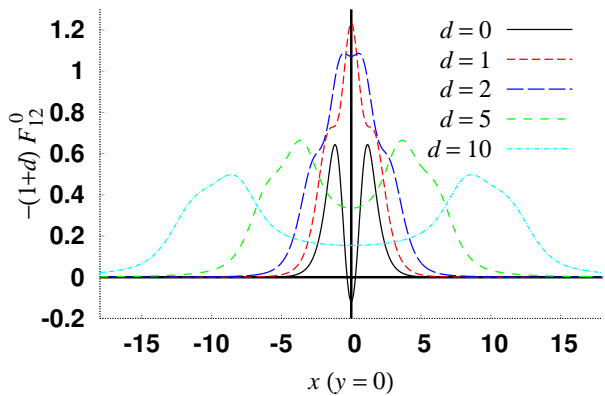}}}
\caption{The Abelian magnetic field strength density, rescaled by
  $(1+d)$ with $2d$ being the separation distance as function of 
  $d$. Here the couplings are $\kappa=4$ and $\mu=2$, while we have
  set $\xi=2$. } 
\label{fig:frac_kgr_ba_profile}
\end{center}
\end{figure}
\begin{figure}[!tbp]
\begin{center}
\mbox{
\subfigure[]{\resizebox{!}{5.2cm}{\includegraphics{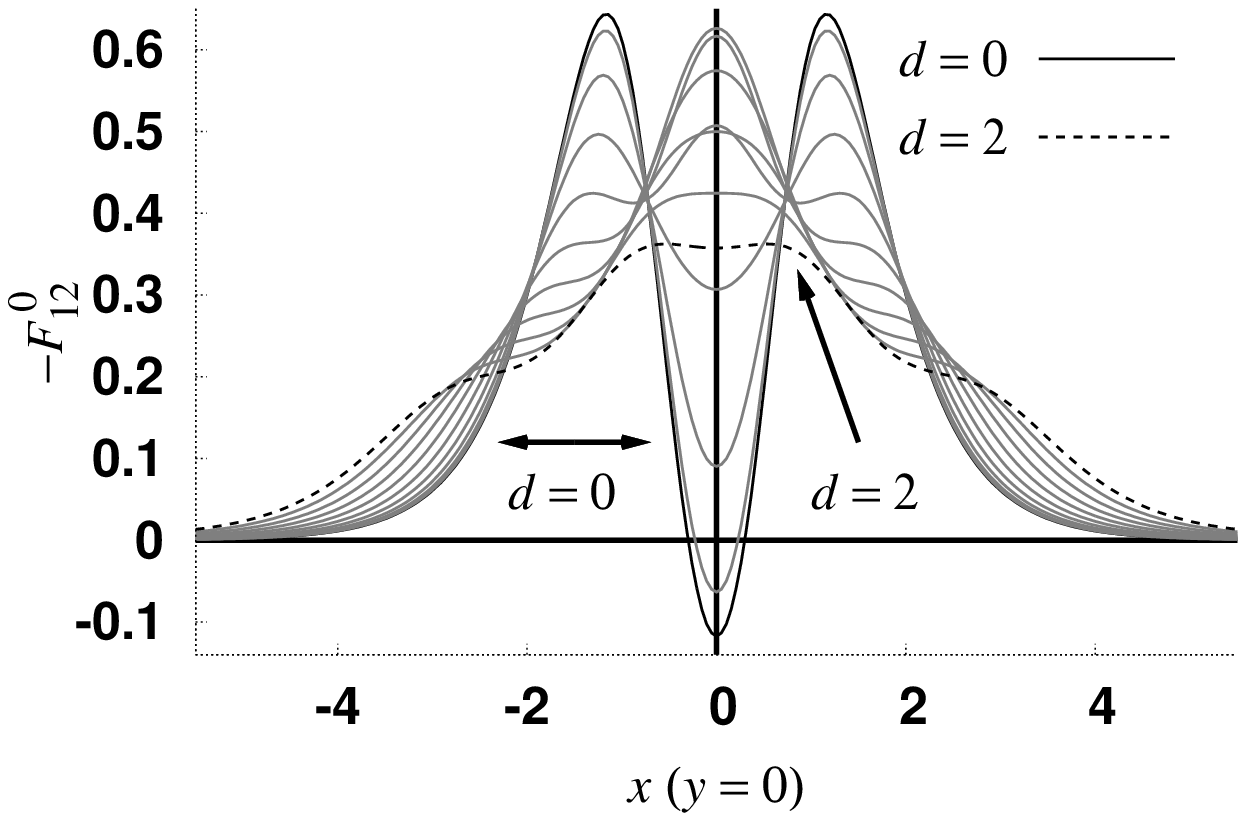}}}\quad
\subfigure[]{\resizebox{!}{5.2cm}{\includegraphics{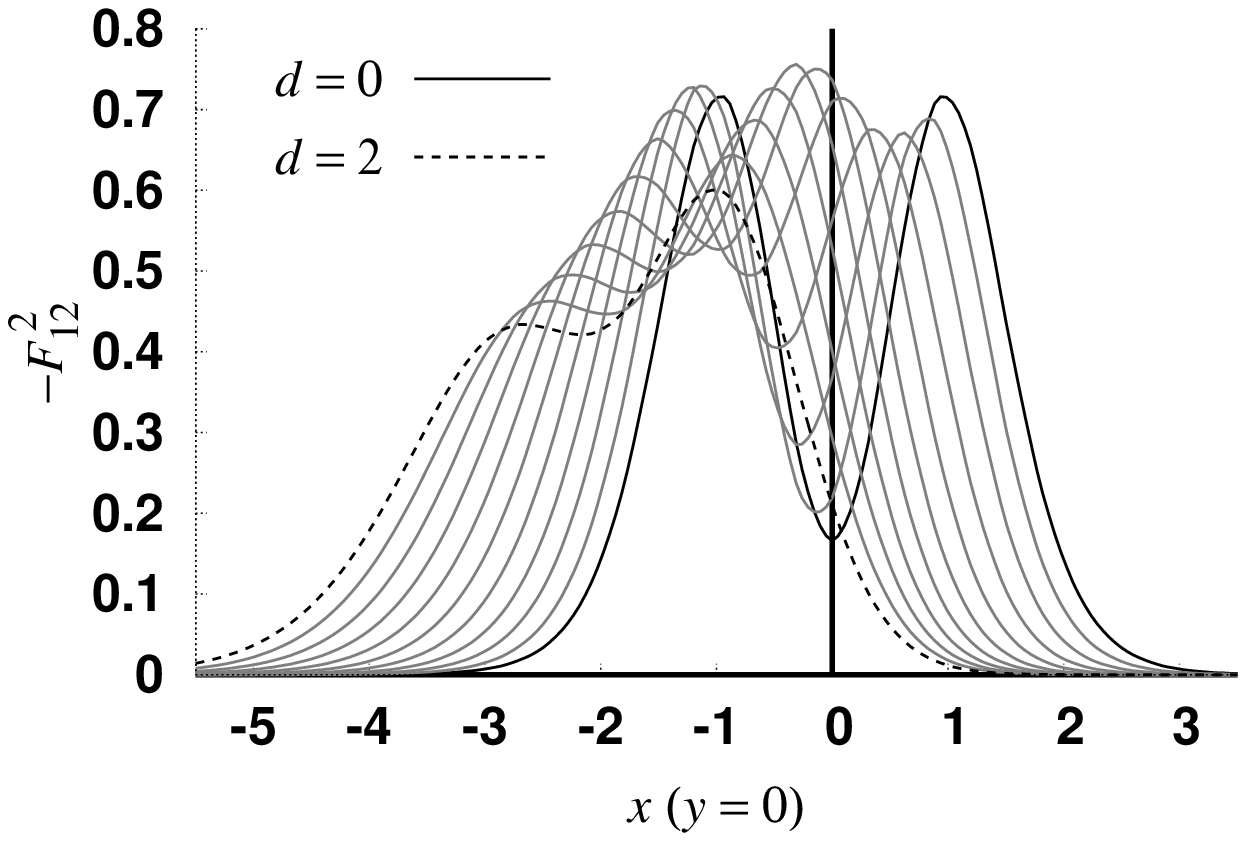}}}}
\caption{The (a) Abelian and (b) non-Abelian magnetic field strength
  density with $2d$ being the separation distance as function of  
  $d=\{0,0.2,0.4,0.6,0.8,1,1.2,1.4,1.6,1.8,2\}$. Here the couplings 
  are $\kappa=4$ and $\mu=2$, while we have set $\xi=2$. }
\label{fig:frac_kgr_bn2a_profile_details}
\end{center}
\end{figure}
In Fig.~\ref{fig:frac_kgr_ba_profile} is shown the Abelian magnetic
flux density on the slice $(x,0)$ of the configuration for different
values of the separation modulus parameter $d=\{0,1,2,5,10\}$, while
in Fig.~\ref{fig:frac_kgr_bn2a_profile_details} are shown the Abelian
and non-Abelian magnetic flux densities in detail for various values
of $d$ ranging from $d=0$ to $d=2$. 
We observe a situation quite similar to the case with equal
couplings, except from the fact that the Abelian magnetic flux density
starts out being negative at the origin for $d=0$, while the
non-Abelian one remains positive at the origin. The end result for
large separation $2d$ is analogous to the equal coupling case.

\subsection{Positive Abelian magnetic flux density at the origin:
  $\kappa<\mu$} 

Let us now turn to the case with $\kappa=1,\mu=2$.
\begin{figure}[!tbp]
\begin{center}
\mbox{\resizebox{!}{5.2cm}{\includegraphics{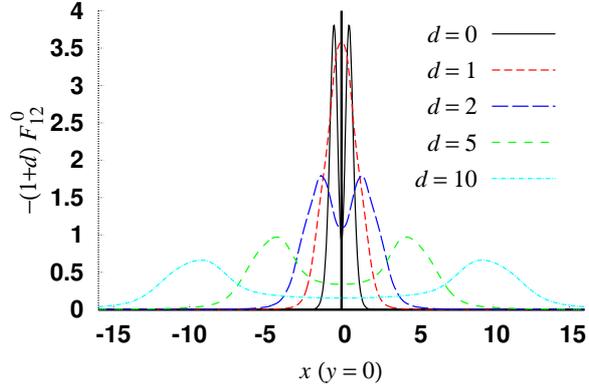}}}
\caption{The Abelian magnetic field strength density, rescaled by
  $(1+d)$ with $2d$ being the separation distance as function of
  $d$. Here the couplings are $\kappa=1$ and $\mu=2$, while we have
  set $\xi=2$. }
\label{fig:frac_kle_ba_profile}
\end{center}
\end{figure}
\begin{figure}[!tbp]
\begin{center}
\mbox{
\subfigure[]{\resizebox{!}{5.2cm}{\includegraphics{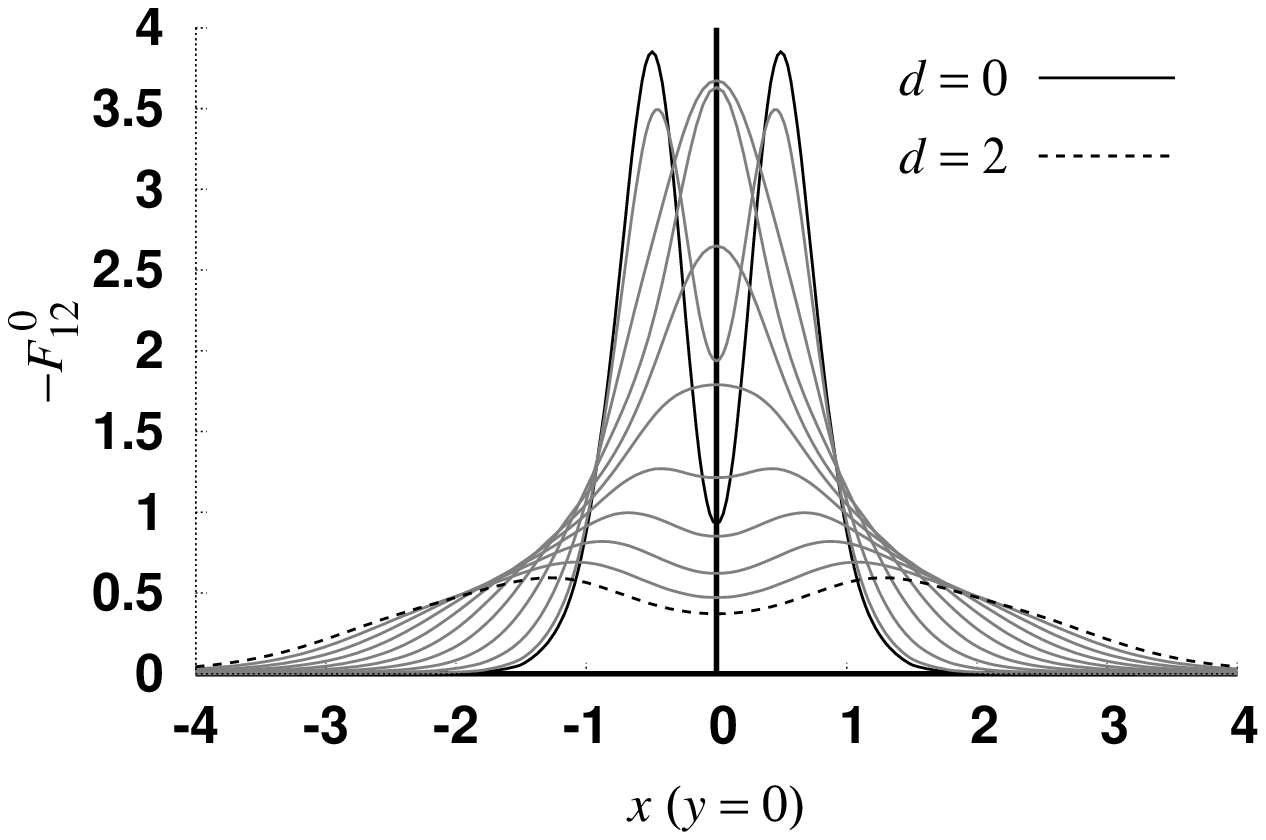}}}\quad
\subfigure[]{\resizebox{!}{5.2cm}{\includegraphics{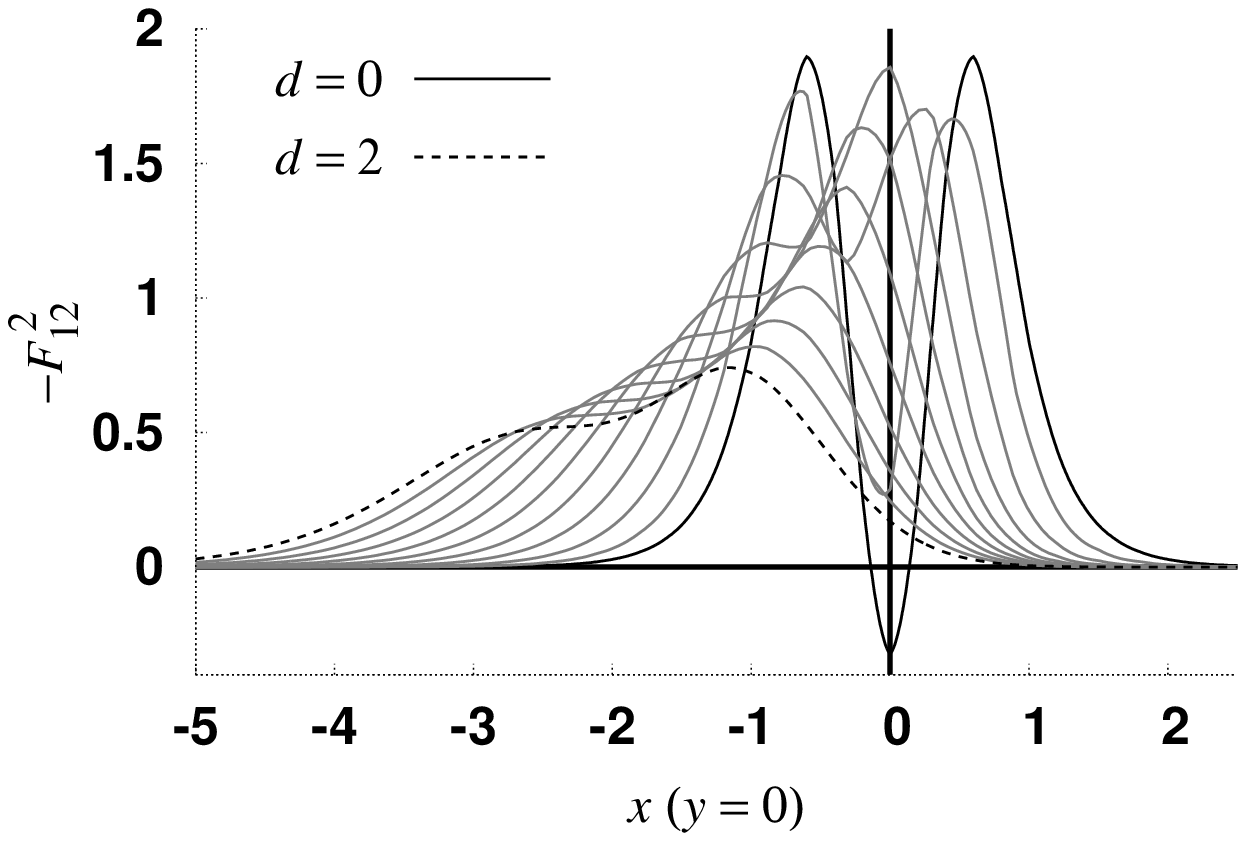}}}} 
\caption{The (a) Abelian and (b) non-Abelian magnetic field strength
  density with $2d$ being the separation distance as function of
  $d=\{0,0.2,0.4,0.6,0.8,1,1.2,1.4,1.6,1.8,2\}$. Here the couplings 
  are $\kappa=1$ and $\mu=2$, while we have set $\xi=2$. }
\label{fig:frac_kle_bn2a_profile_details}
\end{center}
\end{figure}
In Fig.~\ref{fig:frac_kle_ba_profile} is shown the Abelian magnetic
flux density on the slice $(x,0)$ of the configuration for different
values of the separation modulus parameter $d=\{0,1,2,5,10\}$, while
in Fig.~\ref{fig:frac_kle_bn2a_profile_details} are shown the Abelian
and non-Abelian magnetic flux densities in details for various values
of $d$ ranging from $d=0$ to $d=2$. 
We observe a situation quite similar to the case with equal
couplings, except from the fact that the Abelian magnetic flux density
starts out being positive at the origin for $d=0$, while the
non-Abelian one is negative at the origin. The end result for
large separation $2d$ is analogous to the equal coupling case.

\subsection{Bell-like structures: $-\kappa=\mu=2$}

The last case studied in Ref.~\cite{Gudnason:2009ut}, was the vortex
with negative Abelian Chern-Simons coupling and positive non-Abelian
coupling: $k=-2$ and $\mu=2$. 

\begin{figure}[!tbp]
\begin{center}
\includegraphics[height=2.9cm]{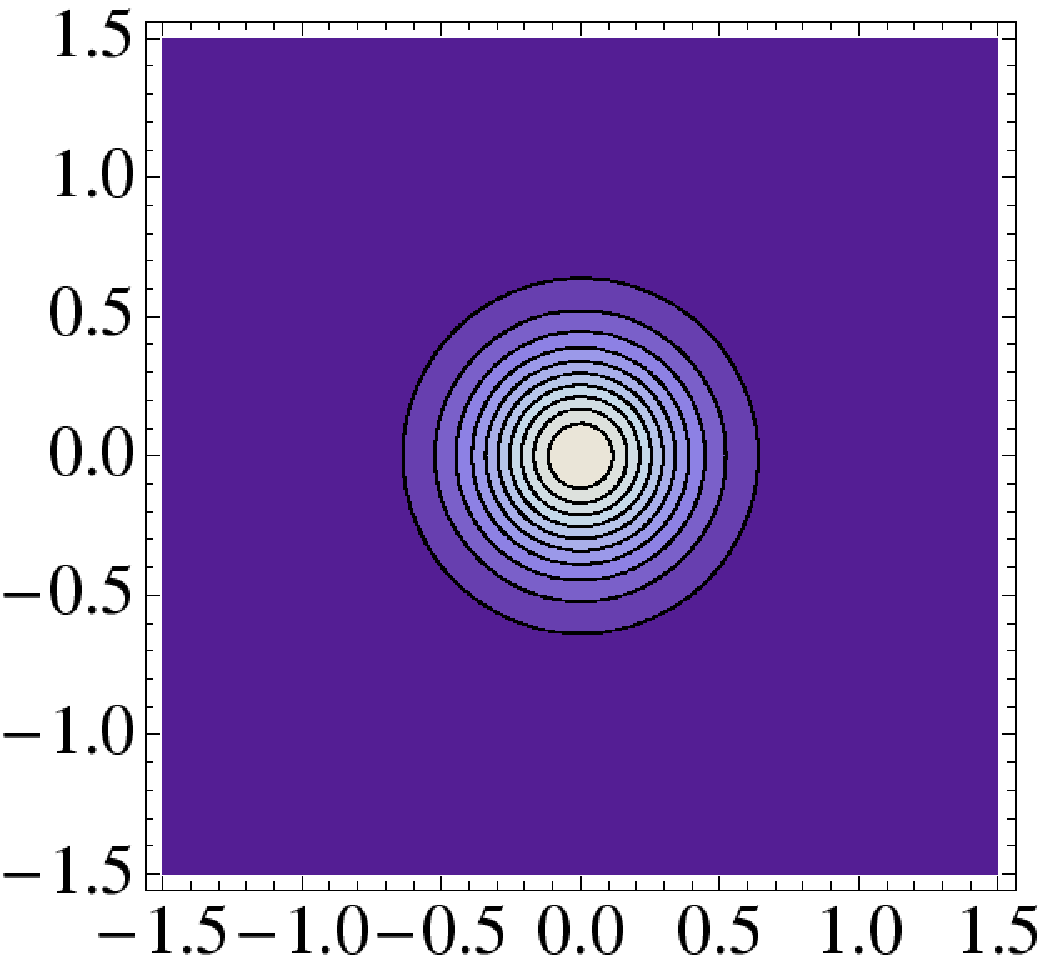}
\includegraphics[height=2.9cm]{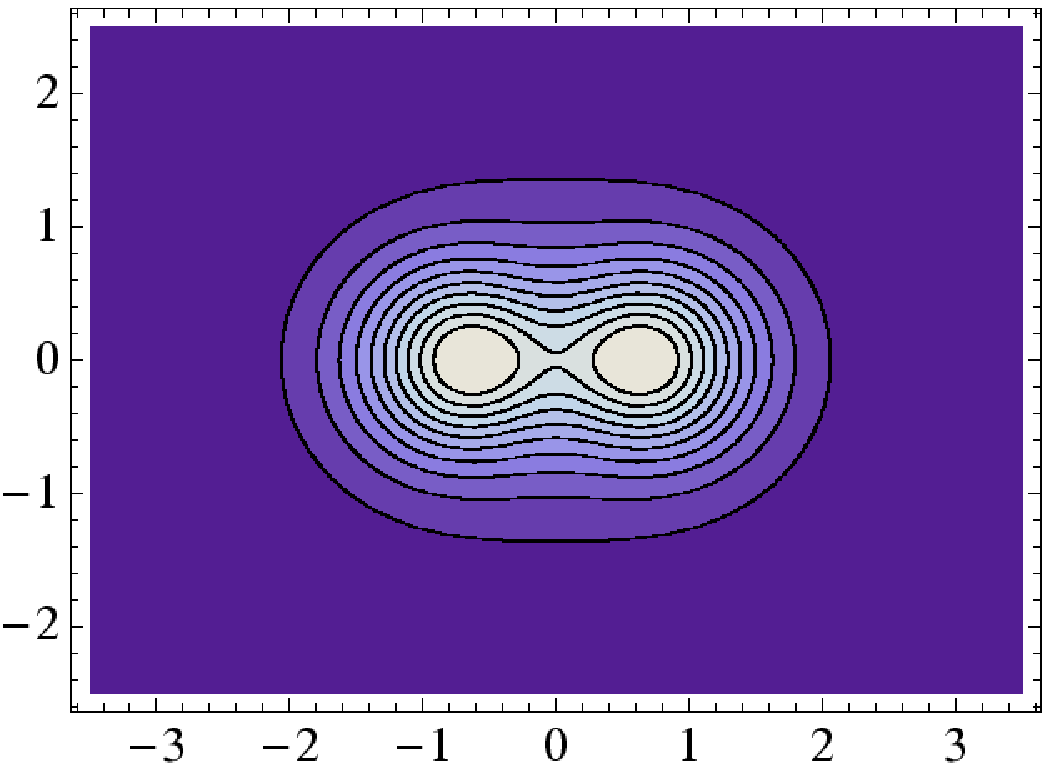}
\includegraphics[height=2.9cm]{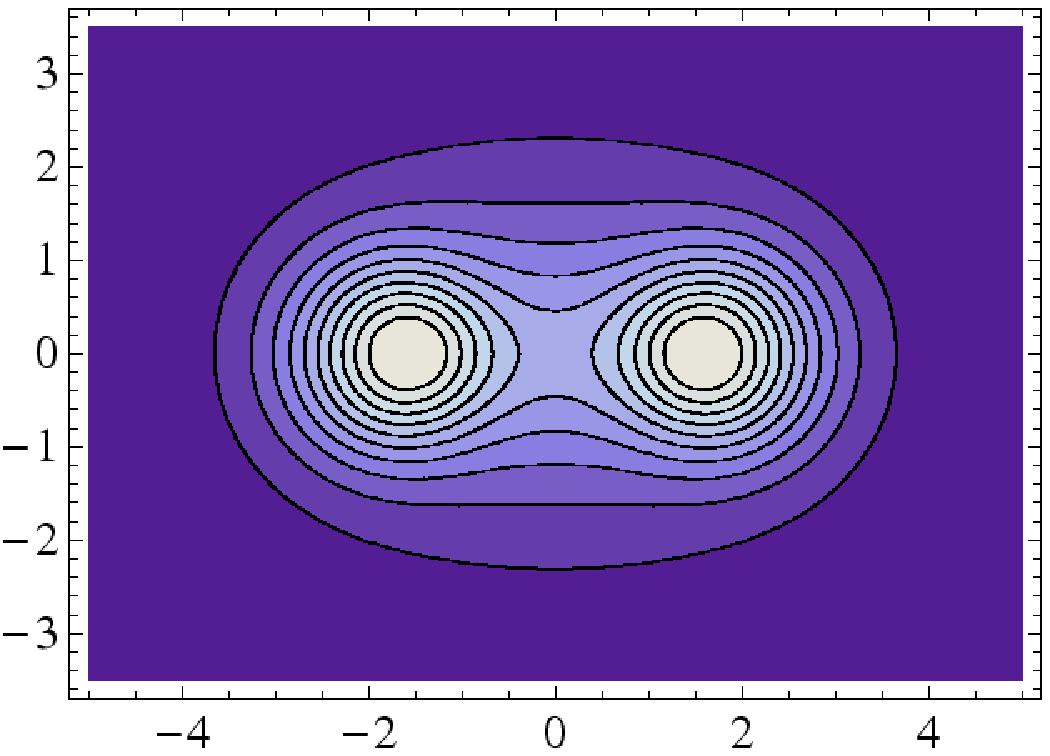}
\includegraphics[height=2.9cm]{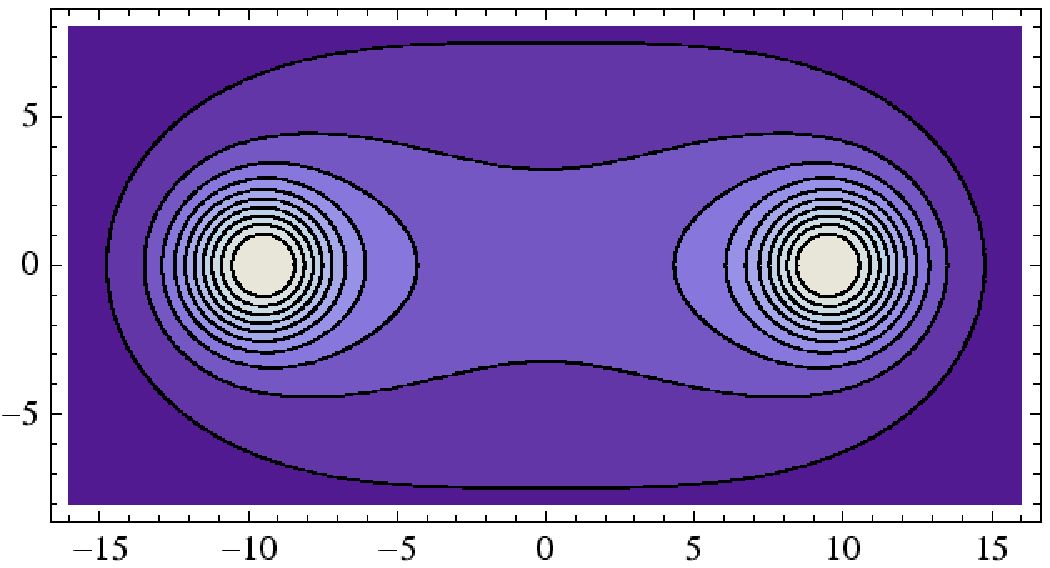}\\
\includegraphics[height=2.9cm]{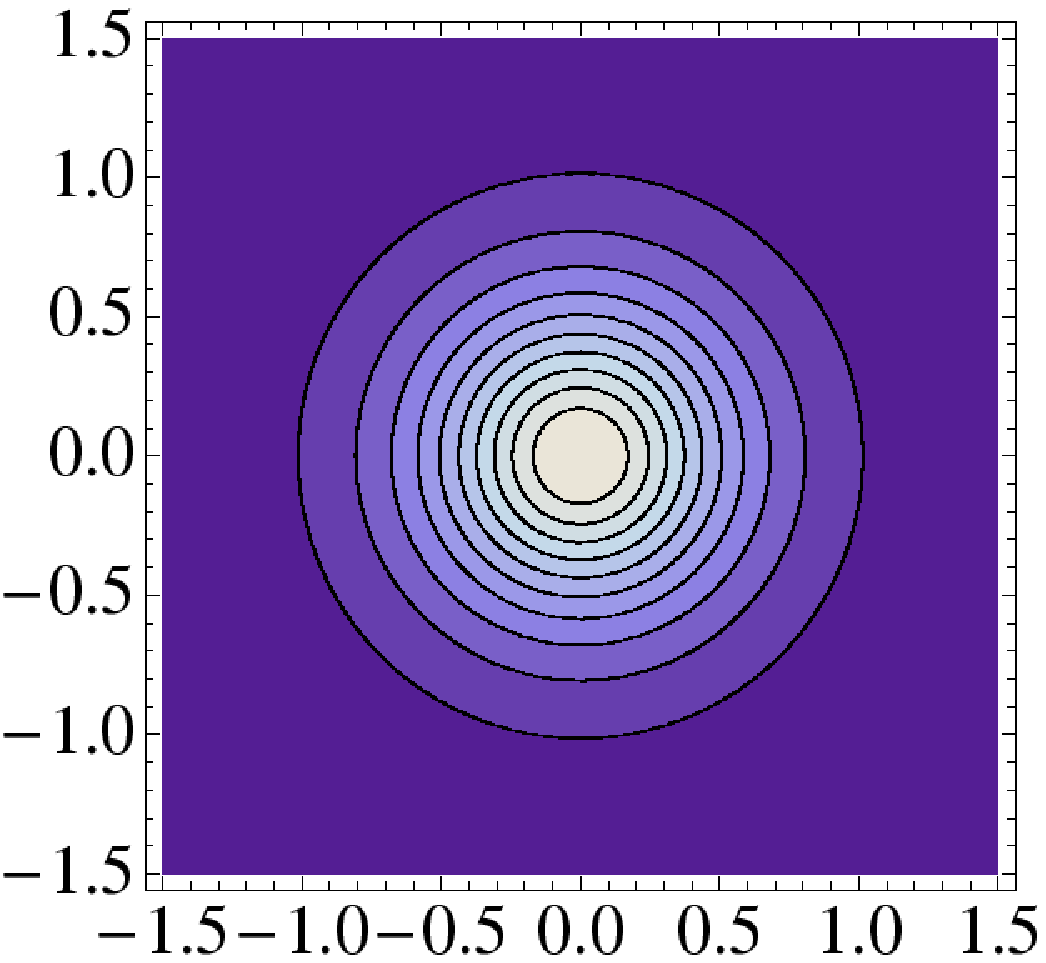}
\includegraphics[height=2.9cm]{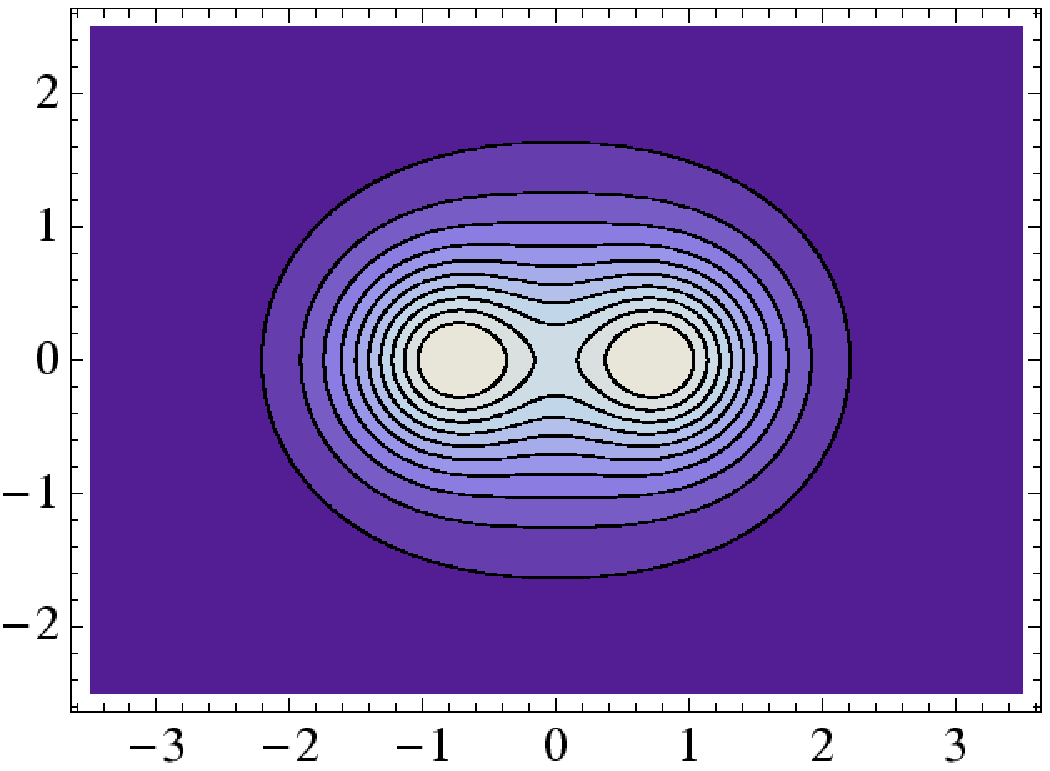}
\includegraphics[height=2.9cm]{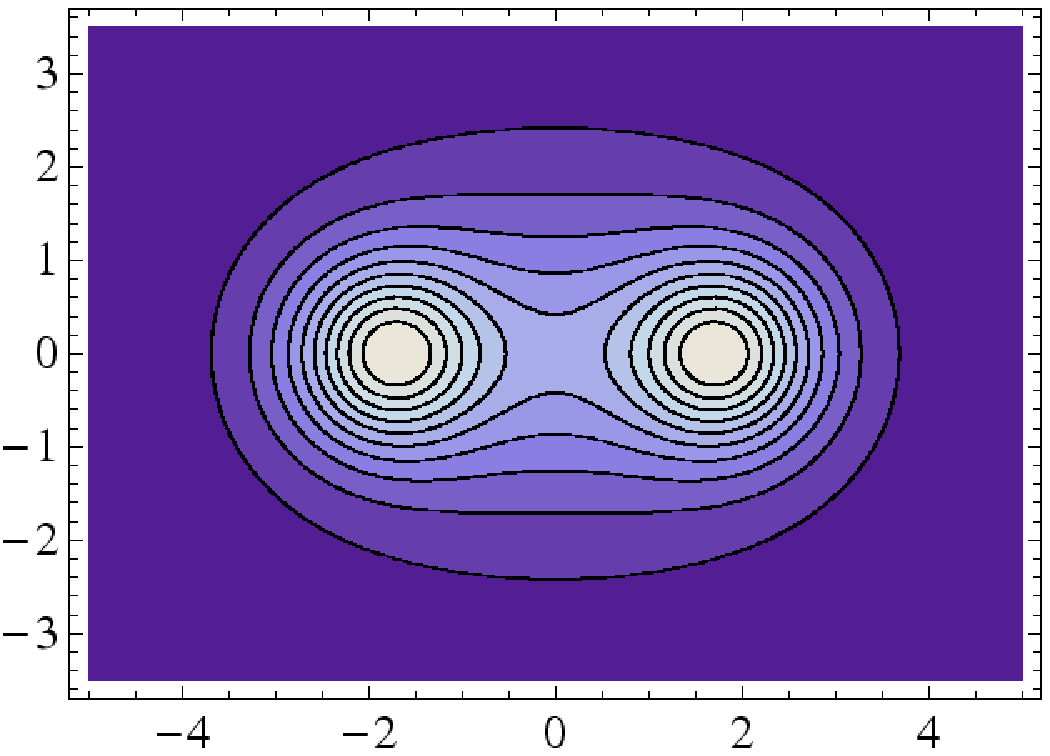}
\includegraphics[height=2.9cm]{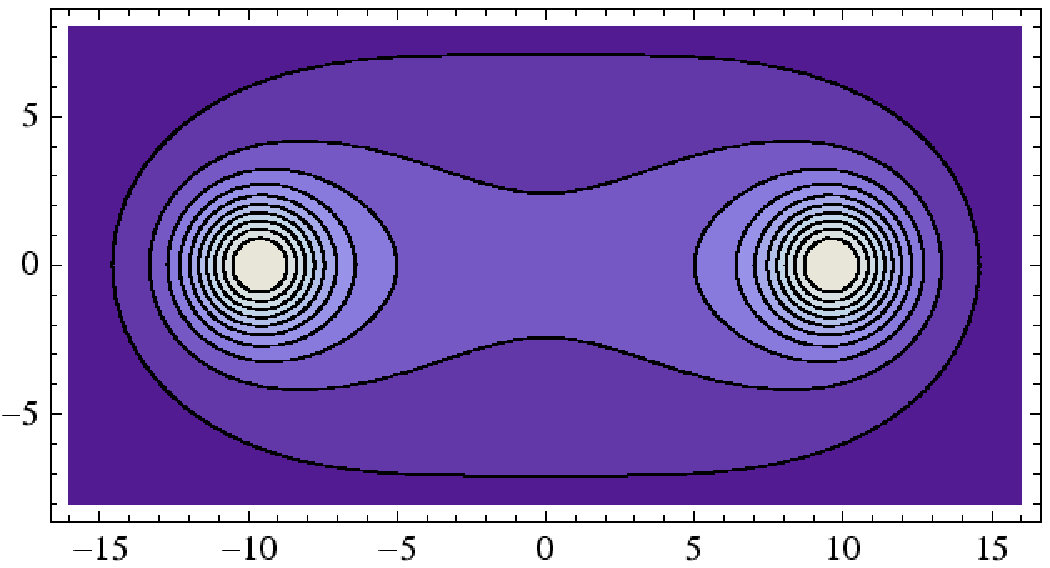}\\
\includegraphics[height=2.9cm]{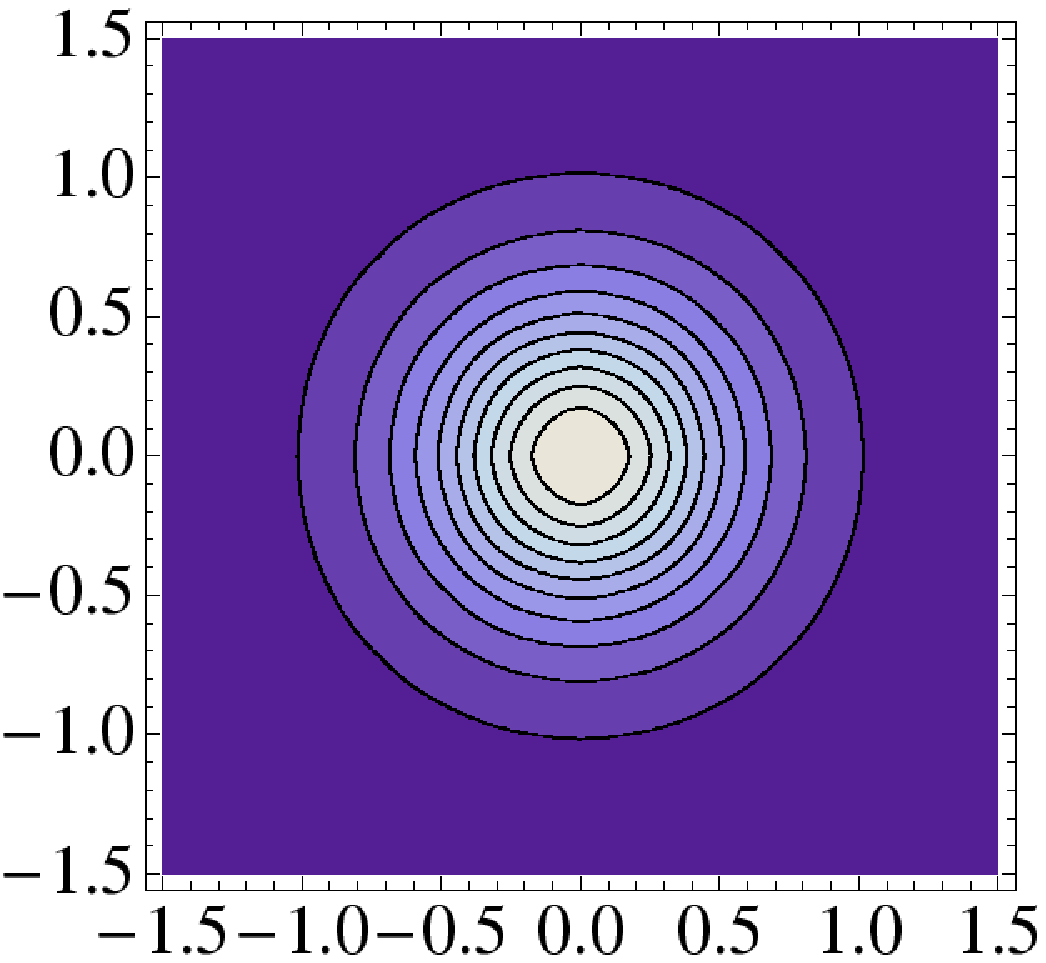}
\includegraphics[height=2.9cm]{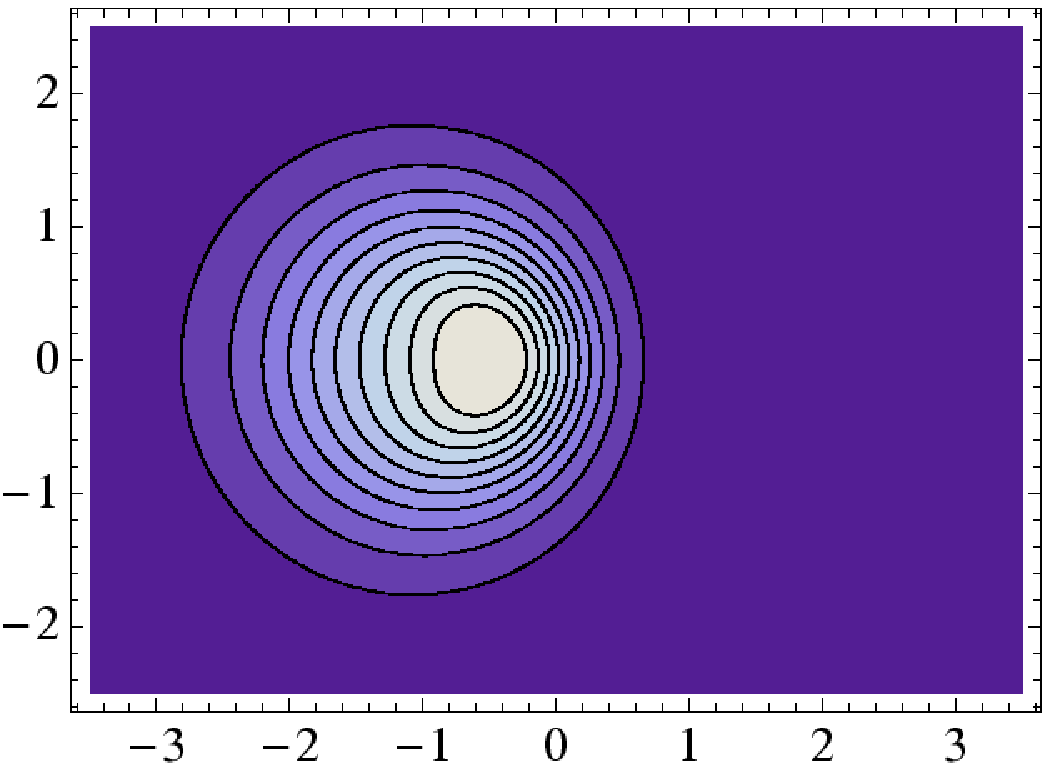}
\includegraphics[height=2.9cm]{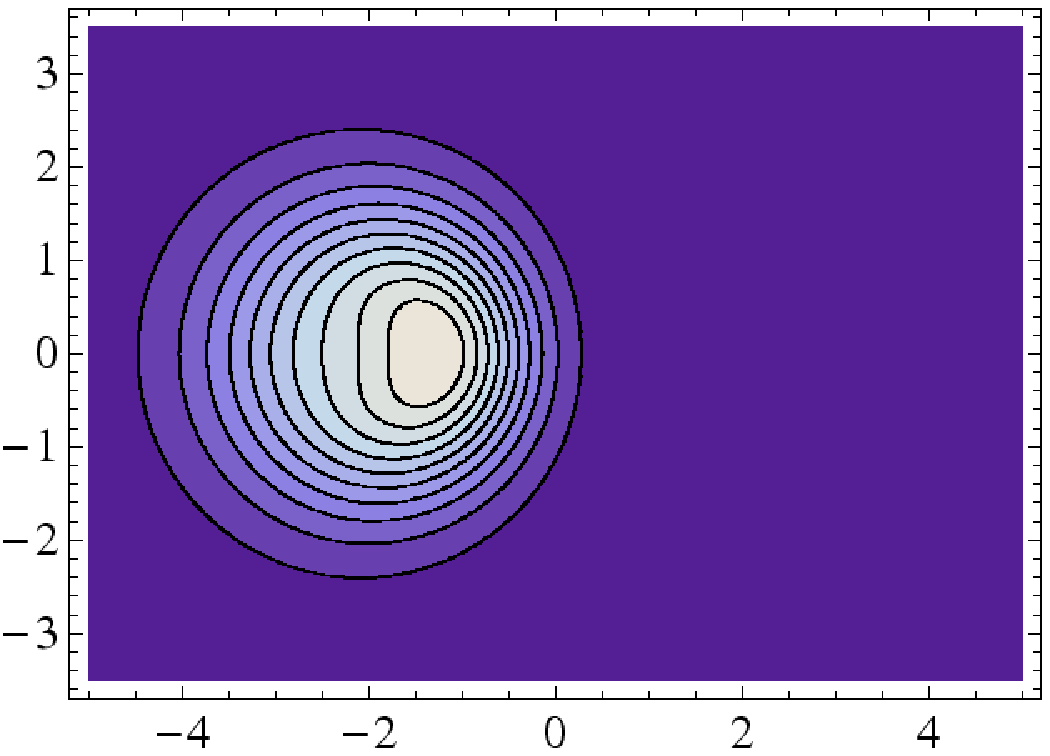}
\includegraphics[height=2.9cm]{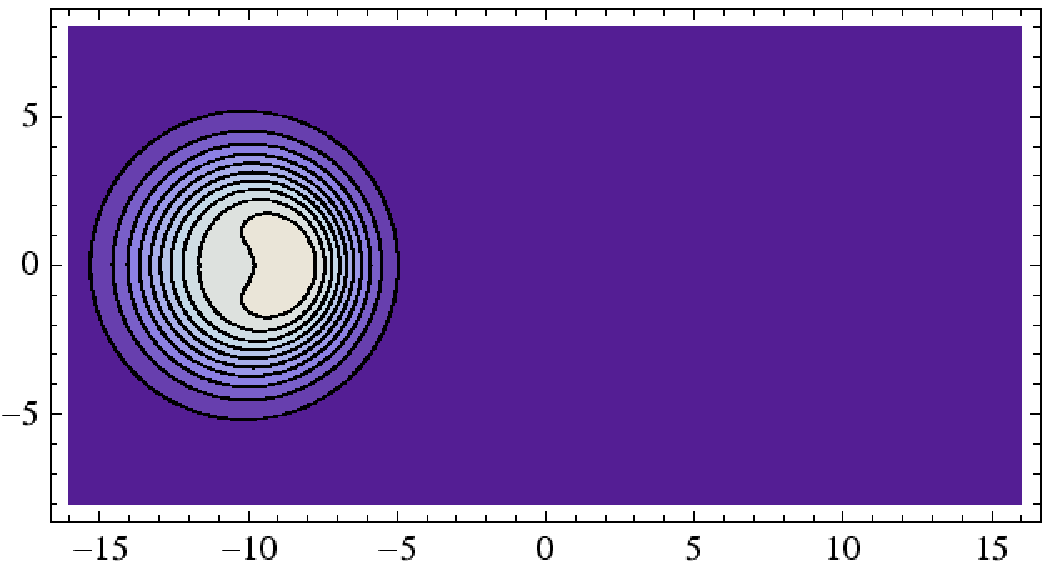}\\
\includegraphics[height=2.9cm]{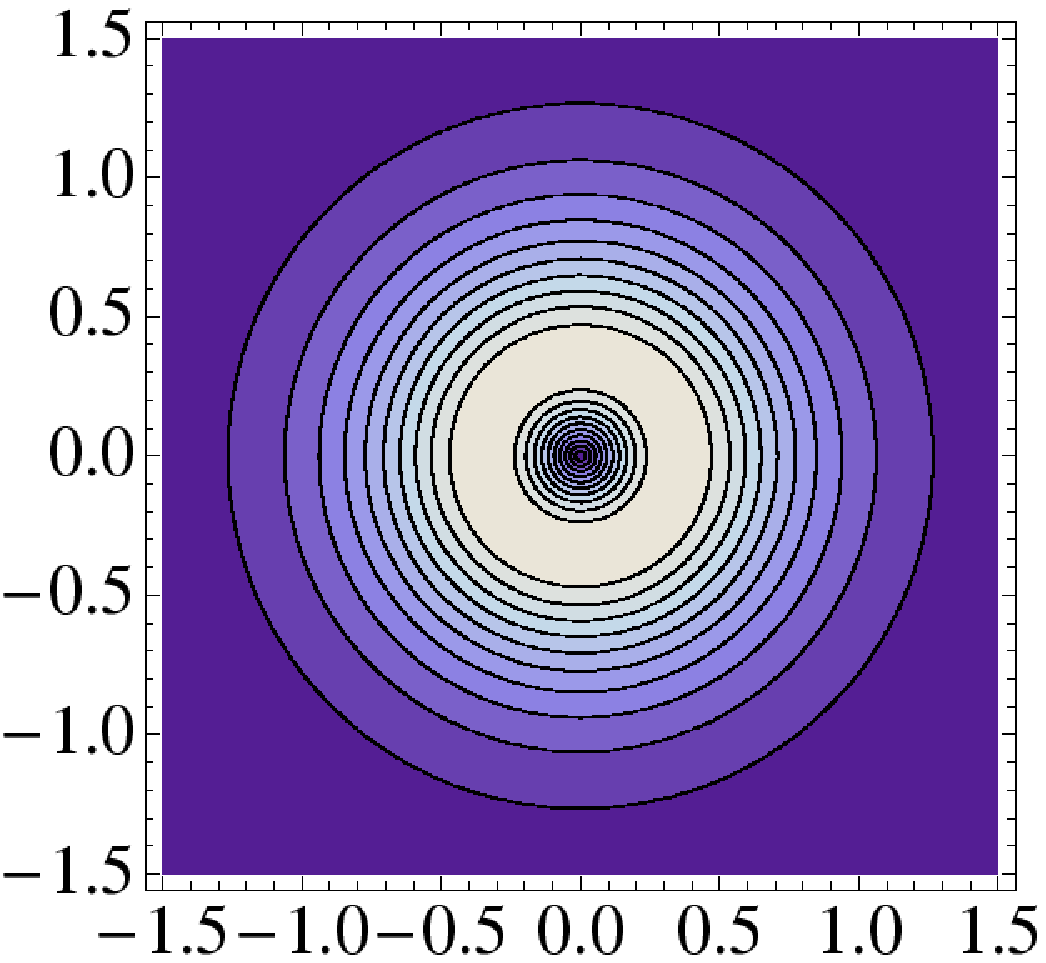}
\includegraphics[height=2.9cm]{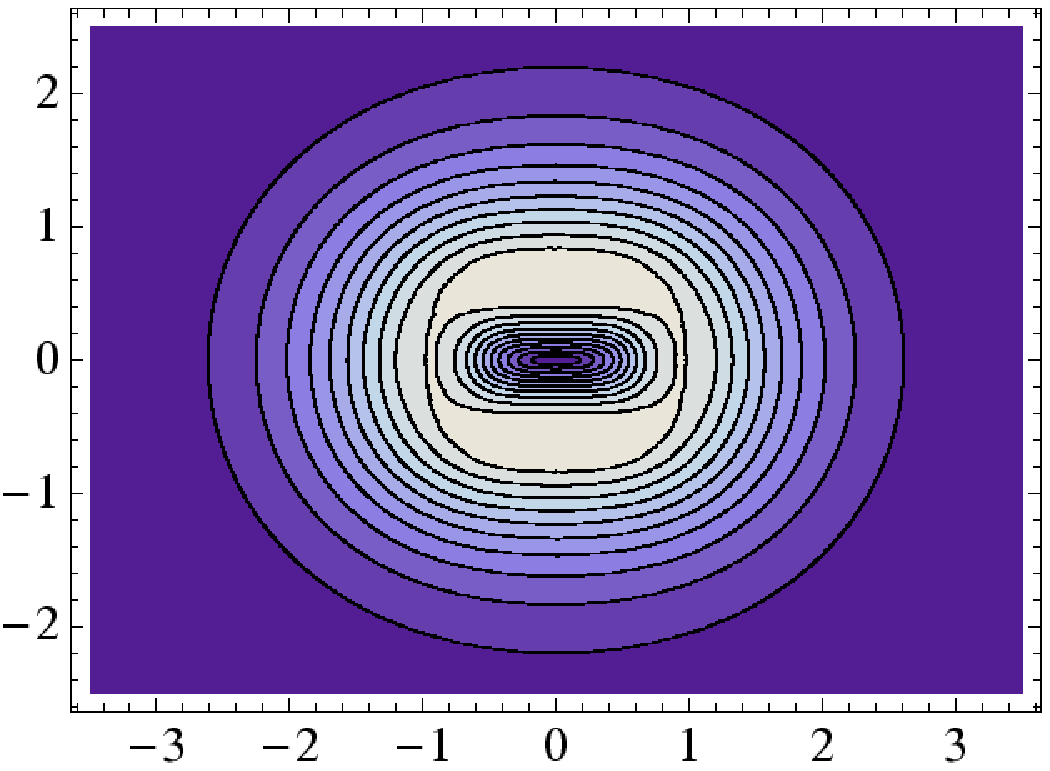}
\includegraphics[height=2.9cm]{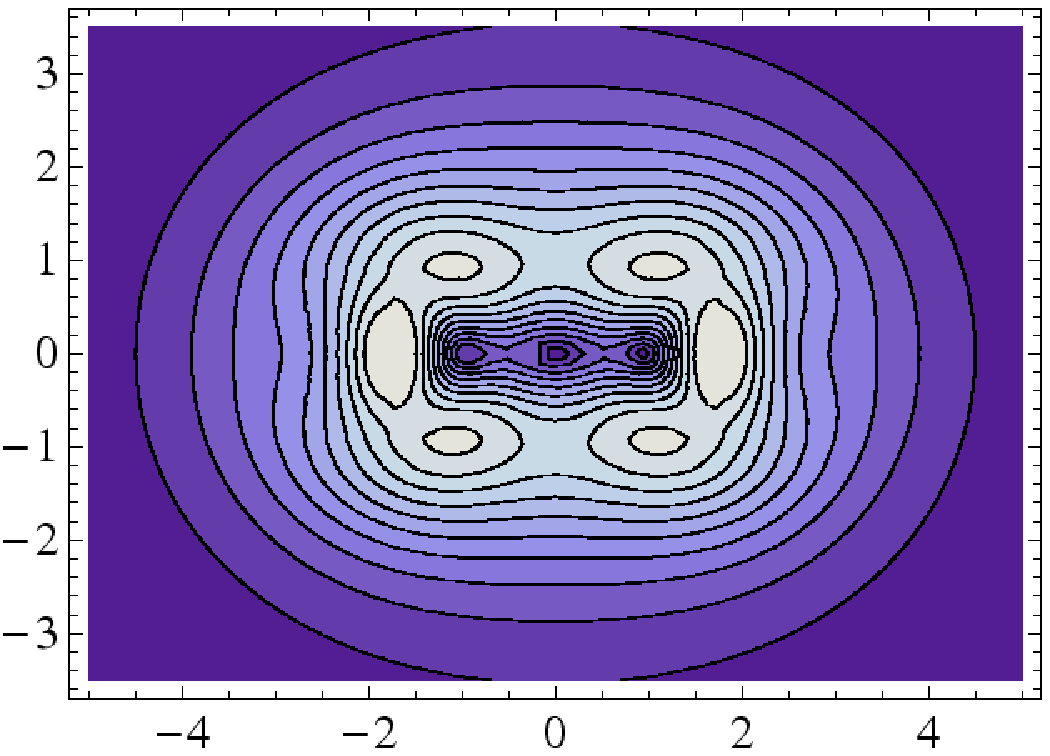}
\includegraphics[height=2.9cm]{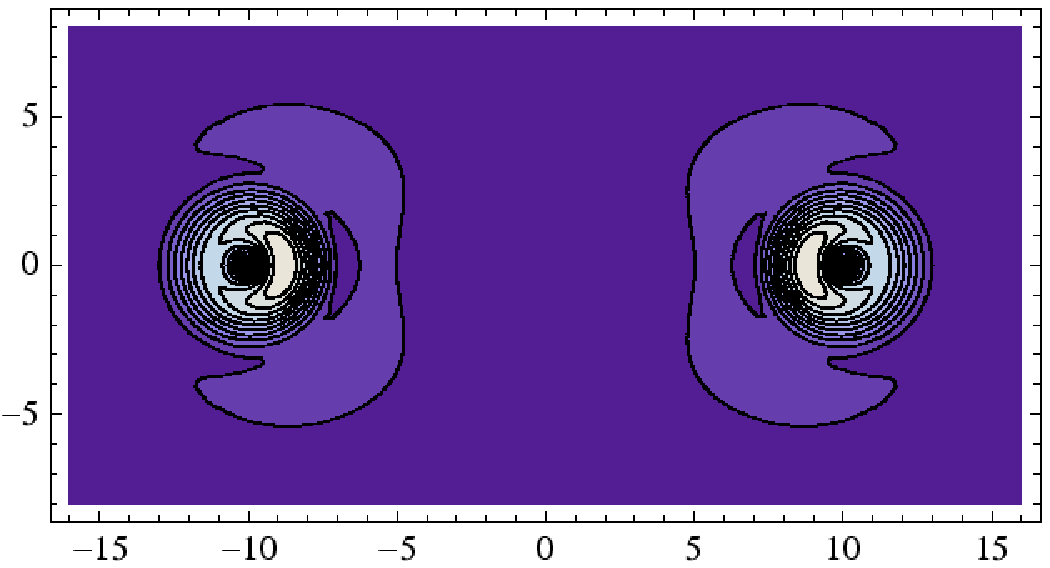}\\
\includegraphics[height=2.9cm]{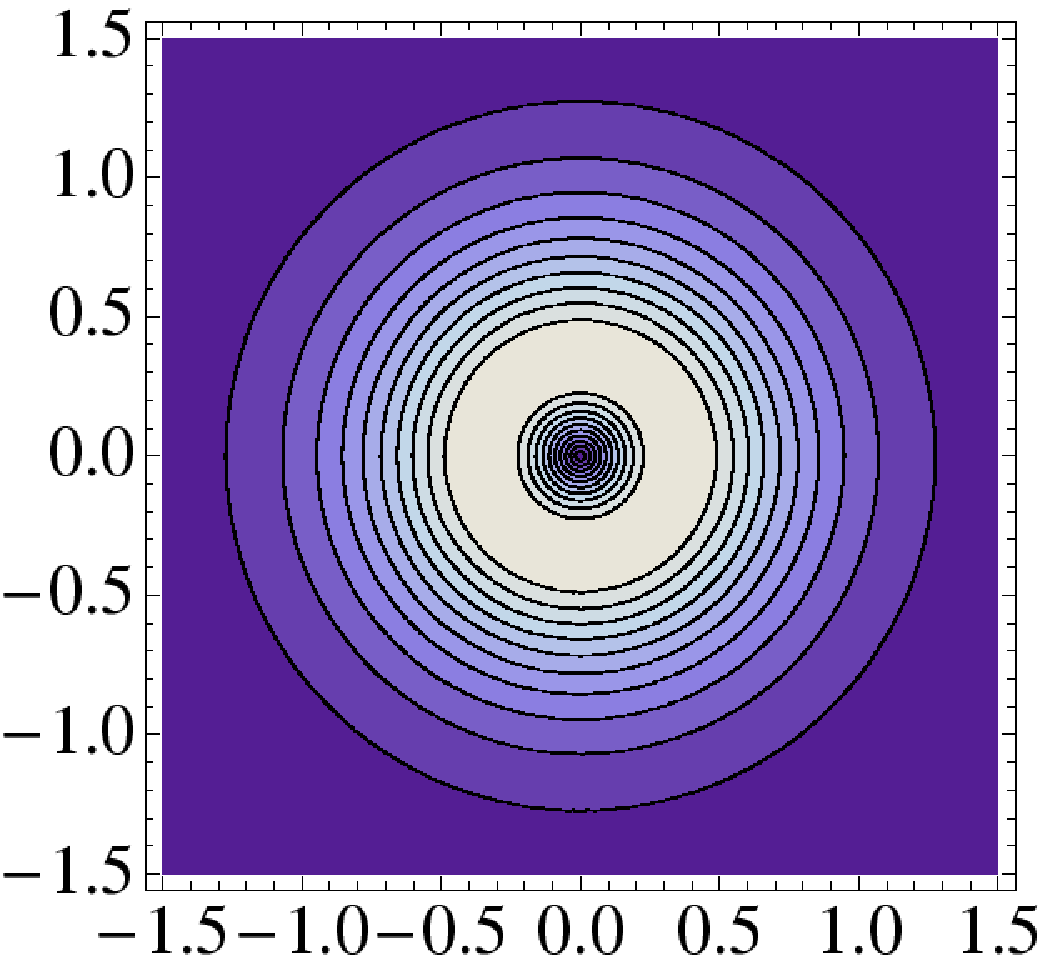}
\includegraphics[height=2.9cm]{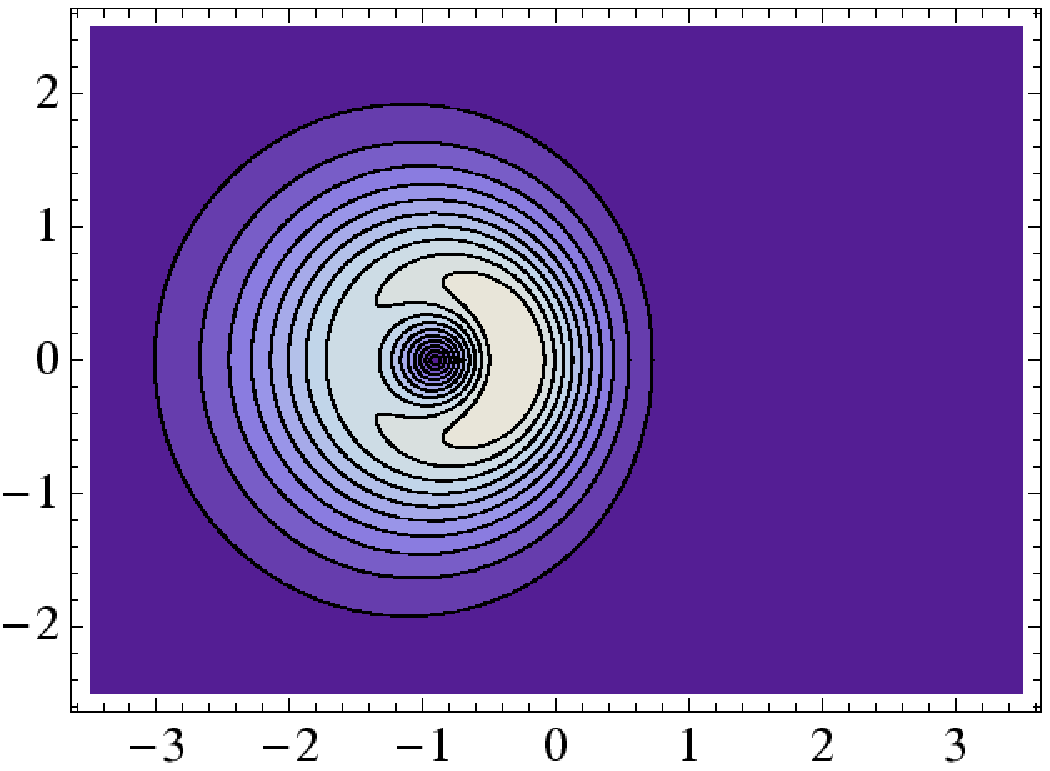}
\includegraphics[height=2.9cm]{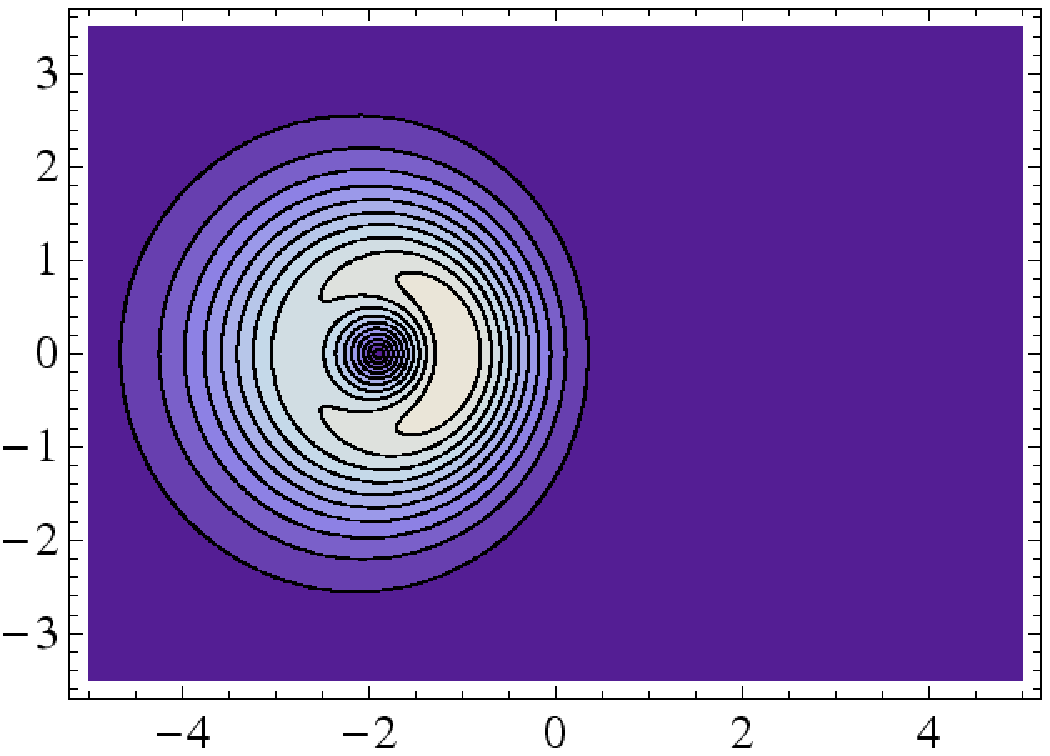}
\includegraphics[height=2.9cm]{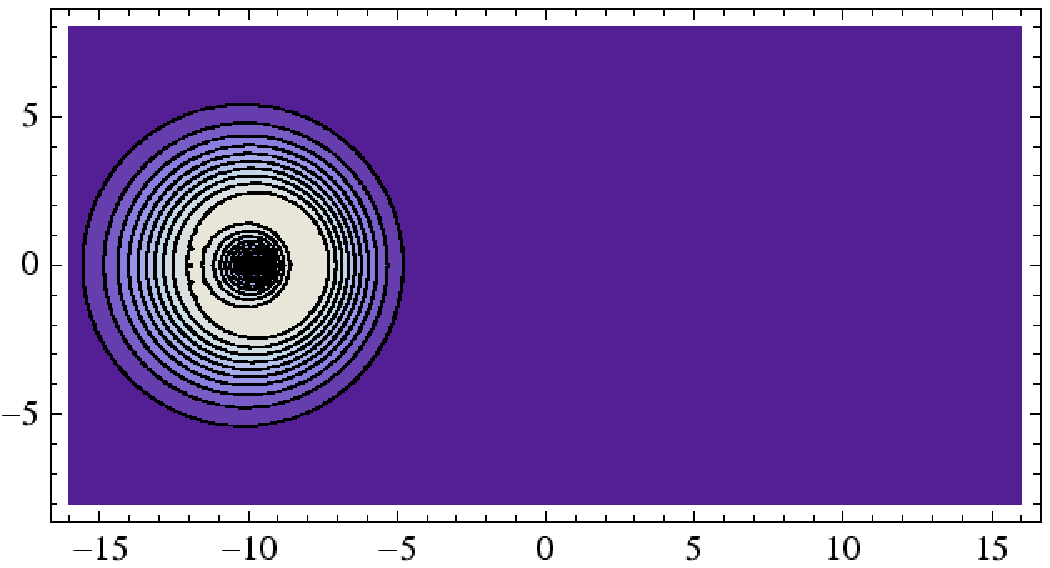}
\end{center}
\caption{The non-Abelian Chern-Simons fractional vortex with
  $G'=SO(4)$ and 
  $G'=USp(4)$ for $-\kappa=\mu=2$, where the rows of the Figure
  correspond to the energy density, the Abelian magnetic flux
  $F_{12}^0$, the non-Abelian magnetic flux $F_{12}^2$, the magnitude
  of the Abelian electric field $|E_i^0|$ and finally the magnitude of
  the non-Abelian electric field $|E_i^2|$, whereas the columns of the
  Figure correspond to the separation distance $2d=2\{0,1,2,10\}$. We
  have set $\xi=2$.}
\label{fig:frac_kneg}
\end{figure}
In Fig.~\ref{fig:frac_kneg} is shown a matrix of graphs of
the non-Abelian Chern-Simons vortex configuration for couplings with
opposite signs: $-\kappa=\mu=2$, where the rows correspond to the
energy density, the Abelian magnetic flux $F_{12}^0$, the non-Abelian
magnetic flux $F_{12}^2$, the magnitude of the Abelian electric field
$|E_i^0|$ and finally the magnitude of the non-Abelian electric field
$|E_i^2|$, whereas the columns correspond to the separation distance
$2d=2\{0,1,2,10\}$. 
The centers of the fractional sub-peaks are placed at $z_1=d$ and
$z_2=-d$, respectively. Note that the symmetry of the configuration
has allowed us to show only one of the non-Abelian fields,
i.e.~$F_{12}^2$ for which the other field is given by the reflection
in the $x$-axis $F_{12}^1(x,y)=F_{12}^2(-x,y)$ and analogously for the
non-Abelian electric field strengths. 
Unlike the case with equal couplings, there is basically not changing
much when separating the fractional sub-peaks in this case with
opposite signs of the couplings. The energy density and the Abelian 
magnetic flux density are both single peaked bells which split into
two sub-peaks for non-zero separation modulus parameter $d$. The
electric field strengths on the other hand retain their structure as
rings. The non-Abelian one simply moves with $d$ (though it gets a bit
distorted), while the Abelian density has a transition where it gets
distorted until it splits into two ring-like structures. For the
intermediate value $d=2$, the Abelian electric field strength
density has in fact six maxima, demonstrating the complexity of this
soliton.

%%%%%%%%%%%%%

\section{Semi-local vortices\label{sec:semilocal}}

In this Section, we will study the moduli matrix (\ref{eq:H0-SOUSp4})
with $\mathbf{A}=\mathbf{B}=\mathbf{0}$ and only have a non-vanishing
$\mathbf{C}$. We will thus denote this vortex a purely semi-local
vortex. We will however restrict ourselves to configurations
with a single size modulus $c$. For both $SO(2M)$ and $USp(2M)$ with 
$M=2m,\;m\in\mathbb{Z}_{>0}$, we can parametrize the moduli matrix
(\ref{eq:H0-SOUSp4}) in this case as follows 
\beq 
H_0(z) = \begin{pmatrix}
z\mathbf{1}_{2m}  & \mathbf{C}_{S,A} \\
\mathbf{0} & \mathbf{1}_{2m}
\end{pmatrix} \ , \quad
\mathbf{C}_{S,A} = c J_{2m} = c
\begin{pmatrix}
0 & \mathbf{1}_m\\
\epsilon\mathbf{1}_m & 0
\end{pmatrix} \ ,
\eeq
with $\epsilon=+1$ for $SO$ and $\epsilon=-1$ for $USp$ as usual. 
For $USp(4)$ it is the most generic case for $\mathbf{C}$, while it is
a simplified case for $SO(4)$ as the diagonal of $\mathbf{C}$ does not
have to vanish. 
For $m>1$ the above matrix is a simplified example, but it turns out
that the equations have a very simple dependence on $m$, hence we will
keep $m$ as a parameter. 
In the case of $SO(2M=4m+2)$, we can use the parametrization as
follows 
\beq \mathbf{C} = c\mathbf{1}_{M} \ . \eeq
In both cases it turns out that we can still use a diagonal Ansatz 
\beq \Omega'=\diag\left(e^\chi \mathbf{1}_{M} ,
  e^{-\chi}\mathbf{1}_{M}\right) \ , \eeq
even though $\Omega_0$ no longer is diagonal due to the fact that the
master equations remain diagonal. 
The master equations are thus
\begin{align}
\bar{\p}\p\chi &= -\frac{\pi^2}{\kappa\mu}
\left[\left(|z|^2+|c|^2\right)e^{-\psi-\chi} + e^{-\psi+\chi}
-\frac{\xi}{M}\right]
\left[\left(|z|^2+|c|^2\right)e^{-\psi-\chi} - e^{-\psi+\chi}\right]
 \non
&\phantom{=\ }
-\frac{\pi^2}{\mu^2}\left[
\left(\left(|z|^2+|c|^2\right)e^{-\psi-\chi}\right)^2 - 
\left(e^{-\psi+\chi}\right)^2\right] \ , 
\label{eq:semilocal-masterequation1} \\
\bar{\p}\p\psi &= -\frac{\pi^2}{\kappa^2}
\left[\left(|z|^2+|c|^2\right)e^{-\psi-\chi} + e^{-\psi+\chi}\right]
\left[\left(|z|^2+|c|^2\right)e^{-\psi-\chi} + e^{-\psi+\chi}
  -\frac{\xi}{M}\right] \non
&\phantom{=\ }
-\frac{\pi^2}{\kappa\mu}
\left[\left(|z|^2+|c|^2\right)e^{-\psi-\chi} - e^{-\psi+\chi}\right]^2
\ , \label{eq:semilocal-masterequation2}
\end{align}
which has the typical form of semi-local equations.
Note that the dependence of $M$ can be eliminated by a rescaling of
$\xi$.

The boundary conditions for $|z|\to\infty$, which are also the lump
solution, that is, the solution in the weak coupling limit
$\kappa=\mu\to 0$, read  
\begin{align}
\psi^{\infty} = \log\left\{\frac{2M}{\xi}\sqrt{|z|^2+|c|^2}\right\}
\ , \quad 
\chi^{\infty} = \log\sqrt{|z|^2+|c|^2} \ .
\label{eq:semilocal-boundary-conditions}
\end{align}
The Abelian and non-Abelian magnetic field strengths read respectively
\beq
F_{12}^{0} = -4\sqrt{M}\bar{\p}\p\psi \ , \quad
F_{12}^a t^a \equiv F_{12}^{\rm NA} t = -4\sqrt{M}\bar{\p}\p\chi\, t \ , 
\eeq
where the Abelian one contributes together with a boundary term to the
energy density as
\beq \mathcal{E} = 2\xi\bar{\p}\p\psi
+2M\bar{\p}\p\left[\left(|z|^2+|c|^2\right)e^{-\psi-\chi} + 
  e^{-\psi+\chi}\right] \ . \eeq
The Abelian and non-Abelian electric field strengths are, respectively
\begin{align}
E_r^0 &= \frac{2\sqrt{M}\pi}{\kappa}
\p_r\left[\left(r^2+|c|^2\right)e^{-\psi-\chi} + e^{-\psi+\chi}\right]
\ , \\
E_r^a t^a &\equiv E_r^{\rm NA} t = \frac{2\sqrt{M}\pi}{\mu}
\p_r\left[\left(r^2+|c|^2\right)e^{-\psi-\chi} -
  e^{-\psi+\chi}\right]\, t
\ ,
\end{align}
where the radial coordinate is $r\equiv|z|$ and we have defined the
normalized generator 
\beq t\equiv \frac{1}{2\sqrt{M}}
\diag\left(\mathbf{1}_{M},-\mathbf{1}_{M}\right) \ . \eeq

Let us now make some qualitative calculations. We consider some
small fluctuations around the boundary conditions
(\ref{eq:semilocal-boundary-conditions}) as follows
\beq \chi = \chi^{\infty} + \delta\!\chi \ , \quad
\psi = \psi^{\infty} + \delta\psi \ . \eeq
Plugging them into the master equations
(\ref{eq:semilocal-masterequation1})-(\ref{eq:semilocal-masterequation2}) 
yields to linear order
\begin{align}
\bar{\p}\p\,\delta\!\chi + \bar{\p}\p\chi^{\infty} &= 
\frac{m_\mu^2}{4}\delta\!\chi
\ ,
\label{eq:chifluc_semilocal}
\\
\bar{\p}\p\delta\psi + \bar{\p}\p\psi^{\infty} &= 
\frac{m_\kappa^2}{4}\delta\psi \ .
\label{eq:psifluc_semilocal}
\end{align}
Expanding in small $|c|/|z|$, we obtain
\begin{align}
  \bar{\p}\p\psi^{\infty} = \bar{\p}\p\chi^{\infty} \simeq 
\frac{|c|^2}{2|z|^4} - \frac{|c|^4}{|z|^6} \ ,
\end{align}
which has the consequence that both the field fluctuations attain a
power-law behavior
\begin{align}
\delta\!\chi &= \frac{2|c|^2}{m_\mu^2}|z|^{-4}
+\frac{4|c|^2}{m_\mu^2}\left(\frac{8}{m_\mu^2} - |c|^2\right)
|z|^{-6} + \mathcal{O}\left(|z|^{-8}\right) \ , 
\label{eq:chiflucsol_semilocal} \\
\delta\psi &= \frac{2|c|^2}{m_\kappa^2}|z|^{-4}
+\frac{4|c|^2}{m_\kappa^2}\left(\frac{8}{m_\kappa^2} - |c|^2\right)
|z|^{-6} + \mathcal{O}\left(|z|^{-8}\right) \ .
\label{eq:psiflucsol_semilocal}
\end{align}
This behavior is the typical and well-known behavior of a semi-local
non-Abelian vortex. 
Next we will show that in the case of equal Chern-Simons couplings
$\kappa=\mu$, the Abelian magnetic flux and the non-Abelian
magnetic flux are equal as well as that the Abelian electric field
equals the non-Abelian electric field. Subtracting the two master
equations
(\ref{eq:semilocal-masterequation1})-(\ref{eq:semilocal-masterequation2}),
we obtain
\begin{align}
\bar{\p}\p\left(\psi-\chi\right) &=
-\frac{\pi^2}{\kappa^2}\bigg[
\left(1-\frac{\kappa^2}{\mu^2}\right)
\left(\left(|z|^2+|c|^2\right) 
e^{-\psi-\chi}\right)^2
+\left(1+\frac{\kappa}{\mu}\right)^2\left(e^{-\psi+\chi}\right)^2
\label{eq:psi-chi}
\\ &\phantom{= - \frac{\pi^2}{\kappa^2}\bigg[}
+2\left(1-\frac{\kappa}{\mu}\right)\left(|z|^2+|c|^2\right)e^{-\psi-\chi}
\left(e^{-\psi+\chi} - \frac{\xi}{2M}\right)
+\left(1+\frac{\kappa}{\mu}\right)\frac{\xi}{M}e^{-\psi+\chi}
\bigg] \ , \nonumber
\end{align}
which in the case $\kappa=\mu$ clearly reduces to 
\begin{align}
\bar{\p}\p\left(\psi-\chi\right) &=
-\frac{4\pi^2}{\kappa^2}e^{-\psi+\chi}\left(
e^{-\psi+\chi} - \frac{\xi}{2M}\right) \ , 
\end{align}
being independent of $z,\bar{z}$ and thus is satisfied by the vacuum
solution 
\beq e^{-\psi+\chi} = \frac{\xi}{2M} \label{eq:vac_kappa=mu} \ . \eeq
Now it is easy to show that 
\begin{align}
F_{12}^0 = F_{12}^{\rm NA} =
\frac{8\sqrt{M}\pi^2}{\kappa^2}\left(|z|^2+|c|^2\right)e^{-\psi-\chi}
\left(\left(|z|^2+|c|^2\right)e^{-\psi-\chi} - \frac{\xi}{2M}\right)
\ , 
\end{align}
and we can readily observe that for vanishing size modulus $c=0$, the
magnetic fields are zero at the center of the vortex (the local
case) \cite{Gudnason:2009ut}. 
This statement gets modified by the presence of the size
modulus. The exact value of the magnetic fields at the center of the
semi-local vortex is not so easy to calculate and we demonstrate
numerically, that it is indeed non-vanishing at the center of the
vortex. 
Inserting the vacuum solution (\ref{eq:vac_kappa=mu}) into the
electric field strengths, it is easily seen that the Abelian and the
non-Abelian one are equal in the case of $\kappa=\mu$
\begin{align}
E_r^0 = E_r^{\rm NA} =
\frac{2\sqrt{M}\pi}{\kappa}
\p_r\left[\left(r^2+|c|^2\right)e^{-\psi-\chi}\right] \ . 
\end{align}
It furthermore turns out that the Abelian and non-Abelian field
strengths are equal also in the opposite coupling case. This can be
seen from equation (\ref{eq:psi-chi}) by inserting $\mu=-\kappa$
giving rise to
\begin{align}
\bar{\p}\p\left(\psi-\chi\right) &=
-\frac{4\pi^2}{\kappa^2}
\left(|z|^2+|c|^2\right)e^{-\psi-\chi}
\left(e^{-\psi+\chi} - \frac{\xi}{2M}\right) \ ,
\end{align}
which does depend on $z,\bar{z}$ but still allows the vacuum solution
(\ref{eq:vac_kappa=mu}) and the consequence is then the same as in the
case of $\kappa=\mu$; the Abelian magnetic and electric fields equal
their non-Abelian counterparts. 

\begin{figure}[!p]
\begin{center}
\mbox{\resizebox{!}{5.2cm}{\includegraphics{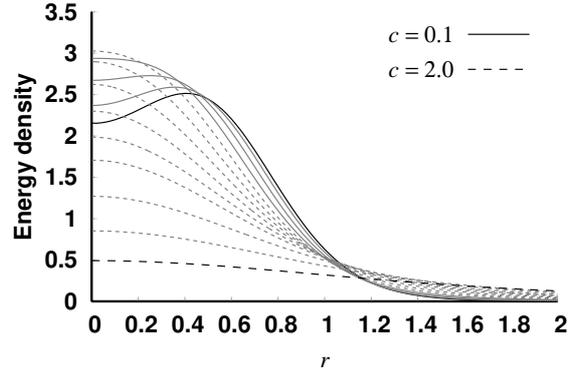}}}
\caption{The energy density for the semi-local vortex solution for
  various values of the semi-local modulus
  $c=\{0.1,0.2,0.3,0.4,0.5,0.6,0.7,0.8,0.9,1.0,1.2,1.5,2.0\}$. The
  couplings are chosen as $\kappa=\mu=2$ and $\xi=2$. }
\label{fig:energy1}
\end{center}
\end{figure}

\begin{figure}[!p]
\begin{center}
\mbox{
\subfigure[]{\resizebox{!}{5.2cm}{\includegraphics{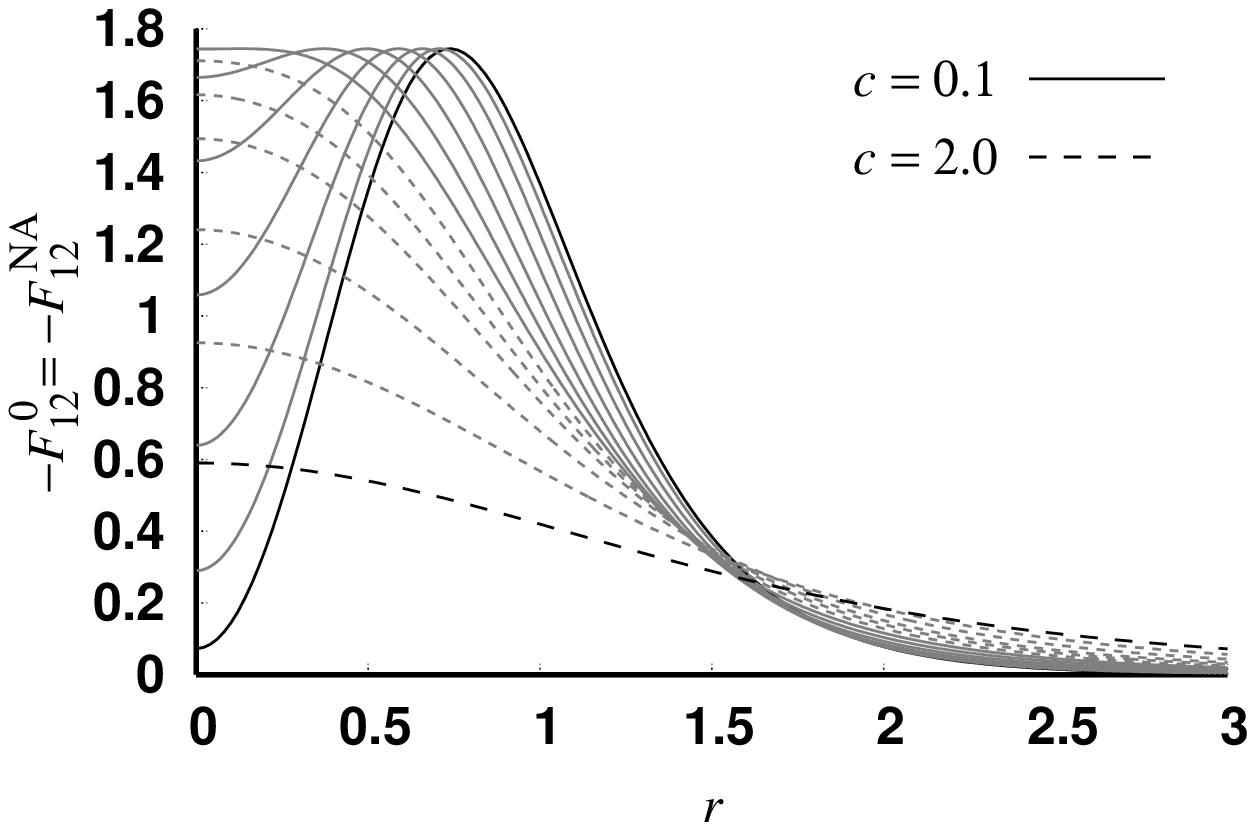}}}\quad
\subfigure[]{\resizebox{!}{5.2cm}{\includegraphics{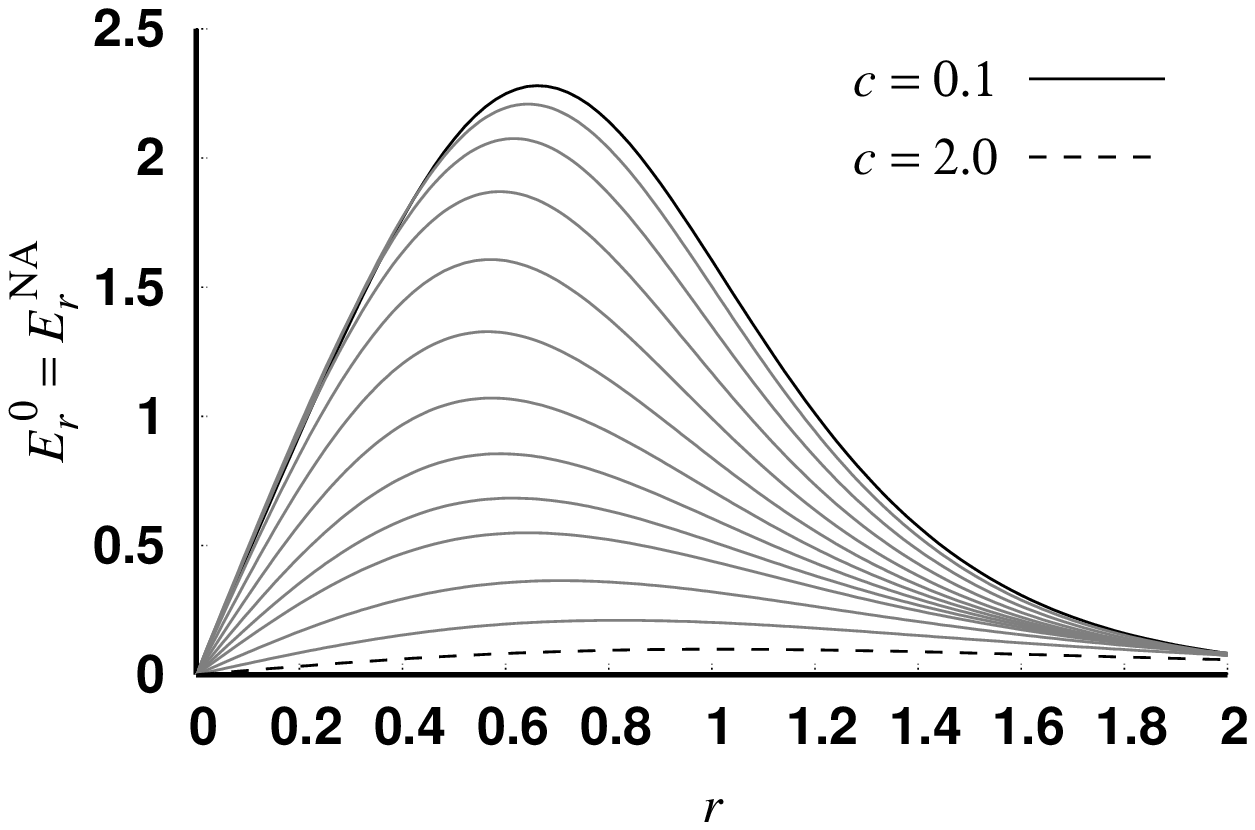}}}}
\caption{The Abelian and non-Abelian (a) magnetic flux density
  and (b) electric field density for the semi-local vortex solution
  for various values of the semi-local modulus 
  $c=\{0.1,0.2,0.3,0.4,0.5,0.6,0.7,0.8,0.9,1.0,1.2,1.5,2.0\}$. The
  couplings are chosen as $\kappa=\mu=2$ and $\xi=2$. }
\label{fig:magelecflux1}
\end{center}
\end{figure}

Let us now compute some numerical solutions as examples of semi-local
non-Abelian vortices. First we take the equal coupling case and show
in Fig.~\ref{fig:energy1} the energy density for thirteen different
values 
of the semi-local size parameter. In Fig.~\ref{fig:magelecflux1} are
shown the magnetic flux densities and electric field densities for the
Abelian fields (which equals the non-Abelian ones as demonstrated
above). We observe that there is a transition from the ring-like
magnetic flux into a Gauss-bell-like magnetic flux. The electric field
does not change qualitatively, but only spreads out as the vortex size 
increases. 


\subsection{Negative magnetic flux density at the origin}

\begin{figure}[!p]
\begin{center}
\mbox{
\subfigure[]{\resizebox{!}{5.2cm}{\includegraphics{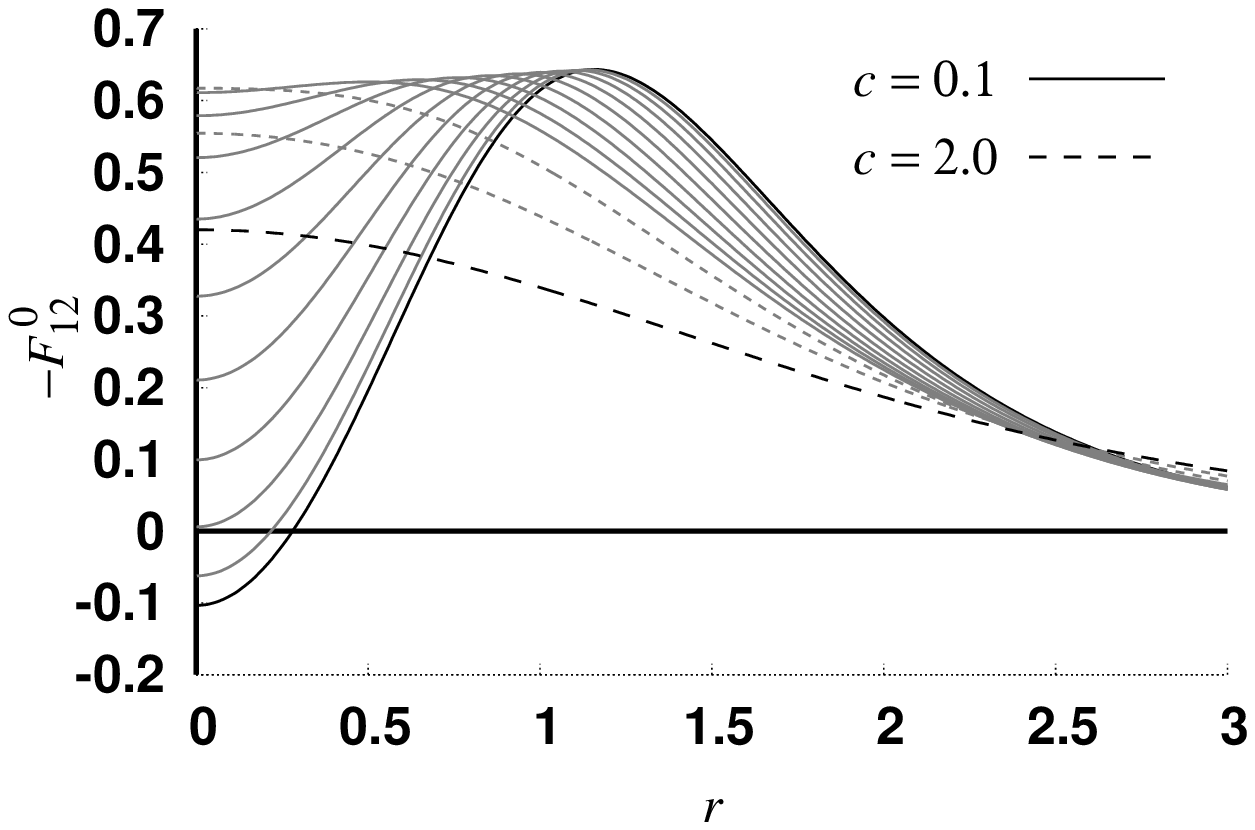}}}\quad
\subfigure[]{\resizebox{!}{5.2cm}{\includegraphics{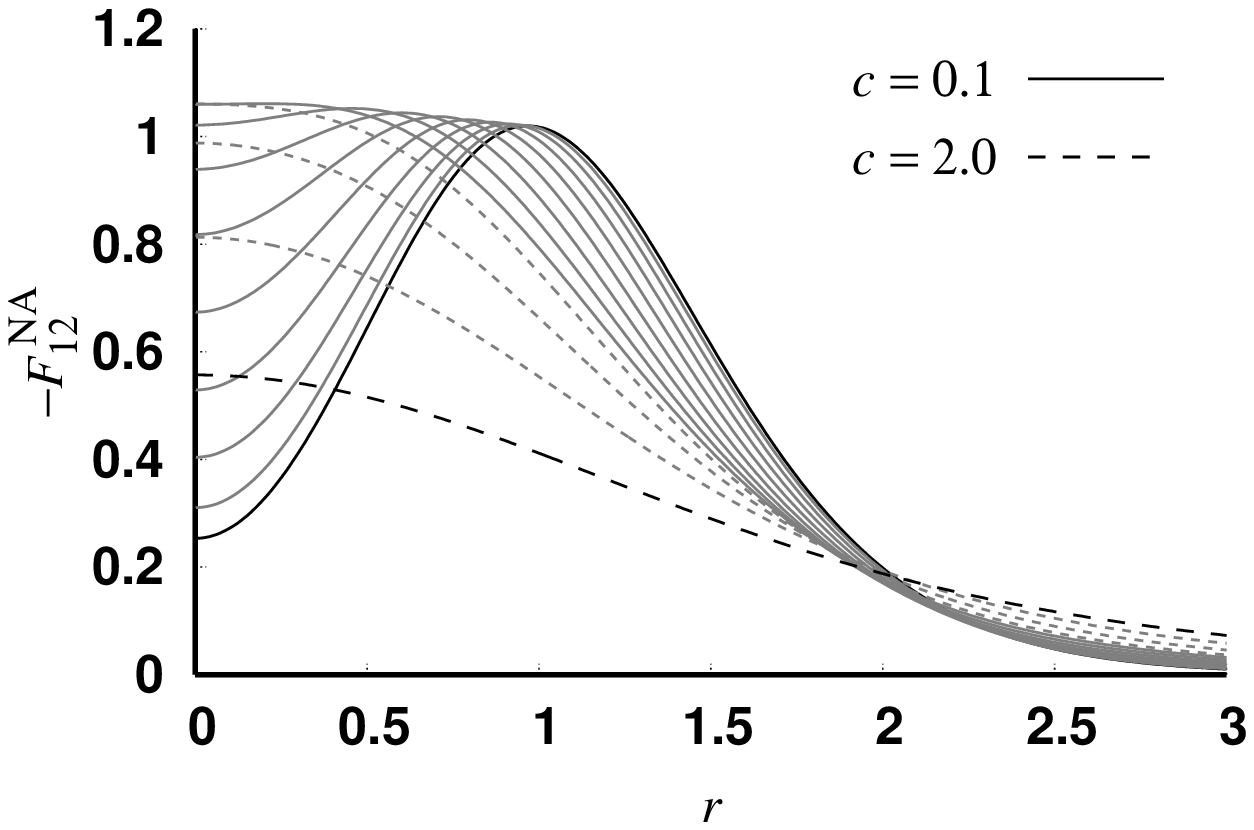}}}}
\caption{The (a) Abelian and (b) non-Abelian magnetic flux
  density for the semi-local vortex solution for various values of the
  semi-local modulus  
  $c = \{$ $0.1,$ $0.2,$ $0.3,$ $0.4,$ $0.5,$ $0.6,$ $0.7,$ $0.8,$
  $0.9,$ $1.0,$ $1.2,$ $1.5,$ $2.0$ $\}$. The
  couplings are chosen as $\kappa=4,\mu=2$ and $\xi=2$. }
\label{fig:magflux2}
\end{center}
\end{figure}

For completeness, let us repeat the calculation with the different
Chern-Simons couplings as we did in the fractional case, considering 
first $\kappa=4,\mu=2$. In Fig.~\ref{fig:magflux2} are shown the
magnetic field strengths; Abelian in (a) and non-Abelian in (b), as
function of the size modulus $c$. It is observed that the effect of
negative Abelian magnetic flux at the origin of the vortex vanishes
quickly as the size is turned on. This can be understood from the fact
that both the Abelian and the non-Abelian fluxes become bell-like
structures. Furthermore, they become equal even for different
Chern-Simons couplings $\kappa\neq\mu$ in the limit of large
$|c|$. This can qualitative be shown in terms of the lump solution
(\ref{eq:semilocal-boundary-conditions}) by calculating the field
strengths in the limit $|c|\gg 1/(g\sqrt{\xi})$ (i.e.~the size modulus 
being much larger than the local size of the vortex)
\beq F_{12}^0 = F_{12}^{\rm NA} = 
-2\sqrt{M}\frac{|c|^2}{\left(|z|^2+|c|^2\right)^2}
\ . \eeq 

\begin{figure}[!p]
\begin{center}
\mbox{
\subfigure[]{\resizebox{!}{5.2cm}{\includegraphics{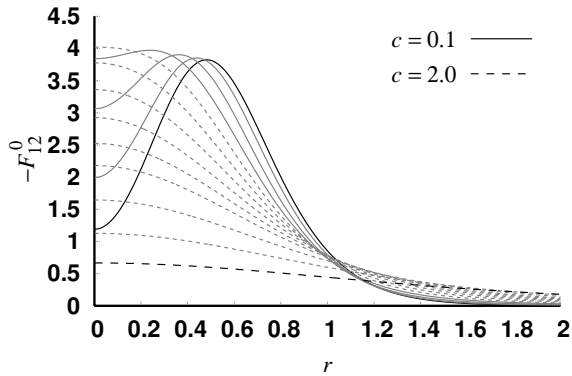}}}\quad
\subfigure[]{\resizebox{!}{5.2cm}{\includegraphics{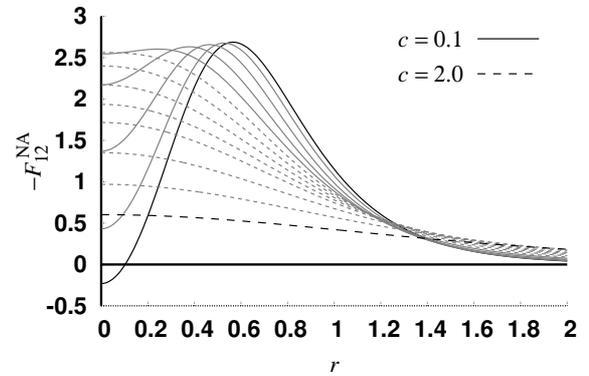}}}}
\caption{The (a) Abelian and (b) non-Abelian magnetic flux
  density for the semi-local vortex solution for various values of the
  semi-local modulus  
  $c=\{$ $0.1,$ $0.2,$ $0.3,$ $0.4,$ $0.5,$ $0.6,$ $0.7,$ $0.8,$
  $0.9,$ $1.0,$ $1.2,$ $1.5,$ $2.0$ $\}$. The
  couplings are chosen as $\kappa=1,\mu=2$ and $\xi=2$. }
\label{fig:magflux3}
\end{center}
\end{figure}

Now we set $\kappa=1,\mu=2$ and show the corresponding magnetic flux
densities in Fig.~\ref{fig:magflux3}.

\subsection{Bell-like structures: $-\kappa=\mu=2$}

\begin{figure}[!tbp]
\begin{center}
\mbox{\resizebox{!}{5.2cm}{\includegraphics{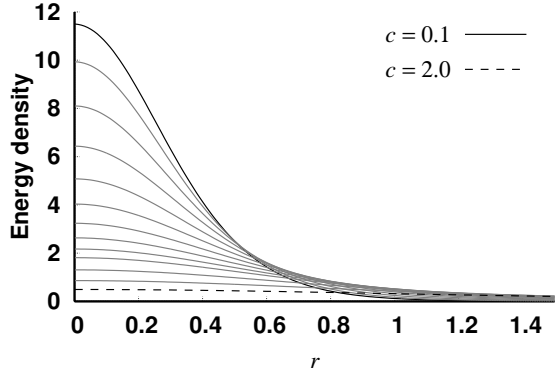}}}
\caption{The energy density for the semi-local vortex solution for
  various values of the semi-local modulus
  $c=\{0.1,0.2,0.3,0.4,0.5,0.6,0.7,0.8,0.9,1.0,1.2,1.5,2.0\}$. The
  couplings are chosen as $-\kappa=\mu=2$ and $\xi=2$. }
\label{fig:energy4}
\end{center}
\end{figure}

\begin{figure}[!tbp]
\begin{center}
\mbox{
\subfigure[]{\resizebox{!}{5.2cm}{\includegraphics{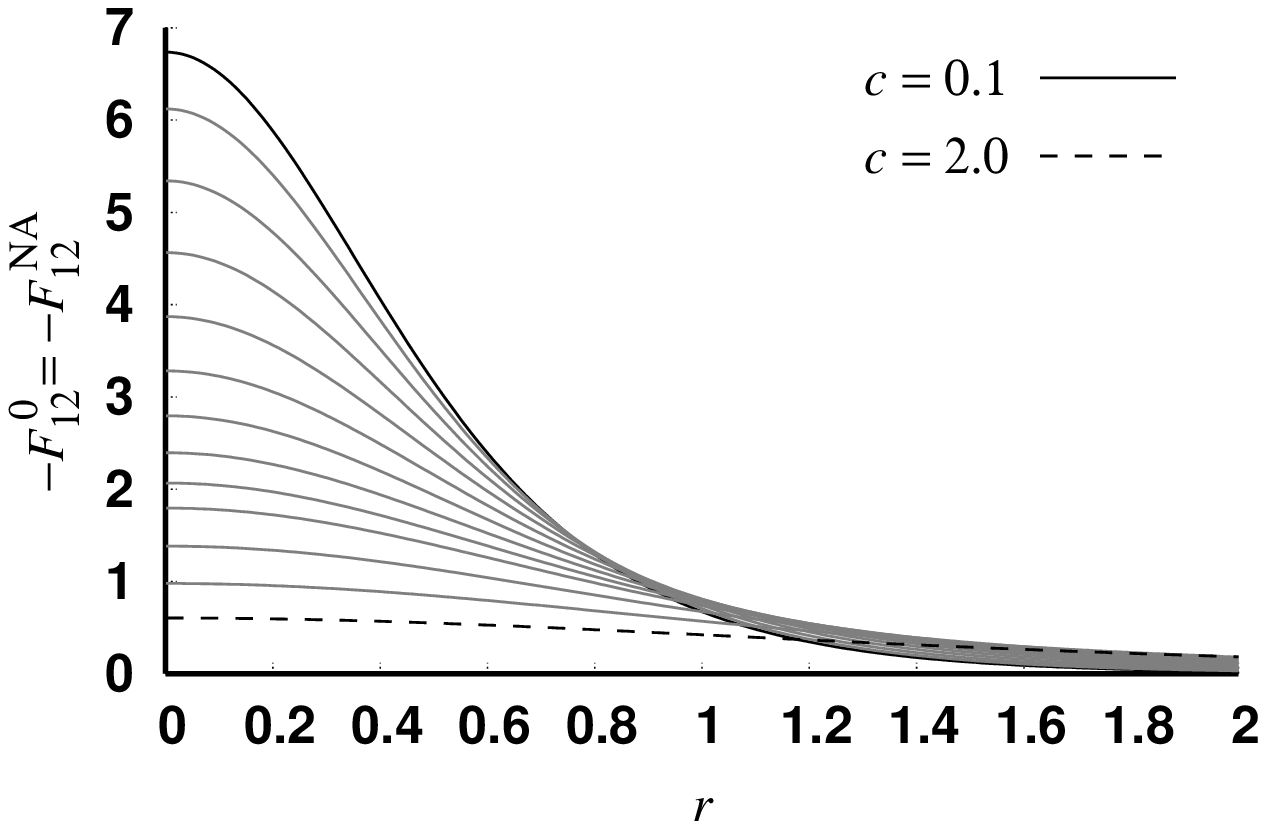}}}\quad
\subfigure[]{\resizebox{!}{5.2cm}{\includegraphics{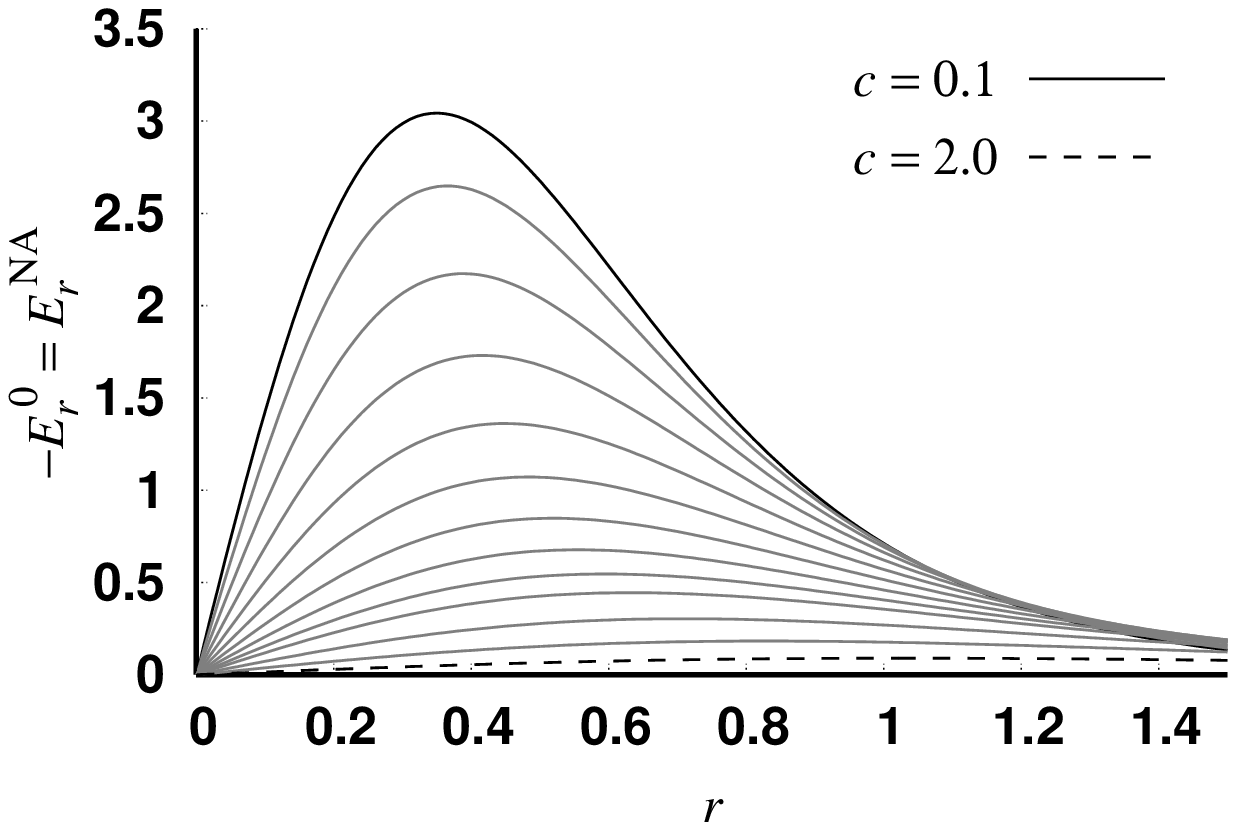}}}}
\caption{The Abelian and non-Abelian (a) magnetic flux density
  and (b) electric field density for the semi-local vortex solution
  for various values of the semi-local size modulus 
  $c=\{0.1,0.2,0.3,0.4,0.5,0.6,0.7,0.8,0.9,1.0,1.2,1.5,2.0\}$. The
  couplings are chosen as $-\kappa=\mu=2$ and $\xi=2$. }
\label{fig:magelecflux4}
\end{center}
\end{figure}

Finally, we will consider the case of opposite signs of the
Chern-Simons couplings $\kappa=-\mu$. The energy is shown in
Fig.~\ref{fig:energy4}, the magnetic flux densities in
Fig.~\ref{fig:magelecflux4}a and the  electric field densities in
Fig.~\ref{fig:magelecflux4}b, all as function of the semi-local
size modulus. We can confirm numerically that the Abelian magnetic
field equals the non-Abelian one as well as the Abelian electric field
equals minus the non-Abelian one (as shown above). 


\section{Semi-local fractional vortices\label{sec:frac-semilocal}}

We will now consider the fractional vortices with
non-zero size parameter $\mathbf{C}$ in the moduli matrix
(\ref{eq:H0-SOUSp4}) turned on. It turns out that it is not possible
to obtain diagonal master equations in this case for $USp$. Hence, we
will consider only $G'=SO(2M)$ here and take the moduli matrix as the
one of Refs.~\cite{Eto:2008qw,Eto:2009bz}, namely
\beq H_0 = \left(
\begin{array}{ccc|ccc}
z-z_1 & & & c_1 \\
 & \ddots & & & \ddots \\
& & z-z_M & & & c_M \\
\hline
0 & & & 1  \\
& \ddots & & & \ddots \\
& & 0 & &  & 1
\end{array}\right)\ ,
\eeq
while we still use the Ansatz for the moduli matrix field of
Eq.~(\ref{eq:omegapAnsatz}). This leads to the master equations which
remain diagonal
\begin{align}
\bar{\p}\p\chi_m &= -\frac{\pi^2}{M\kappa\mu}
\left[\sum_{n=1}^M\left(\left|z-z_n\right|^2+|c_n|^2\right)e^{-\psi-\chi_n}
+\sum_{n=1}^M e^{-\psi+\chi_n}-\xi\right]
\non & \phantom{=\ -\frac{\pi^2}{M\kappa\mu}}
\times
\left[\left(\left|z-z_m\right|^2+|c_m|^2\right)e^{-\psi-\chi_m}
 -e^{-\psi+\chi_m}\right]\non
&\phantom{=\ }
-\frac{\pi^2}{\mu^2}\left[\left(\left|z-z_m\right|^2+|c_m|^2\right)^2
  \left(e^{-\psi-\chi_m}\right)^2
-\left(e^{-\psi+\chi_m}\right)^2\right] \ , \quad 
m=1,2,\ldots,M \ , 
\label{eq:semilocal-frac-masterequation1}\\
\bar{\p}\p\psi &= -\frac{\pi^2}{M^2\kappa^2}
\left[\sum_{n=1}^M\left(\left|z-z_n\right|^2+|c_n|^2\right)e^{-\psi-\chi_n}
+ \sum_{n=1}^M e^{-\psi+\chi_n}-\xi\right]
\non & \phantom{=\ -\frac{\pi^2}{M^2\kappa^2}}
\times
\left[\sum_{n'=1}^M\left(\left|z-z_{n'}\right|^2+|c_{n'}|^2\right)e^{-\psi-\chi_{n'}}
+\sum_{n'=1}^M e^{-\psi+\chi_{n'}}\right]
\non &\phantom{=\ }
-\frac{\pi^2}{M\kappa\mu}
\sum_{n=1}^M\left[\left(\left|z-z_n\right|^2+|c_n|^2\right)e^{-\psi-\chi_n}
 -e^{-\psi+\chi_n}\right]^2
\ . 
\label{eq:semilocal-frac-masterequation2}
\end{align}
Setting to zero all $c_n$ we get back the fractional master equations
(\ref{eq:frac-masterequation1})-(\ref{eq:frac-masterequation2}).
The boundary conditions which are also the lump solution
(i.e.~corresponding to the weak coupling limit $\kappa\to 0$ and
$\mu\to 0$) read  
\begin{align}
\psi^{\infty} = 
\log\left(\frac{2}{\xi}\sum_{n=1}^M\sqrt{\left|z-z_n\right|^2
+|c_n|^2}\right)
\ , \quad
\chi_m^{\infty} = \log\sqrt{\left|z-z_m\right|^2+|c_m|^2} \ .
\label{eq:semilocal-frac-boundary-conditions}
\end{align}
The magnetic fluxes remain those of
Eqs.~(\ref{eq:frac-Abelianmagflux}) and
(\ref{eq:frac-nonAbelianmagflux}), respectively, while the energy
density changes according to
\begin{align}
\mathcal{E} &=
2\xi\bar{\p}\p\psi + 
2\sum_{n=1}^M\bar{\p}\p\left[\left(\left|z-z_n\right|^2 +|c_n|^2\right)
e^{-\psi-\chi_n} + e^{-\psi+\chi_n}\right] \ .
\end{align}
The Abelian and non-Abelian electric field strengths also change and
are now, respectively
\begin{align}
E_i^0 &= \frac{2\pi}{\kappa\sqrt{M}}\sum_{n=1}^M\p_i
\left[\left(\left|z-z_n\right|^2+|c_n|^2\right)
e^{-\psi-\chi_n} + e^{-\psi+\chi_n}\right] \ , \\
E_i^{m} &= \frac{2\pi}{\mu}\p_i
\left[\left(\left|z-z_{m}\right|^2 +|c_m|^2\right) e^{-\psi-\chi_{m}} 
- e^{-\psi+\chi_{m}}\right] \ , \quad m=1,\ldots,M\ , 
\end{align}
where the relevant generators $t^m$ are those of
Eq.~(\ref{eq:frac_generators}). 

Let us now repeat some qualitative calculations in this case. We
again consider some small fluctuations around the boundary conditions 
(\ref{eq:semilocal-frac-boundary-conditions}) as follows
\beq \chi_m = \chi_m^{\infty} + \delta\!\chi_m \ , \quad
\psi = \psi^{\infty} + \delta\psi \ . \eeq
Plugging them into the master equations
(\ref{eq:semilocal-frac-masterequation1})-(\ref{eq:semilocal-frac-masterequation2})  
yields to linear order 
\begin{align}
\bar{\p}\p\delta\!\chi_m +\bar{\p}\p\chi_m^{\infty} &= \frac{M^2m_\mu^2}{4}
\frac{|z-z_m|^2+|c_m|^2}{\left(\sum_{n=1}^{M}\sqrt{|z-z_n|^2+|c_n|^2}\right)^2}\;\delta\!\chi_m
\ ,
\label{eq:chifluc_semilocal-frac}
\\
\bar{\p}\p\delta\psi + \bar{\p}\p\psi^{\infty} &= 
\frac{m_\kappa^2}{4}\delta\psi \ .
\label{eq:psifluc_semilocal-frac}
\end{align}
We first expand the Laplacian of the lump solution in
$z^{-1},\bar{z}^{-1}$ as
\begin{align}
\bar{\p}\p\chi_m^{\infty} &\simeq 
%\frac{1}{2}\bar{\p}\p
%\log\left(1-\frac{z_m}{z}-\frac{\bar{z}_m}{\bar{z}}
%+\frac{|z_m|^2+|c_m|^2}{|z|^2}
%+\mathcal{O}\left(|z|^{-3}\right)\right) 
%\simeq \frac{|c_m|^2}{2|z|^4} + 
%\mathcal{O}\left(|z|^{-5}\right) \ , \\
\frac{|c_m|^2}{2|z|^4}
+\left(\frac{z_m}{z}+\frac{\bar{z}_m}{\bar{z}}\right)
  \frac{|c_m|^2}{|z|^4} + \mathcal{O}\left(|z|^{-6}\right) \ , \\ 
\bar{\p}\p\psi^{\infty} &\simeq \frac{1}{4M}\Bigg(
\sum_{n=1}^M \left(|z_n|^2+2|c_n|^2\right)
-\frac{1}{M}\left|\sum_{n=1}^M z_n\right|^2\Bigg) |z|^{-4} \\
&+\frac{1}{4M}\Bigg[\frac{1}{2}\sum_{n=1}^M\left(|z_n|^2+4|c_n|^2\right)
  \left(\frac{z_n}{z}+\frac{\bar{z}_n}{\bar{z}}\right)
-\frac{1}{2M}\left(\sum_{n=1}^M\frac{z_n^2}{z}\sum_{n'=1}^M\bar{z}_{n'}
  +\sum_{n=1}^M z_n\sum_{n'=1}^M\frac{\bar{z}_{n'}^2}{\bar{z}}\right)
\non
&\phantom{+\frac{1}{4M}\Bigg[}
  +\frac{1}{M}\Bigg(\sum_{n=1}^M\left(|z_n|^2+2|c_n|^2\right)
  -\frac{1}{M}\left|\sum_{n=1}^M z_n\right|^2\Bigg)
  \sum_{n'=1}^M
  \left(\frac{z_{n'}}{z}+\frac{\bar{z}_{n'}}{\bar{z}}\right)
\Bigg] |z|^{-4}
+ \mathcal{O}\left(|z|^{-6}\right) \ , \nonumber
\end{align}
and then we expand also the right hand side of
Eq.~(\ref{eq:chifluc_semilocal-frac}) to obtain the following
asymptotic solutions
\begin{align}
\delta\!\chi_m &= 
\frac{2|c_m|^2}{m_\mu^2} |z|^{-4} 
+\frac{2|c_m|^2}{m_\mu^2}
\left[3\left(\frac{z_m}{z}+\frac{\bar{z}_m}{\bar{z}}\right)
-\frac{1}{M}\sum_{n=1}^M\left(\frac{z_n}{z}+\frac{\bar{z}_n}{\bar{z}}\right)
\right] |z|^{-4}
+ \mathcal{O}\left(|z|^{-6}\right) \ , 
\label{eq:chifluc_frac-semilocal} \\
\delta\psi &= \frac{1}{M m_\kappa^2}\Bigg(
\sum_{n=1}^M\left(|z_n|^2+2|c_n|^2\right)
-\frac{1}{M}\left|\sum_{n=1}^M z_n\right|^2\Bigg) |z|^{-4} 
\label{eq:psifluc_frac-semilocal}\\
&+\frac{1}{M m_\kappa^2}\Bigg[\frac{1}{2}\sum_{n=1}^M\left(|z_n|^2+4|c_n|^2\right)
  \left(\frac{z_n}{z}+\frac{\bar{z}_n}{\bar{z}}\right)
-\frac{1}{2M}\left(\sum_{n=1}^M\frac{z_n^2}{z}\sum_{n'=1}^M\bar{z}_{n'}
  +\sum_{n=1}^M z_n\sum_{n'=1}^M\frac{\bar{z}_{n'}^2}{\bar{z}}\right)
\non
&\phantom{+\frac{1}{4M}\Bigg[}
  +\frac{1}{M}\Bigg(\sum_{n=1}^M\left(|z_n|^2+2|c_n|^2\right)
  -\frac{1}{M}\left|\sum_{n=1}^M z_n\right|^2\Bigg)
  \sum_{n'=1}^M
  \left(\frac{z_{n'}}{z}+\frac{\bar{z}_{n'}}{\bar{z}}\right)
\Bigg] |z|^{-4}
+ \mathcal{O}\left(|z|^{-6}\right) \ , \nonumber
%\frac{1}{M m_\kappa}\left[
%\sum_{n=1}^{M}\left(|z_n|^2+2|c_n|^2\right)
%-\frac{1}{M}\left|\sum_{n=1}^M z_n\right|^2\right] |z|^{-4}
%+ \mathcal{O}\left(|z|^{-5}\right) \ .
\end{align}
The solution $\delta\psi$ is similar to that of the fractional case
(\ref{eq:psifluc_frac}) now with a dependence on $c_m$. Interestingly,
it was already noted in Eq.~(\ref{eq:chifluc_frac}) that $\chi_m$ in
the ``purely'' fractional case is kind of ``local'' and in fact we now
observe the power-behaved tail depending only on $c_m$. This is
consistent with the results of \cite{Eto:2008qw,Eto:2009bz}, viz.~the 
lump solution becomes singular if any of $c_m$ vanishes. We can also
see that if the fractional vortex centers coincide, $z_n=z_0, \,
\forall n$, then the solution (\ref{eq:psifluc_frac-semilocal})
coincides to lowest order with that of the purely semilocal case,
Eq.~(\ref{eq:psifluc_semilocal}) by identification of $|c|^2$ and
$\sum_{n=1}^M|c_n|^2/M$. 

\begin{figure}[!p]
\begin{center}
\includegraphics[height=2.9cm]{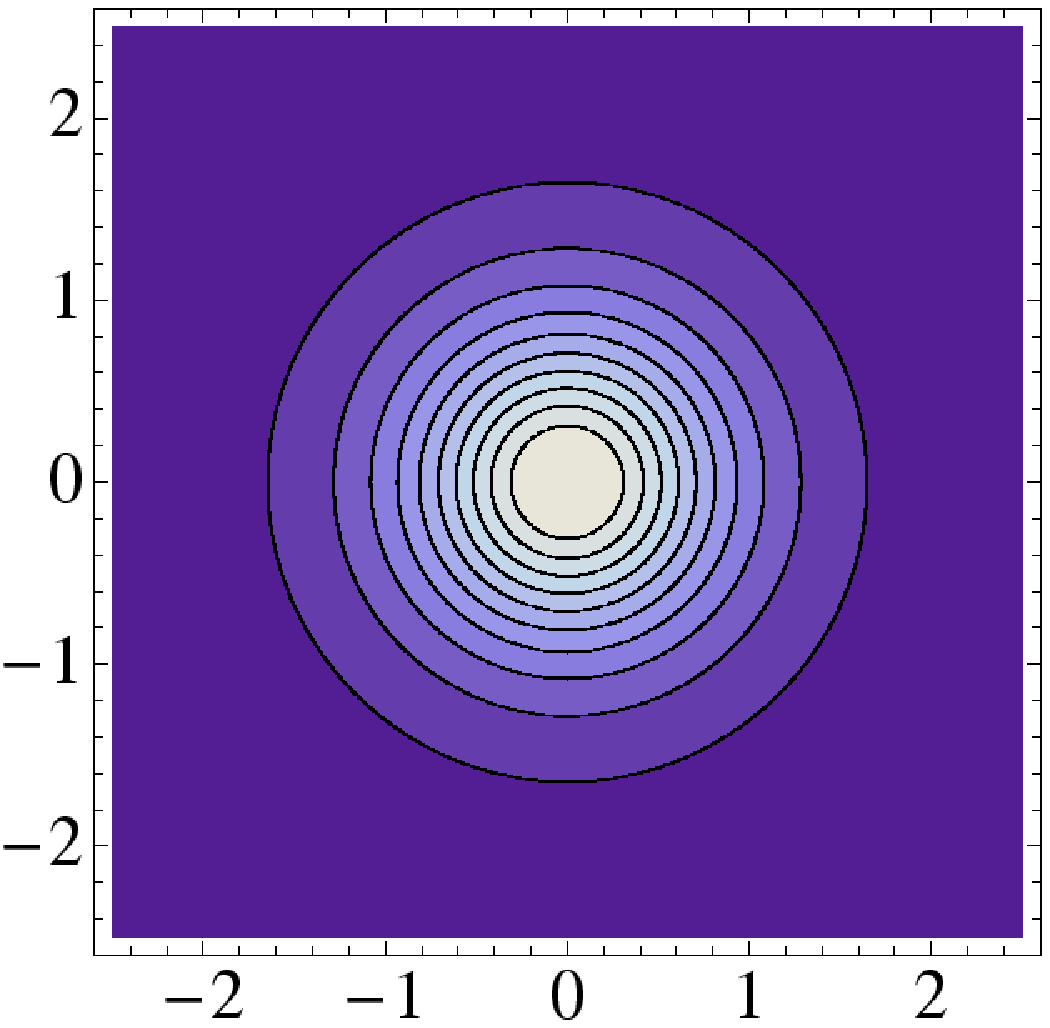}
\includegraphics[height=2.9cm]{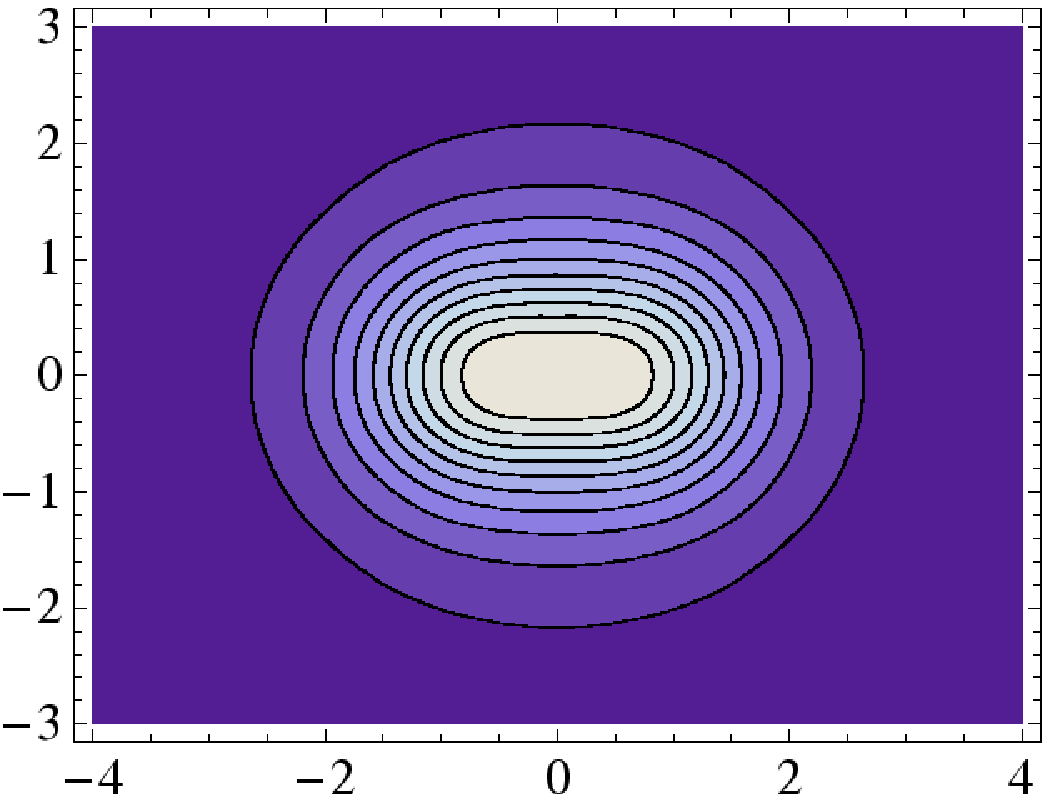}
\includegraphics[height=2.9cm]{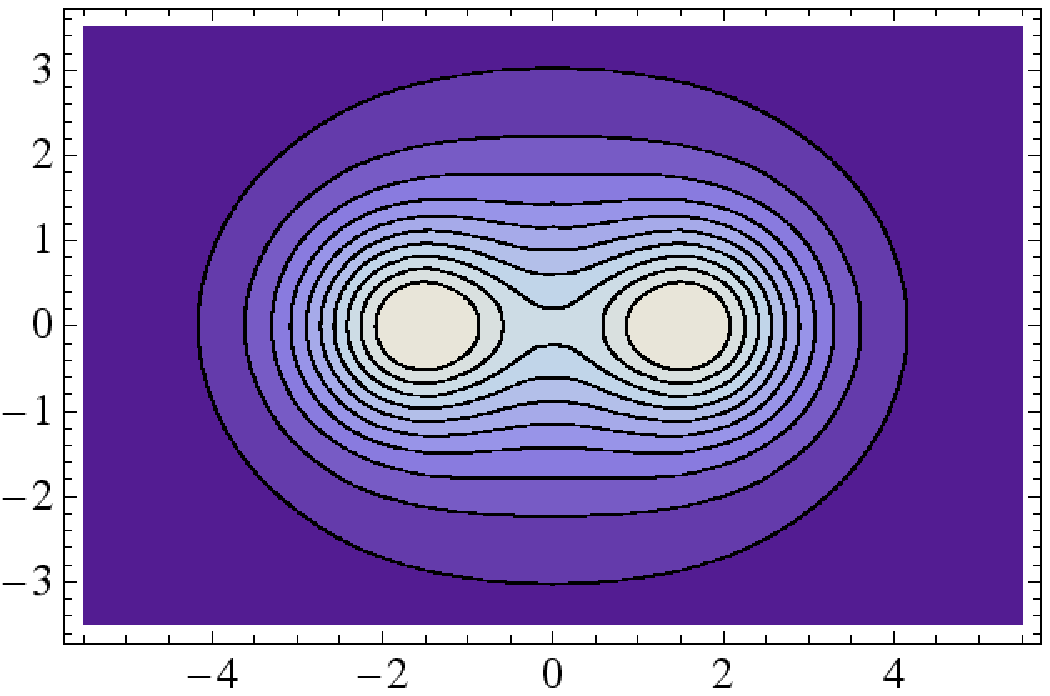}
\includegraphics[height=2.9cm]{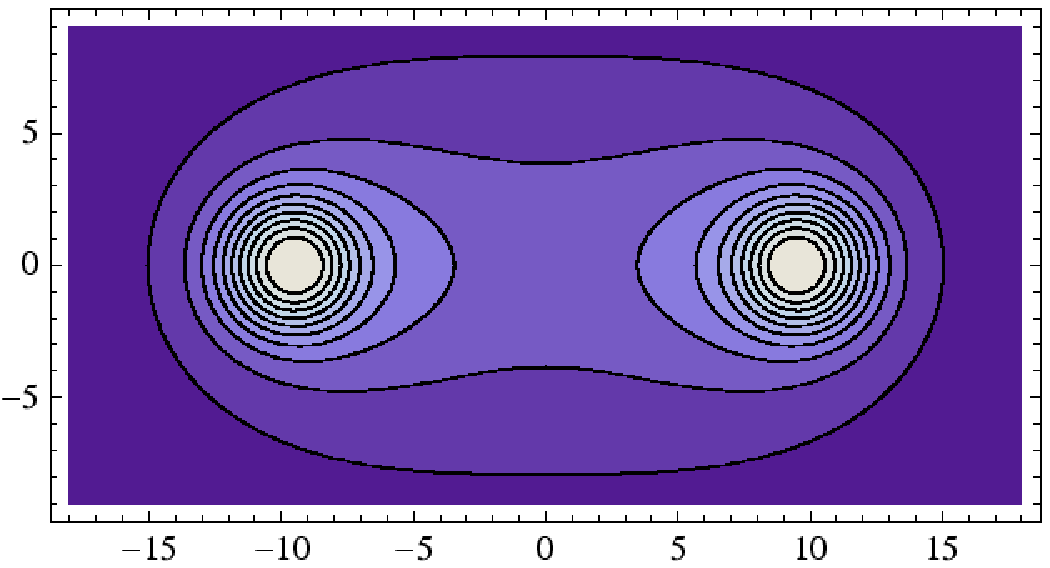}\\
\includegraphics[height=2.9cm]{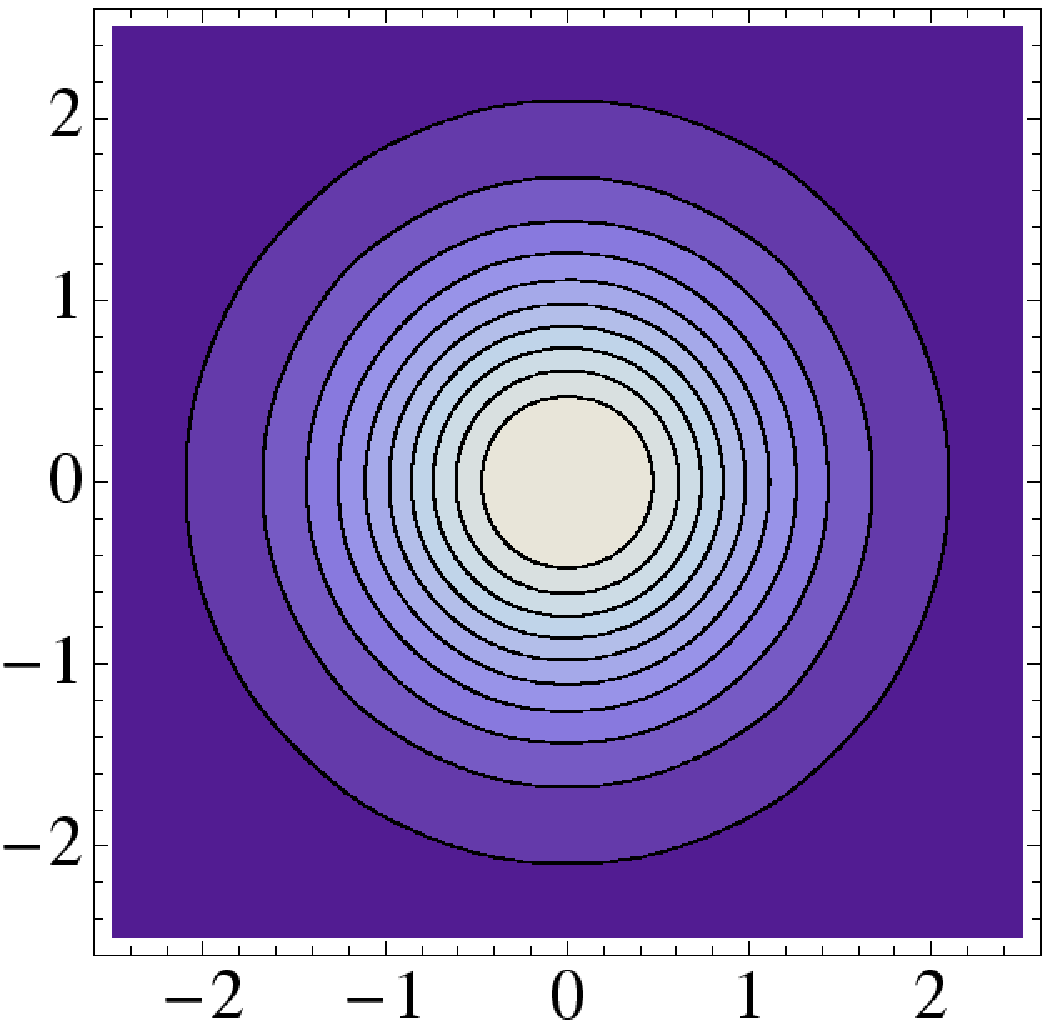}
\includegraphics[height=2.9cm]{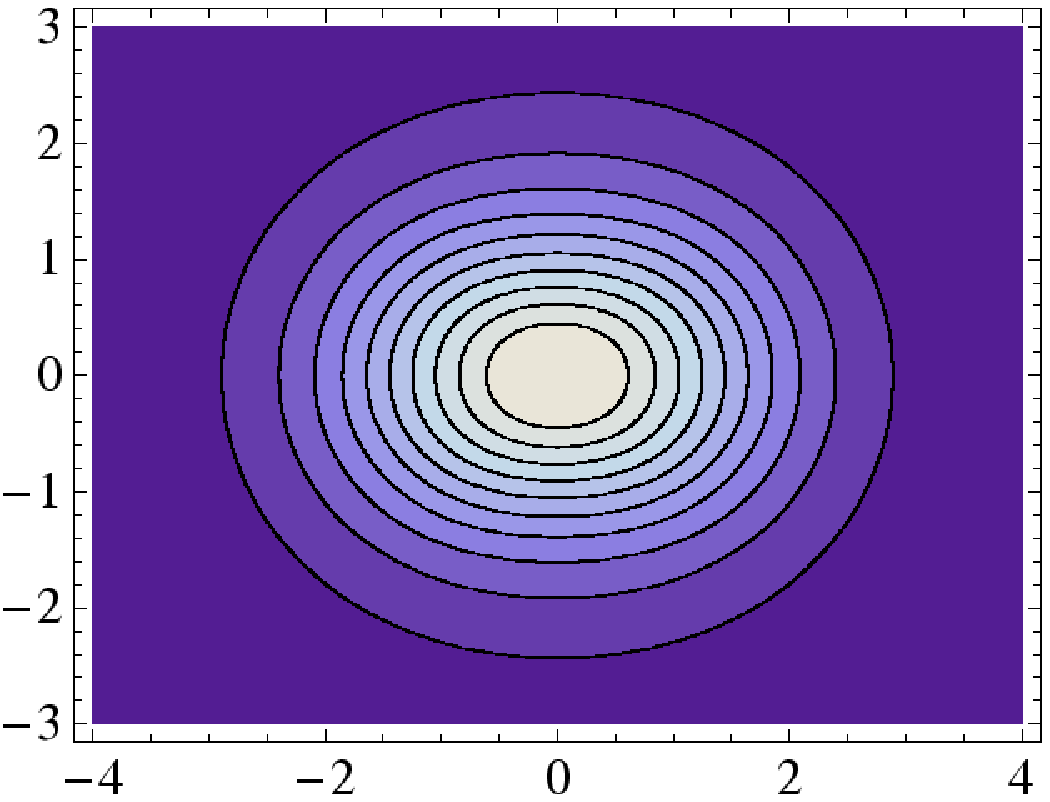}
\includegraphics[height=2.9cm]{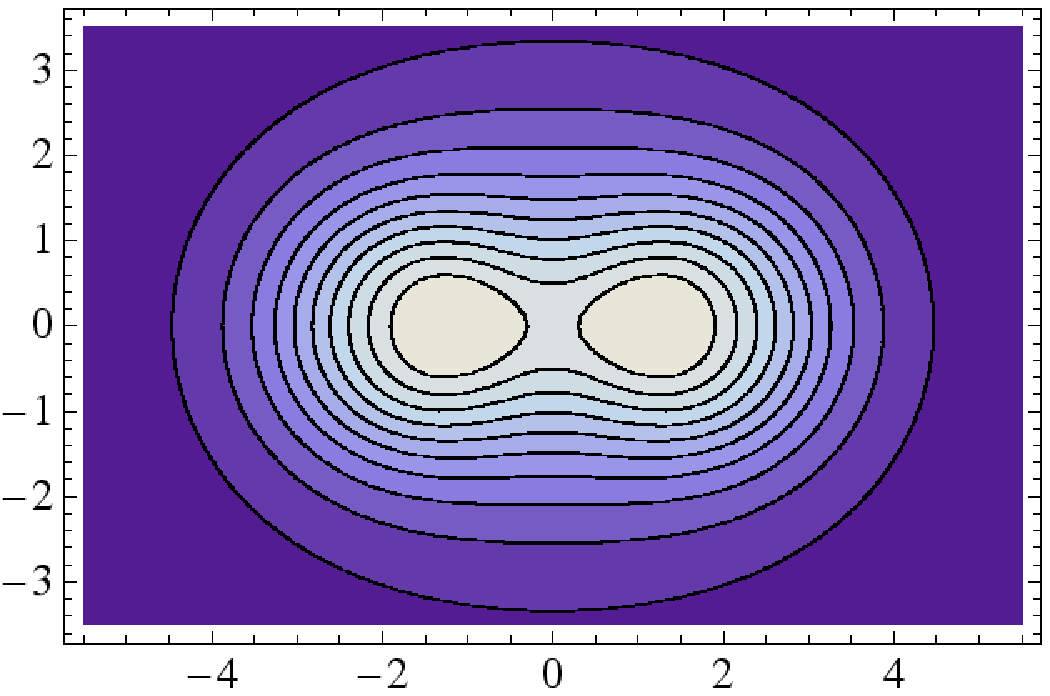}
\includegraphics[height=2.9cm]{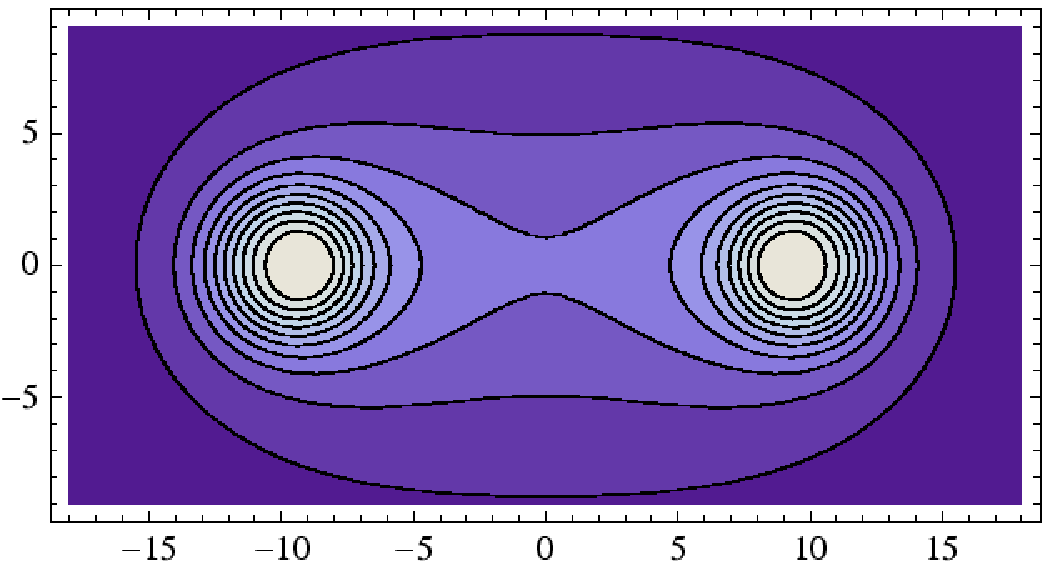}\\
\includegraphics[height=2.9cm]{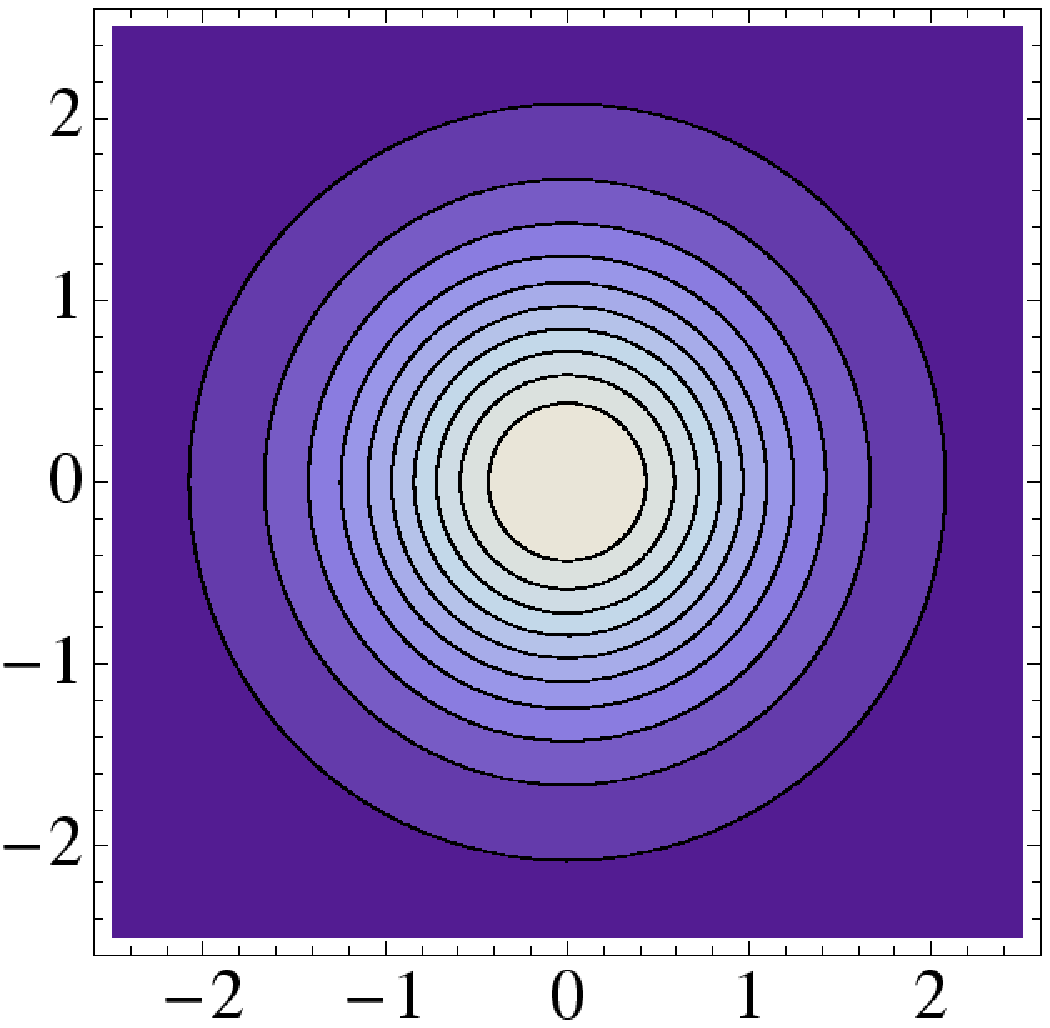}
\includegraphics[height=2.9cm]{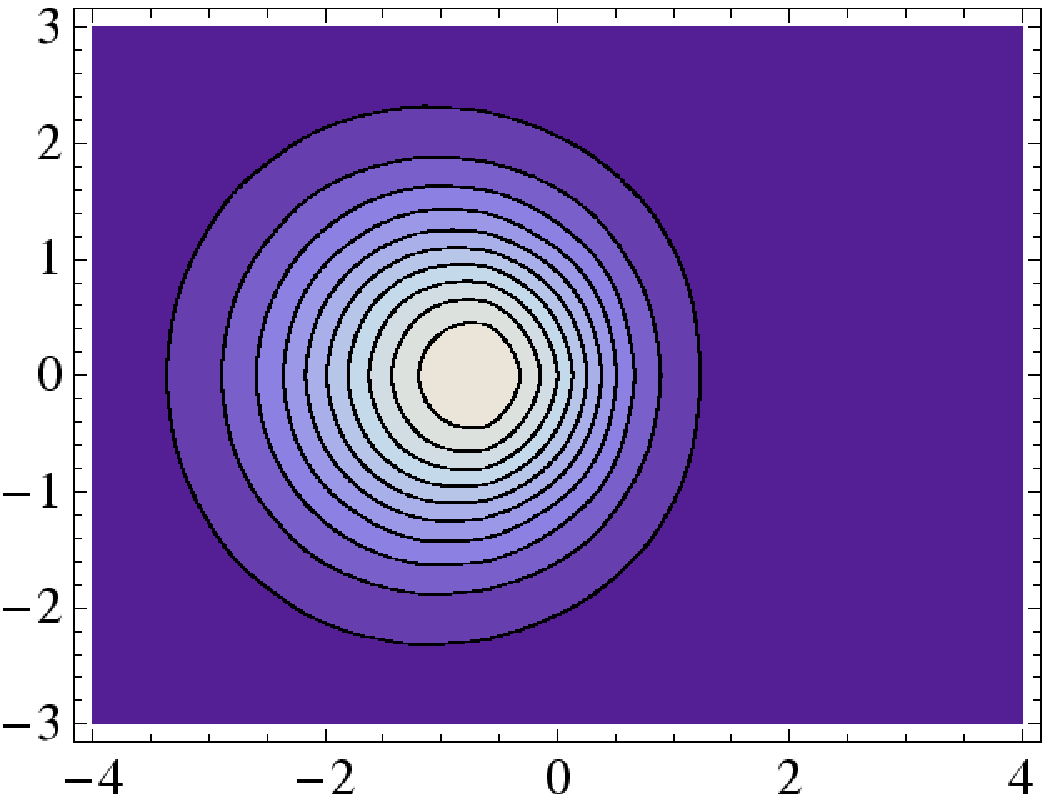}
\includegraphics[height=2.9cm]{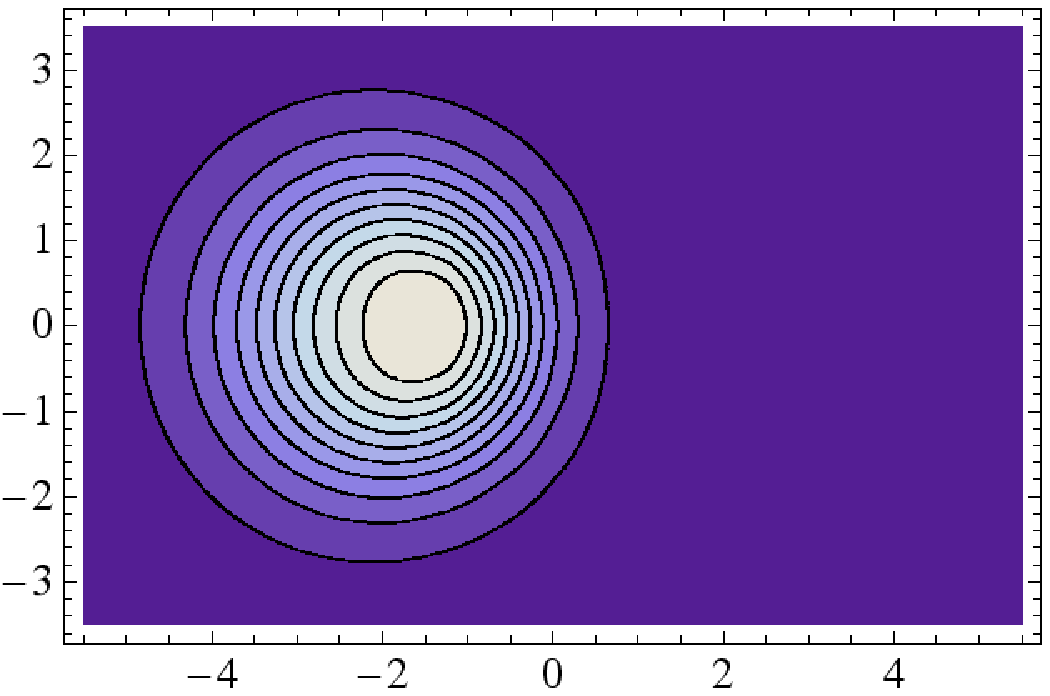}
\includegraphics[height=2.9cm]{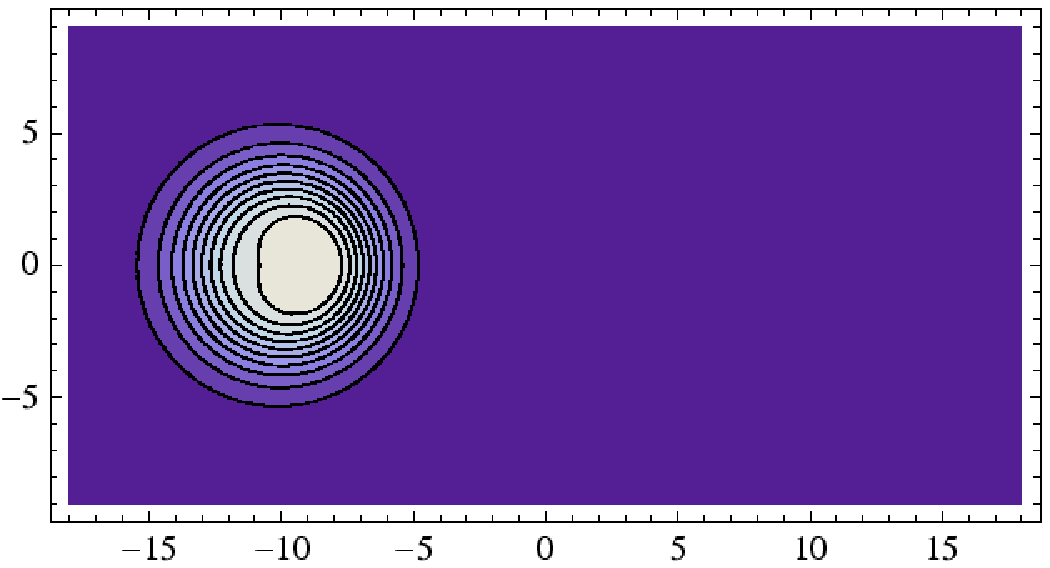}\\
\includegraphics[height=2.9cm]{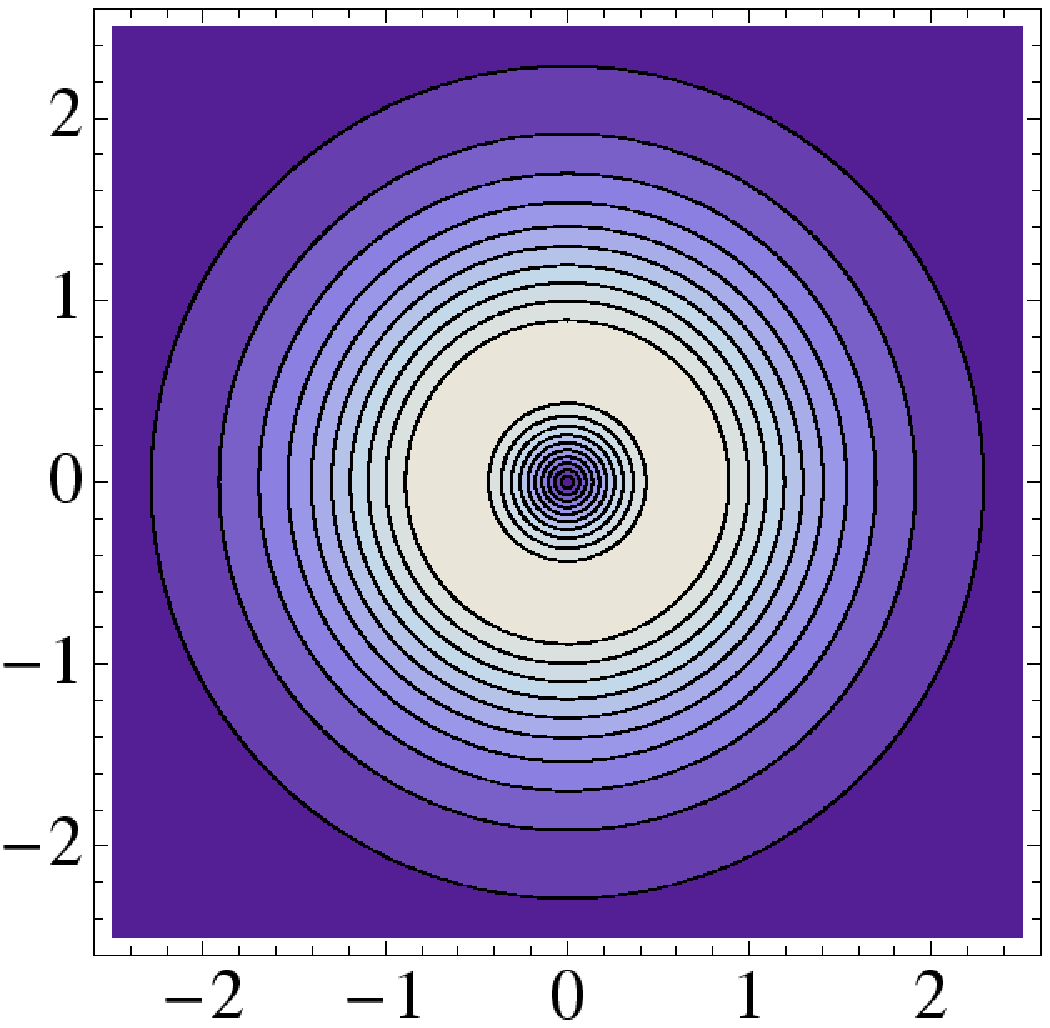}
\includegraphics[height=2.9cm]{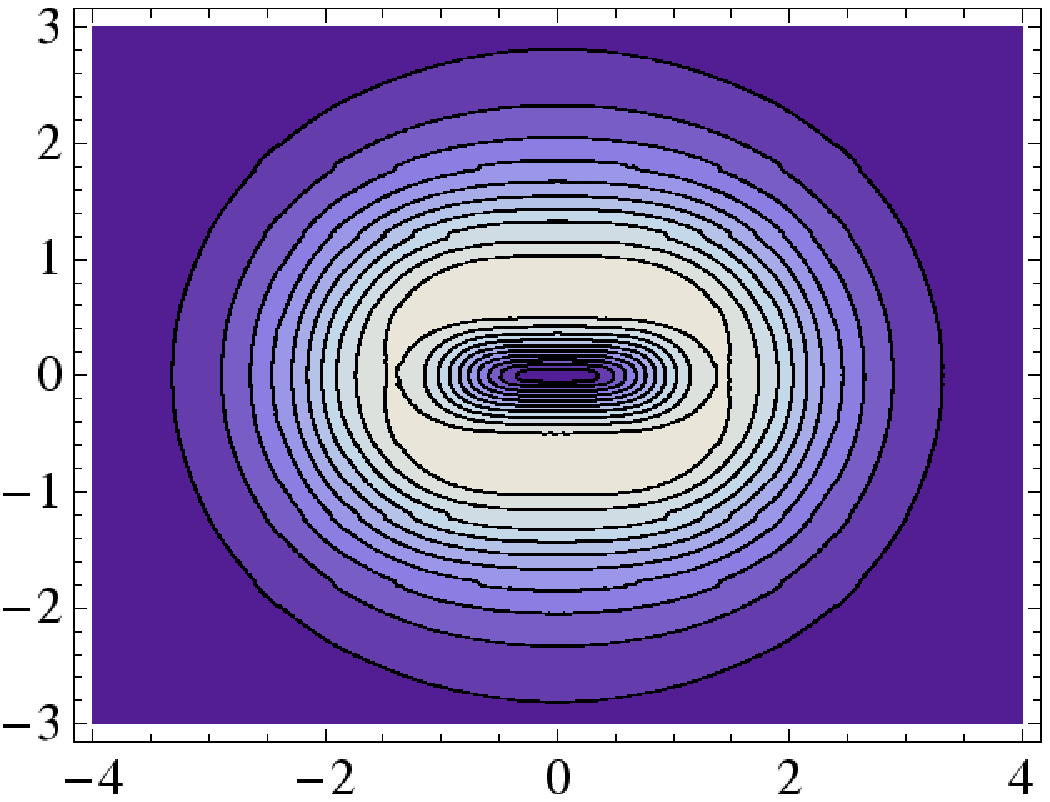}
\includegraphics[height=2.9cm]{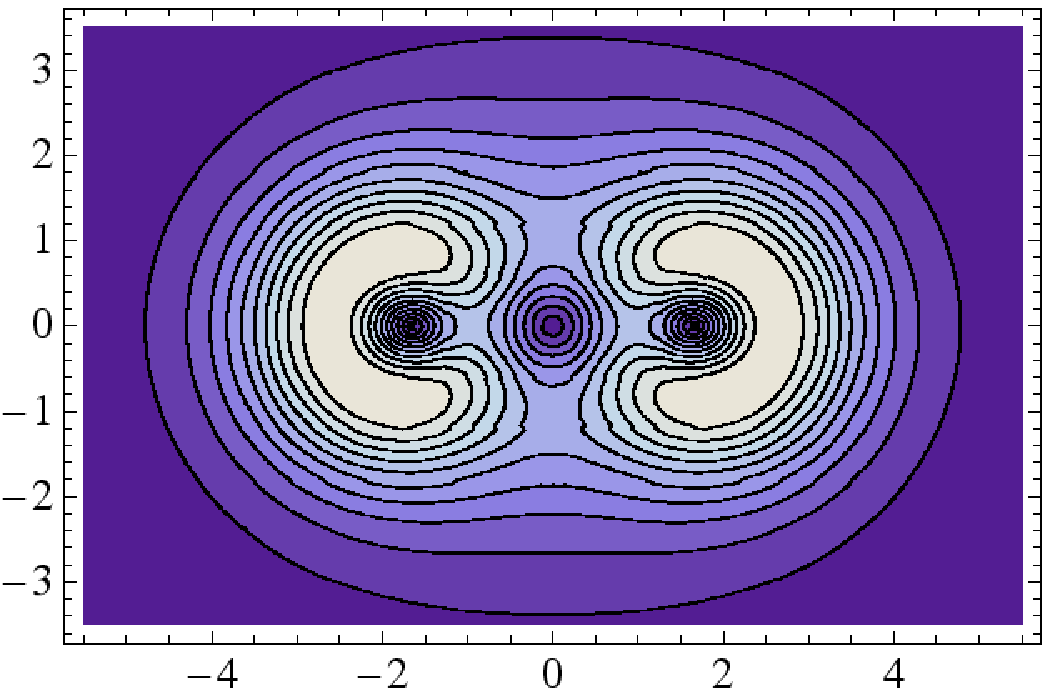}
\includegraphics[height=2.9cm]{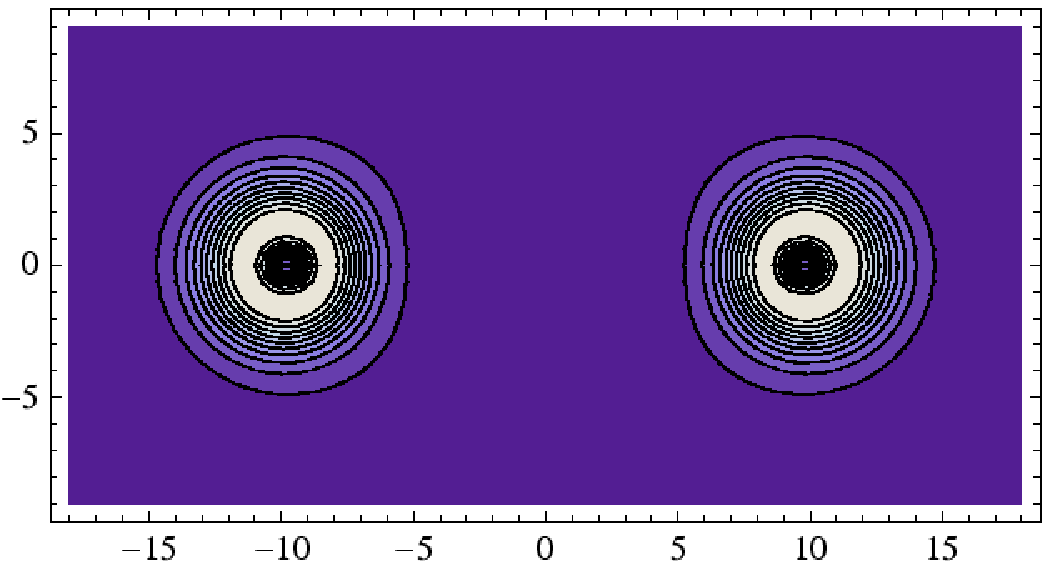}\\
\includegraphics[height=2.9cm]{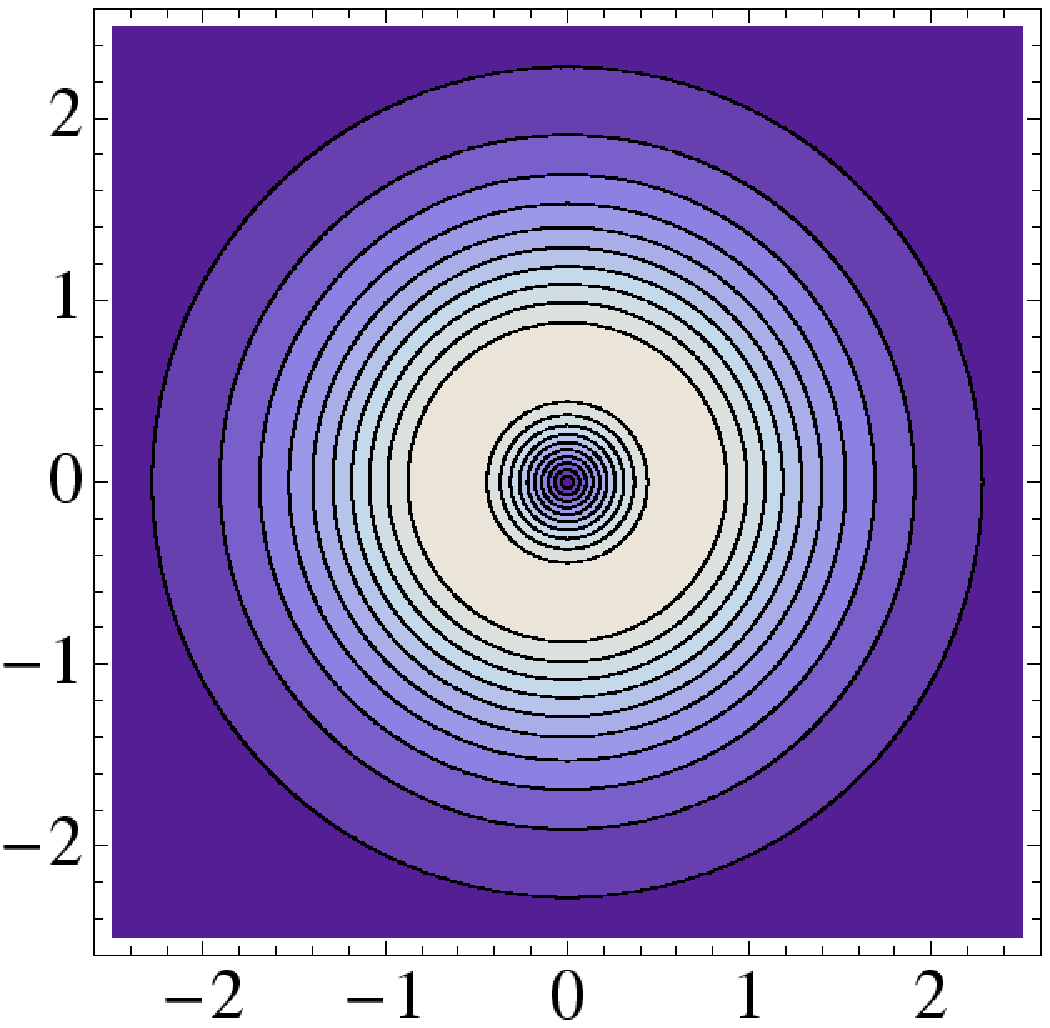}
\includegraphics[height=2.9cm]{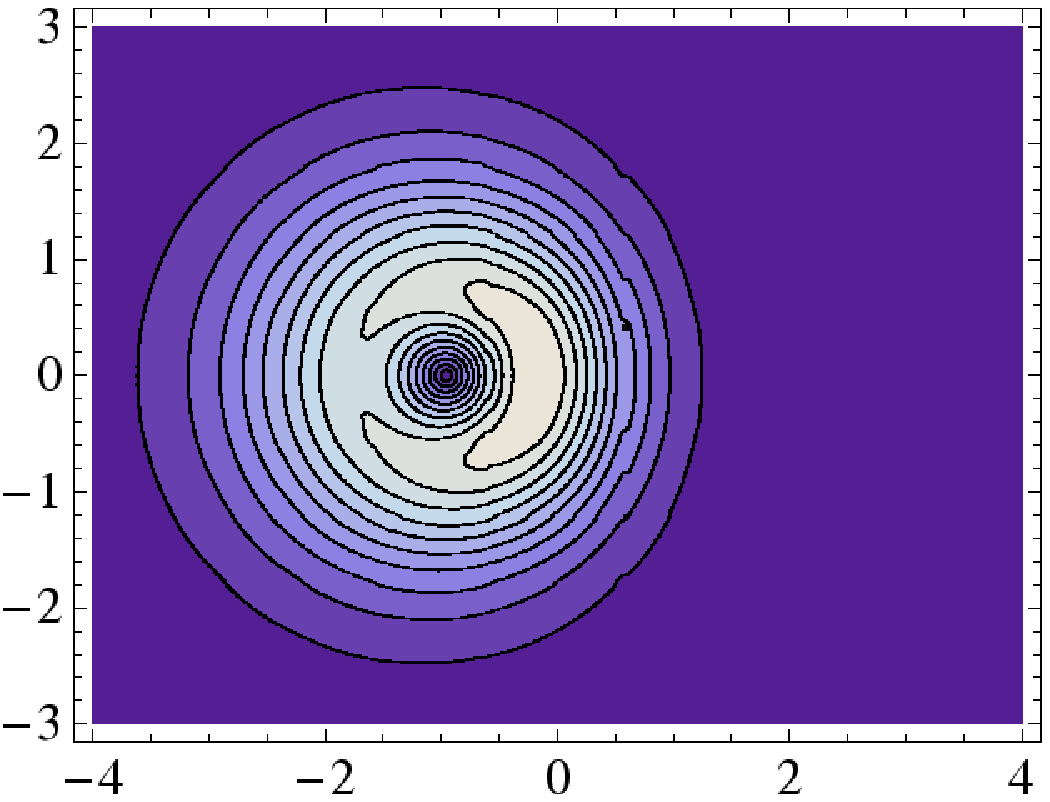}
\includegraphics[height=2.9cm]{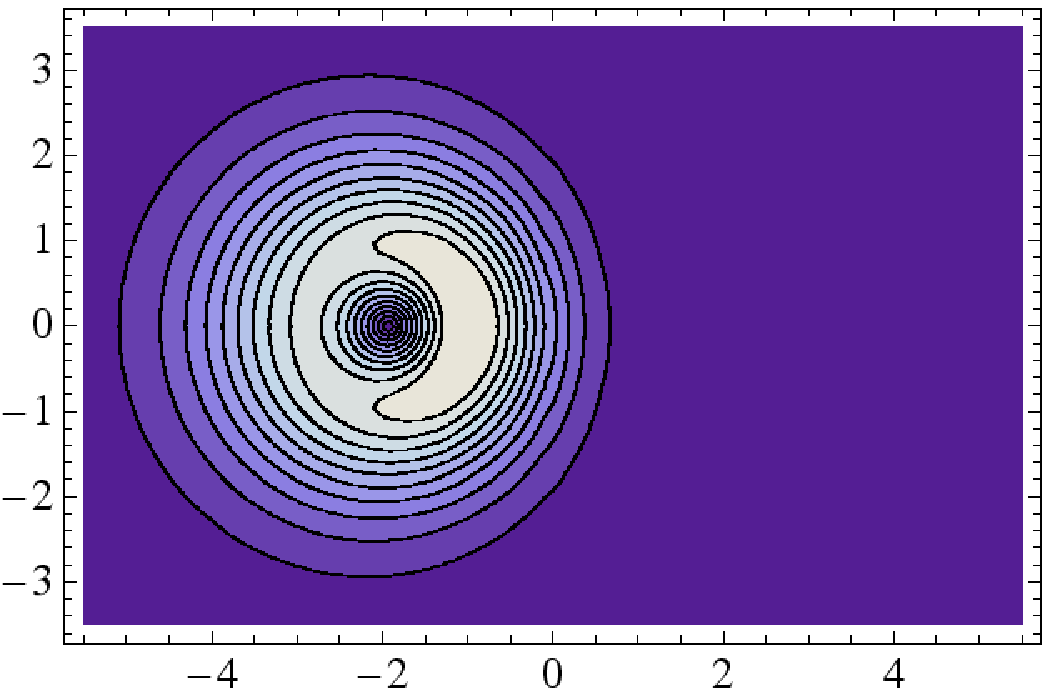}
\includegraphics[height=2.9cm]{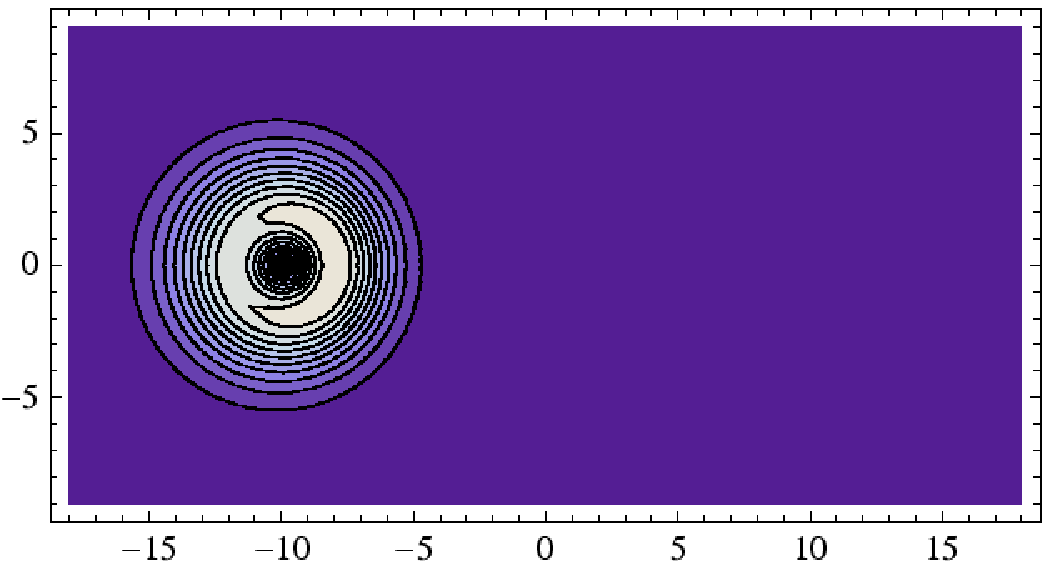}
\end{center}
\caption{The non-Abelian Chern-Simons semi-local fractional vortex
  with $G'=SO(4)$ and $G'=USp(4)$ for $\kappa=\mu=2$ with sizes
  $c_1=c_2=1$, where the rows of the figure
  correspond to the energy density, the Abelian magnetic flux
  $F_{12}^0$, the non-Abelian  magnetic flux $F_{12}^2$, the magnitude
  of the Abelian electric field $|E_i^0|$ and finally the magnitude of
  the non-Abelian electric field $|E_i^2|$, whereas the columns of the
  figure correspond to the separation distance $2d=2\{0,1,2,10\}$. We
  have set $\xi=2$. } 
\label{fig:frac-semilocal_equalcouplings}
\end{figure}

Let us close this section with showing this semi-local fractional
vortex in Fig.~\ref{fig:frac-semilocal_equalcouplings} for equal
Chern-Simons couplings $\kappa=\mu=2$ and size parameters 
$c_1=c_2=1$. The graph is a matrix as in the previous sections with
the columns representing the energy density, the Abelian magnetic flux
$F_{12}^0$, the non-Abelian magnetic flux $F_{12}^2$, the magnitude of
the Abelian electric field $|E_i^0|$ and finally the magnitude of the
non-Abelian electric field $|E_i^2|$, whereas the rows represent
the relative distance $2d=2\{0,1,2,10\}$. 
The difference between this semi-local fractional vortex with equal
couplings and is purely fractional counterpart (i.e.~with $c=0$) is
obviously quite big. The magnetic fluxes have a completely different
nature for vanishing relative distance $d=0$, whereas the difference is
almost absent for large $d$. On the other hand, there is seemingly not
much difference between the pure fractional vortex ($c=0$) with
opposite couplings $-\kappa=\mu$ and the semi-local fractional vortex
with equal couplings, except some differences in the electric field
densities. We will not exhaust the reader with further graphs but just
mention what happens by changing the individual sizes $c_n$. Each
sub-peak, when well separated has the size controlled by its
corresponding size modulus.

\section{Discussion\label{sec:disc}}

In this paper we have studied the fractional as well as semi-local
Chern-Simons vortices in $G=U(1)\times SO(2M)$ and $G=U(1)\times
USp(2M)$ theories. It is known that the fractional-vortex positions
are semi-local moduli and it is shown that unless they are all
coincident (local case) the ring-like flux structure, characteristic
of Chern-Simons vortices, will become bell-like fluxes -- just as
those of the standard Yang-Mills vortices. The asymptotic profile
functions have been calculated and it is shown that the vortex becomes
(a kind of) semi-local for non-coincident positions of the fractional
vortices. 
The calculation was repeated for the purely semi-local vortex
(i.e.~with non-vanishing size modulus but coincident zeroes of the
squark fields),
which however differs in the fact that the non-Abelian profile
function $\chi$ approaches its VEV exactly as the Abelian profile
function $\psi$, whereas in for the fractional vortex the non-Abelian
profile function $\chi$ remains approximately local. Note however,
that in the classic way of describing the squark fields with a matrix
$q = \diag(f,g)$, we should identify the profile functions as
\beq f^2 = \diag \left(|z-z_1|^2e^{-\psi-\chi_1},\ldots, 
|z-z_M|^2e^{-\psi-\chi_M}\right) \ , \quad
g^2 = \diag\left(e^{-\psi+\chi_1},\ldots,e^{-\psi+\chi_M}\right) \ ,
\eeq
hence it suffices that just $\psi$ has a power-law behavior in order
for $f,g$ to both have it as well. 

In order to study the fractional vortices we have solved the (BPS)
vortex equations (master equations) numerically in the plane, as the
rotational symmetry obviously is lost when the fractional position
moduli are non-coincident. 
A peculiar effect of the non-Abelian Chern-Simons vortices was
discovered in Ref.~\cite{Gudnason:2009ut} which is observed only for
different Chern-Simons couplings $\kappa\neq\mu$ where the local
vortex possesses a negative (positive) Abelian and positive (negative)
non-Abelian magnetic flux for $\kappa>\mu$ ($\kappa<\mu$). This effect
is destroyed by the semi-local size moduli which is easily seen from
the lump solution that has both the Abelian and the non-Abelian flux
concentrated at the origin in same amounts. The fractional vortices
have the same fate for large relative distance as each fractional peak
can be thought of as an effective semi-local vortex having power-like
tails in its profile functions. 

Comparing the asymptotic expansions
(\ref{eq:chiflucsol_semilocal})-(\ref{eq:psiflucsol_semilocal}) of the
semi-local vortex to the asymptotic expansions of the semi-local
fractional vortex
(\ref{eq:chifluc_frac-semilocal})-(\ref{eq:psifluc_frac-semilocal}),
we can identify the effective size of a $k=1$ semi-local fractional
Chern-Simons vortex with sizes $c_n$
\begin{align}
|c_{\rm effective}|^2 =
\frac{1}{2M}\sum_{n=1}^M\left(|z_n|^2+2|c_n|^2\right) 
-\frac{1}{2M^2}\left|\sum_{n=1}^M z_n\right|^2 \ ,
\end{align}
valid for the Abelian fields, while the non-Abelian fields have the
effective size $c_n$. If we now set $c_n$ to zero on the right hand
side, we see the effective size of the Abelian field explicitly in
terms of the fractional vortex centers $z_n$ (i.e.~the case of the
pure fractional Chern-Simons vortex). Note that the next-to-leading
order is not radially symmetric. 

In order to understand this type of fractional vortex better, one
should understand the second homotopy group of the moduli space of
vacua in the theory at hand \cite{Eto:2008qw}. This has not been
pursued in this paper and remains as a future task.  

The substructures of the non-Abelian vortices are very
interesting. For many reasons it would be very interesting to study
the non-Abelian vortices in a symmetric phase (i.e.~the Chern-Simons
phase) -- these vortices are known as non-topological vortices. They
have been studied in
Refs.~\cite{Khare:1990bd,Bazeia:1990ee,Khare:1991zt,Bazeia:1991ee} in
the Abelian case and are still not very well-understood in the
following sense. Their moduli have not been written down
explicitly. Their existence has been proven in
Ref.~\cite{Spruck:1992yy} and their topology on compact manifolds has
been studied in Ref.~\cite{Horvathy:1999qj}. To the best of our
knowledge, the closest attempt to classify the moduli has been done by
Lee-Min-Rim in Ref.~\cite{Lee:1991td} where the non-topological moduli
have been written down in the asymptotic profile functions. The
generalization of these studies to the non-Abelian case would be quite
interesting.

\subsection*{Acknowledgments}

The author would like to thank Minoru Eto, Kenichi Konishi and Walter
Vinci for discussions. 

\appendix
\section{Review of the Abelian semi-local
  vortex\label{app:abeliansemilocalreview}} 

Let us recall the asymptotic behavior of the Abelian $G=U(1)$
semi-local vortex. The master equation is
\begin{align}
\bar{\p}\p\psi = - \frac{m^2}{4\xi}
\left[H_0(z)H_0^\dag(\bar{z})e^{-\psi} - \xi\right] \ ,
\end{align}
with $m\equiv \sqrt{\xi}e$.
We can formally calculate the lump solution by sending the mass (the
gauge coupling constant $e$) to infinity and obtain 
\beq \psi_{\infty} =
\log\left[\frac{1}{\xi}H_0(z)H_0^\dag(\bar{z})\right] \ . \eeq  
The lump solution is in the vacuum manifold and can for a finite lump
size be considered as the (approximate) long distance behavior of the 
vortex fields. It is however singular in the local vortex case
(i.e.~the small lump singularity).
If we now consider a small fluctuation around the lump solution
setting $\psi = \psi^{\infty} + \delta\psi$ we have
\begin{align}
\bar{\p}\p\delta\psi + \bar{\p}\p\psi^{\infty} &=
\frac{m^2}{4}\delta\psi \ ,
\end{align}
and we can identify two cases. Since the vacuum manifold is K\"ahler,
the lump solution can formally be a K\"ahler transformation of the
VEV, in which case the lump solution is singular, however the
Laplacian will vanish (apart from some delta functions at each zero of
the squark fields). This is the local vortex case and we can also
calculate the asymptotic profile function as 
\beq \delta\psi = c_{\rm local} K_0(m |z|) \ . \eeq
In semi-local case, in contradistinction, we observe that the
fluctuation will have a power-law behavior. 
For simplicity we consider $\NF=2$ with the following moduli matrix 
\beq H_0(z) = \begin{pmatrix} z & c \end{pmatrix} \ , \eeq
for which the fluctuation equation reads
\begin{align}
\bar{\p}\p\delta\psi + \bar{\p}\p\log\left[|z|^2 + |c|^2\right] &= 
\frac{m^2}{4}\delta\psi \ .
\end{align}
Since the vacuum manifold is K\"ahler, we make a K\"ahler
transformation yielding
\begin{align}
\bar{\p}\p\log\left[|z|^2 + |c|^2\right]
=\bar{\p}\p\log\left[1 + \frac{|c|^2}{|z|^2}\right] 
&\simeq \bar{\p}\p\left[\frac{|c|^2}{|z|^2} - \frac{|c|^4}{2|z|^4} +
    \mathcal{O}\left(|z|^{-6}\right)\right]\ ,
\end{align}
where we are calculating the asymptotic function valid when
$|z|\gg|c|$. Then we find using a series expansion
\begin{align}
\delta\psi &= 
\sum_{n=2}^{\infty} b_n |z|^{-2n} =
\frac{4|c|^2}{m^2}|z|^{-4}
+\left(\frac{64|c|^2}{m^4}-\frac{8|c|^4}{m^2}\right)|z|^{-6} 
+\mathcal{O}\left(|z|^{-8}\right) \ . 
\end{align}
These are of course well-known facts about the Abelian semi-local
vortex solution (see
Refs.~\cite{Vachaspati:1991dz,Achucarro:1992hs,Achucarro:1999it}).

\end{document}